\documentclass[reprint]{revtex4-1}
\usepackage{blindtext}
\usepackage{amsmath, amssymb, graphicx, mathtools, multirow,makecell,subfigure,color,float,stmaryrd}

\makeatletter
\renewcommand{\p@subsection}{}
\renewcommand{\p@subsubsection}{}
\makeatother

\usepackage{hyperref}
\hypersetup{
	colorlinks=true,       
	linkcolor=red,          
	citecolor=blue,        
	filecolor=magenta,      
	urlcolor=blue           
}

\begin{document}
	\title{Emergence of Tension Chains and Active Force Patterning}
	
	\author{Ayan Roychowdhury}
	\thanks{Joint first author}
	\affiliation{Simons Centre for the Study of Living Machines, National Centre for Biological Sciences-TIFR, Bengaluru, India  560065.}
	\author{Saptarshi Dasgupta}
        \thanks{Joint first author}
	\affiliation{Simons Centre for the Study of Living Machines, National Centre for Biological Sciences-TIFR, Bengaluru, India  560065.}
	\author{Madan Rao}
	\thanks{rao.madan@gmail.com}
	\affiliation{Simons Centre for the Study of Living Machines, National Centre for Biological Sciences-TIFR, Bengaluru, India  560065.}
	

\begin{abstract}
Viewed under a fluorescence microscope, the actomyosin cytoskeleton presents vivid streaks of lines together with persistent oscillatory waves. 
Using an active hydrodynamic approach, we show how
a uniform distribution of single or mixture of contractile stresslets spontaneously segregate, followed by the formation of singular structures of high contractility (\textit{tension chains}) in finite time. Simultaneously, the  collection of stresslets exhibit travelling waves and swapping as a consequence of nonreciprocity.
In the finite geometry of the cell, the collection of active tension chains can form an active web held together by specific anchoring at the cell boundary.
On the other hand, preferential wetting at the 
cell boundary can reinforce active segregation in a mixture of stresslets leading to stratification.
\end{abstract}

\maketitle

\section*{Introduction}

The 
distribution of  forces across the scale of the cell is dynamically templated by
the active cytoskeleton, in particular the cell spanning assemblies of a variety of myosins together with actin filaments and their crosslinkers \cite{PollardGoldman2016}.
 Cytoskeletal organisation and remodelling give the cell its dynamical shape and form~\cite{Tanejaetal2020}, as well as its adaptive mechanical 
 response~\cite{Matthewsetal2006,banerjeeetal2020}.
In addition, it sets up a global scaffold for the patterning of mesoscale condensates~\cite{Brangwynne2021} and the relative positioning of subcellular organelles~\cite{Marshall2020}, such as the centrosome position~\cite{jimenezetal2021} and nuclear localization~\cite{Makhija2015,Sunetal2020}.

The active cytoskeleton is put together by the nonequilibrium self-assembly of its active components, that both exert and sense forces,  the latter via their strain dependent turnover  \cite{Kovacsetal2007, Fernandez-Gonzalezetal2009,Mullaetal2022}.
We will refer to these units of mechanotransduction as {\it stresslets}; the spatiotemporal patterning of these stresslets will then mark the patterning of forces. High resolution images 
show distinct actomyosin line patterns (stress fibres) \cite{HotulainenLappalainen2006,Bershadsky2017,Weissenbruchetal2021} and web-like structures \cite{Bershadsky2017,Weissenbruchetal2021}, with different stresslet species displaying different cellular localisations \cite{Beachetal2014,Weissenbruchetal2021}.
More recently, there have been systematic studies of spatial patterning of cytoskeletal structures within cells and tissues~\cite{Vicente-Manzanaresetal2008},
  {\it in vitro} 
 reconstitutions~\cite{silvaetal2011,koenderink-paluch2018}, and in cell extracts on
 micropatterned substrates~\cite{jimenezetal2021,zhangetal2021}.
{\it In vivo}, these structures appear to coexist with persistent waves of oscillation reminiscent of an excitable system \cite{zallen2011,munjal2015,debsankar2017,Lehtimakietal2021}. A physical basis for 
these pervasive phenomena is lacking, as is a dynamical theory for the establishment of the cellular framework, the emergent force patterning and their homeostatic response \cite{Weissenbruchetal2021}.

Here we address these issues using an active hydrodynamic description \cite{Marchettietal2013} for single and mixtures of contractile stresslets, such as myosin IIA and IIB  \cite{Bershadsky2017,Shutovaetal2017,Weissenbruchetal2021}, on an (actin) elastomer, while allowing 
turnover \cite{BanerjeeLiverpoolMarchetti2011,BanerjeeMarchetti2011,debsankar2017}. 
This is the natural setting for experiments such as \cite{Bershadsky2017}.
We next extend the analysis to the case where the cell background is a fluid at long time scales.
In both situations, a slight difference between the contractile activities or turnover rates of the individual stresslets, leads to spontaneous segregation -- stresslets with the higher contractile activity come together, resulting in force   patterns at macroscopic scales much larger than the scale of the stresslets. Unlike conventional segregation driven by gradients in chemical potential \cite{Bray2002}, the spontaneous segregation of the stresslets is driven by an effective elastic stress relaxation. The breaking of time reversal symmetry (TRS) leads to a non-Hermitian dynamical matrix which exhibits striking nonreciprocal features \cite{Fruchartetal2021} such as exceptional points \cite{Kato1984} which presage the travelling wave \cite{Youetal2020} and swap phases \cite{Fruchartetal2021}. At later times, the linearly segregated configurations evolve into well separated, singular structures of enhanced contractility, in striking departure from conventional coarsening. We derive the scaling behaviour of these finite time singular structures \cite{eggersfontelos-book2015} and verify them with careful numerics. The amplification of 
 contractile stresses along singular tension lines recalls the study in~\cite{Roncerayetal2016,Roncerayetal2019}. We find that these tension lines can be static or moving; we derive equations for their mass and force balance and analyse conditions for their merger and phantom crossings. In the finite geometry of the cell, more complex active webs of these tension lines can be shaped and stabilised by cell geometry and cell surface anchors. This is reminiscent of the patterning of actomyosin networks 
 between cadherin-mediated adherens junctions (AJ) and the integrin-mediated focal adhesion (FA),
 with properties that depend on the geometry of anchoring~\cite{GuptonWaterman-Storer2006, GeigerSpatzBershadsky2009}.
Finally,  when coupled with preferential wetting to substrates, such as the cell membrane, this provides a driving force for the differential cellular localisations and stratification of a mixture of contractile stresslets \cite{Vicente-Manzanaresetal2008,Weissenbruchetal2021}.


\section*{Hydrodynamic Equations}

We start with a description of the cellular background as an
	elastomer of mass density $\rho_a$ embedded in the cytosol, whose displacement relative to an unstrained reference state is $\boldsymbol{u}$.
The hydrodynamic  linear momentum balance of the elastomer is, 
	$\rho_a\,\ddot{\boldsymbol{u}}+\Gamma\,\dot{\boldsymbol{u}}=\nabla\cdot\boldsymbol{\sigma}$, where $\Gamma$ is the friction of the elastomer with respect to the fluidic cytosol (whose dynamics we neglect since the volume fraction of the mesh is high), and $\boldsymbol{\sigma}(\rho_a,\{\rho_i\})$ is the total stress in the elastomer, dependent on 
	the densities $\rho_i$ of the bound stresslets.
	

The constitutive equation for the total stress $\boldsymbol{\sigma}$ in the long time limit is the summation of elastic stress $\boldsymbol{\sigma}^e$, the viscous stress $\boldsymbol{\sigma}^d$, and active stress $\boldsymbol{\sigma}^a$:  $\boldsymbol{\sigma}=\boldsymbol{\sigma}^e+\boldsymbol{\sigma}^d+\boldsymbol{\sigma}^a$~\cite{BanerjeeLiverpoolMarchetti2011,BanerjeeMarchetti2011,debsankar2017}. The elastic stress $\boldsymbol{\sigma}^e = \frac{\delta F}{\delta \,\boldsymbol{\epsilon}}$ associated with
the linearized strain $\boldsymbol{\epsilon}:=(\nabla \boldsymbol{u}+\nabla \boldsymbol{u}^T)/2$,
comes from a  free-energy functional  $F(\boldsymbol{\epsilon},\rho_a)
	=\int d^2r f_B$, describing an isotropic, linear elastic material, $f_B=\frac{B}{2}\epsilon^2+\mu\,\text{tr}\,\tilde{\boldsymbol{\epsilon}}^2+C\,\delta\rho_a\,\epsilon+\frac{A}{2}\delta\rho_a^2$,
	where $\epsilon:=\text{tr}\,\boldsymbol{\epsilon}$ and $\tilde{\boldsymbol{\epsilon}}:=\boldsymbol{\epsilon}-(1/d)(\text{tr}\boldsymbol{\epsilon})\mathbf{I}$ are the isotropic and deviatoric strain, respectively, $B>0$ and $\mu>0$ are the passive bulk and shear moduli, respectively, $\delta\rho_a:=\rho_a-\rho_a^0$ is the local deviation of $\rho_a$ from its state value $\rho_a^0$, and $A^{-1}$ is the isothermal compressibility at constant strain.
	The passive viscous stress of the elastomer is 
	$\boldsymbol{\sigma}^d
	=\eta_b\,\dot{\epsilon}\,\mathbf{I}+2\eta_s\,\dot{\tilde{\boldsymbol{\epsilon}}}$, where $\eta_b$ and $\eta_s$  are the  bulk and shear viscosities, respectively.
We take the active contractile stress  $\boldsymbol{\sigma}^a$, to be isotropic and of the form:
	$\boldsymbol{\sigma}^a=\chi(\rho_a)\sum_i \zeta_i\,\rho_i\,\mathbf{I}$, 
	where 
	$\zeta_i>0$;
	and $\chi(\rho_a)$ is a sigmoidal function which captures the dependence of the active stress on the local mesh density, hence, on the local meshwork strain (as $\rho_a$ is enslaved to $\epsilon$). 
	
	The bound active contractile stresslets exert different contractile stresses and undergo different turnovers, which can
 depend on the local strain. Here we have chosen the unbinding to be strain dependent with a Hill form:
	$k_i^{u}(\boldsymbol{\epsilon})=k_{i0}^{u}\, e^{\alpha_i\,\epsilon}$,
	where $k_{i0}^{u}>0$,  are the strain independent parts of the respective rates, and $\alpha_i$ are dimensionless numbers; $\alpha_i>0$ represent {\it catch bonds} where local contraction (extension) will decrease (increase) the unbinding of the stresslets, while $\alpha_i<0$ represent {\it slip bonds} where local extension (contraction) will decrease (increase) the unbinding of the stresslets \cite{Kovacsetal2007,Fernandez-Gonzalezetal2009,Mullaetal2022,debsankar2017}.

For a binary mixture, the physics of segregation, and the subsequent force patterning, is explored by casting the hydrodynamic equations (see SI Sec.\,1D) in terms of the average density
	$\rho:=(\rho_1+\rho_2)/2$, and relative density $\phi:=(\rho_1-\rho_2)/2$ ($\rho_1$ being the more contractile species), which in the overdamped limit reduces to, 
	\begin{subequations}
		\begin{align}
			& \dot{\boldsymbol{u}} = \nabla\cdot\boldsymbol{\sigma}, \label{eqn:main:a}\\
			& \dot{\rho}+\nabla\cdot(\rho\,\dot{\boldsymbol{u}})=D\nabla^2 \rho +1 -\frac{C\,\epsilon}{A} -k^{u}_{\text{avg}}\,\rho-k^{u}_{\text{rel}}\,\phi, \label{eqn:main:b}\\
			& \dot{\phi}+\nabla\cdot(\phi\,\dot{\boldsymbol{u}})=D\nabla^2 \phi + k^{b}_{\text{rel}}\bigg(1-\frac{C\,\epsilon}{A}\bigg) -k^{u}_{\text{avg}}\,\phi-k^{u}_{\text{rel}}\,\rho, \label{eqn:main:c}
		\end{align}
		\label{eqn:main}
	\end{subequations}
made dimensionless by setting the units of time, length and density,
as $t^\star:=1/k^{b}_{\text{avg}}$, $l^\star:=\sqrt{\eta_b/\Gamma}$ and $\rho_a^0$, where 
$k^{b,u}_{\text{avg}}:=(k^{b,u}_1+k^{b,u}_2)/2$, $k^{b,u}_{\text{rel}}:=(k^{b,u}_1-k^{b,u}_2)/2$ are the average and relative binding (unbinding) rates of the stresslets, respectively.

	

With $\rho_a$ enslaved to $\epsilon$, the stress $\boldsymbol{\sigma}^e+
\boldsymbol{\sigma}^a$, can be recast as

	\begin{equation}
		\sigma_0\mathbf{I}+\Big(\tilde{B}\,\epsilon
		+B_2\,\epsilon^2+B_3\,\epsilon^3\Big)\mathbf{I}+2\mu\tilde{\boldsymbol{\epsilon}},
		\label{constitutive-rel}
	\end{equation}
	with a purely {\it active back pressure} $\sigma_0(\rho,\phi):=2\chi(\rho^0_a)(\zeta_{\text{avg}}\,\rho+\zeta_{\text{rel}}\,\phi)$
	where 	$\zeta_{\text{avg}}:=(\zeta_1+\zeta_2)/2>0$, $\zeta_{\text{rel}}:=(\zeta_1-\zeta_2)/2$ are the average and relative contractility, respectively, 
	an activity renormalized bulk modulus $\tilde{B}(\rho,\phi):=B-\frac{C^2}{A}-2\chi'(\rho^0_a)\,\frac{C}{A}(\zeta_{\text{avg}}\,\rho+\zeta_{\text{rel}}\,\phi)$ and activity generated  nonlinear elastic bulk moduli $B_2(\rho,\phi), B_3(\rho,\phi)$ that are linearly dependent on $\rho$ and $\phi$ (see SI Sec.\,1E for details). 
	

\begin{figure*}[t]
		\centering
		\includegraphics[scale=1]{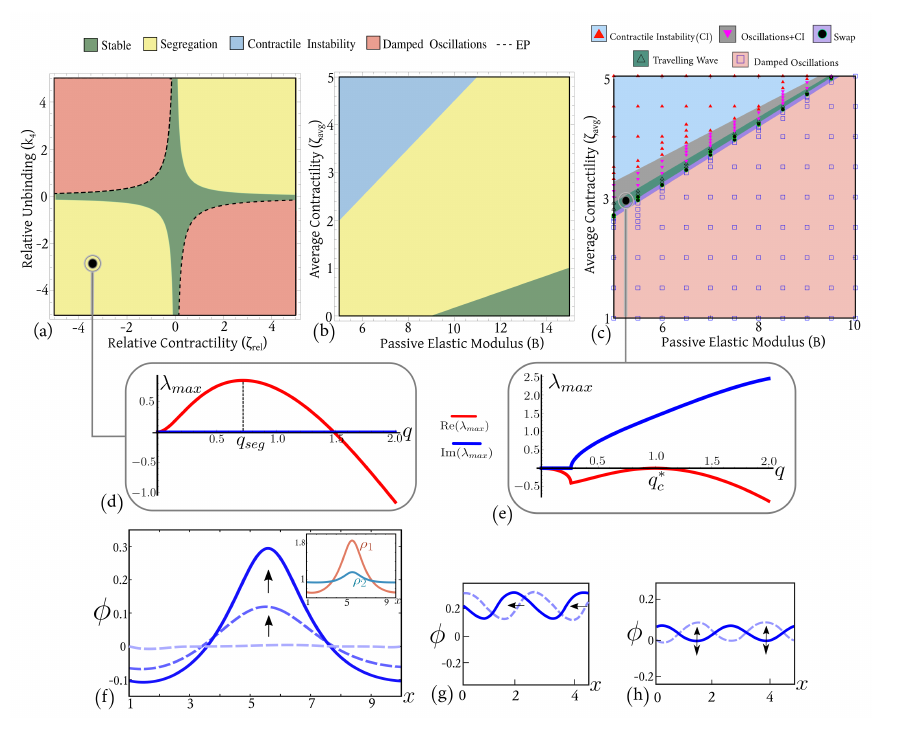}
		\caption{{\bf Nonequilibrium phase diagrams and the nature of the phases in a mixture of stresslets.} (a,b) Linear stability phase diagrams, where we have set $C=A=D=1$, 
		$k_1=k_3=1$, $k_2=0$, $\chi(\rho_a^0)=1$ and $\chi'(\rho_a^0)=1$ throughout.  (a) Phase diagram in relative unbinding versus contractility, with $B=8$, $\zeta_{\text{avg}}=1$, shows segregation (yellow) when $\zeta_{\text{rel}}k_4>0$ (and damped oscillations for $\zeta_{\text{rel}}k_4<0$). The dashed line corresponds to a line of exceptional points (EP). (b) Phase diagram in bare elastic modulus versus average contractility for  $\zeta_{\text{rel}}k_4>0$,  shows segregation emerge from the monotonically stable phase, followed by contractile  instability, with increasing $\zeta_{\text{avg}}$.
		(c) Full phase diagram obtained by numerically solving the scalar version of   \eqref{eqn:main}, for $\zeta_{\text{rel}}k_4<0$, showing successive emergence of swap and travelling waves from the stable phase (with damped oscillations) with increasing average contractility. Symbols denote the state points where the numerical solutions were obtained. Details of numerical solutions are described  in SI Sec.\,6.
			 (d,e) Typical dispersion curves obtained from linear stability analysis for (d) the segregation phase, where  $q_{seg}$ is the fastest growing mode that undergoes segregation; and (e) the travelling wave phase, where the mode $q^\star_c$ is the fastest growing mode that first reaches criticality.
		   (f,g,h) Snapshots of $\phi$-profiles (blue line) and their time-evolution (dashed line with progressively dark hues of blue accompanied by arrows) showing segregation, travelling wave and swap behaviour obtained from a numerical solution of the scalar version of \eqref{eqn:main} with periodic boundary condition (see SI videos). (f) The $\phi$-profile shows segregation starting from an initial uniform configuration of stresslets.  Inset shows co-localisation of the segregating stresslets; the ratio $\rho_1/\rho_2$ within the domain depends on domain activity, turnover and the stress jump across the domain. (g) Travelling wave $\phi$ profile showing movement to the left. (h) Standing wave $\phi$ profile denotes the swap phase.   }
		\label{fig:phasedia-disp}
	\end{figure*}

	The material model in  \eqref{eqn:main} is dynamic, renewable and active, i.e., breaks time reversal symmetry (TRS). As a consequence, the effective elastic constitutive relation  \eqref{constitutive-rel}
		exhibits dynamical compression-weakening and extension-stiffening through differential activity $\zeta_{\text{rel}}$. Indeed,  $\tilde{B}<B_{\text{pass}} \,(>B_{\text{pass}})$ for $\zeta_{\text{rel}}\phi>0(<0)$, where $B_{\text{pass}}:=B-\frac{C^2}{A}$ is the (compressibility renormalized) passive bulk modulus, i.e.,  local accumulation (depletion)  of stronger contractile stresslets   decreases (increases) the bulk modulus, hence, triggering a linear elastic instability ($\tilde{B}< 0$) in the extremely contractile regimes (where $\zeta_{\text{rel}}\phi\gg 1$).
	The material, in turn, actively generates  a stabilising back pressure $\sigma_0$ and nonlinear saturation thresholds $B_{2,3}$ of appropriate signs for this ensuing instability.
	This is reminiscent of other nonlinear, although passive, elastic models for biological fiber networks,
		such as the `bucklable' elastic material which `yields' when the compressive stress locally saturates at a specified threshold
		\cite{Roncerayetal2019}, 
		and a `compression-weakening'  uniaxial elastic material
		with smaller stiffness for contractile strains than extensile strains 
		\cite{Rosakisetal2015, Grekasetal2021}. Interestingly, its extreme dilational softness makes the material highly auxetic~\cite{Keelingetal2017}, i.e., the 2D Poisson's ratio $\nu:=(\tilde{B}-\mu)/(\tilde{B}+\mu)<0$.

		 The linear elastic instability $\tilde{B}< 0$ may manifest either as failure of elliptic elasticity in the bulk when $\tilde{B}+\mu< 0$ \cite{KnowlesSternberg1978}, or as failure of the `complementing condition' for finite bodies when a free boundary is present (this gives rise to surface instability) \cite{SimpsonSpector1985,EgorovPalencia2011}. Hence, with suitable boundary conditions, our active material would exhibit force propagation along the `characteristic' lines in the parabolic and hyperbolic elasticity regimes of extreme contractile activity. This is reminiscent of isostatic elasticity model for cytoskeletal long range force propagation along chains in \cite{Blumenfeld2006}, and compression chains in jammed granular media \cite{Catesetal1998, Bouchaud2002, Otto2003}. Note that, as a consequence of
		 enhanced catch bond response at the sites of 
		 extreme contractile activity, tension gets amplified along the  chains, as in \cite{Roncerayetal2016}.  The connection between non-elliptic elasticity, isostaticity and force chain formation, will be discussed in a forthcoming article~\cite{rct22}.
		 Here, we will see how these force chains emerge in our active elastomer supported by tension, together with excitable behaviour.




	\section*{Linear Stability Analysis} The stability about the homogeneous unstrained steady state of the system (considering perturbation along the strain direction to be purely isotropic), starting from a symmetric mixture of stresslets $\phi=0$,
is described by the linear dynamical system, $\dot{\mathbf{w}}=\mathbf{M}\mathbf{w}$ (see SI Sec.\,2C), where
	$\mathbf{w}=\Big(\hat{\delta \epsilon}(t,{\bf q})\,\,\hat{\delta \rho}(t,{\bf q})\,\,\hat{\delta \phi}(t,{\bf q})\Big)^T$ depends on the wave vector ${\bf q}$ and the dynamical matrix $\mathbf{M}$ is
	non-Hermitian  due to TRS breaking. 	As a consequence, the eigenvectors along which the perturbations propagate are no longer orthogonal to each other and may even co-align for some parameter values, as we will see later.

We make a further simplifying assumption, that the bare (strain independent) unbinding rates and binding rates are identical. With this, the instabilities are determined solely by the maximum eigenvalue

	\begin{equation}
		\lambda_{max}=-\frac{\lambda_a-\sqrt{\lambda_b}}{2(1+q^2)}
		\label{egvalues}
	\end{equation}
	where 
	\begin{subequations}
		\begin{align}
			&\lambda_a:=k_1+\bigg( \tilde{B}_0-\frac{2\zeta_{\text{avg}}}{k_1}+D+k_1\bigg)\,q^2+D\,q^4,\label{lam_a}\\
			&\lambda_b:= \lambda_a^2-4\,q^2(1+q^2)k_1^2\bigg[\tilde{B}_0\,D\,q^2+\nonumber\\
			&\hspace{3mm}k_1\bigg(\tilde{B}_0-2\frac{\zeta_{\text{avg}}}{k_1}\bigg(\frac{C}{A}+\frac{k_3}{k_1}\bigg)-2\frac{\zeta_{\text{rel}}}{k_1}\frac{k_4}{k_1}\bigg)\bigg],\label{lam_b}
		\end{align}
	\end{subequations}
	 $q:=|\mathbf{q}|$, $\tilde{B}_0:=B-\frac{C^2}{A}-2\chi'(\rho^0_a)\,\frac{C}{A}\frac{\zeta_{\text{avg}}}{k_1}$ is the activity renormalized bulk modulus of the homogeneous symmetric mixture,  $k_1$ is the average bare unbinding rate, and $k_3$ and $k_4$ are the average and relative (strain dependent) unbinding rates, respectively.



		\subsection*{Contractile Instability} 
		Starting from a stable elastomer, we see that large enough average activity $\frac{\zeta_{\text{avg}}}{k_1}$
	drives the renormalized bulk modulus $\tilde{B}_0$ of the symmetric mixture to negative values, a linear elastic
	instability (and ellipticity loss) of the underlying elastomer that affects all modes $q\in[0,\infty)$ (see dispersion curve in SI Fig.\,S5(a)). This shows up as self-penetration and subsequent collapse (halted by steric effects) of the uniform contractile  mixture,
	unless constrained by appropriate boundary conditions.

	As we will see, force patterning of the mixture is achieved through entrapping this (system spanning) contractile instability into segregated pockets of the cell body. In the linear theory, this segregation, and other nonequilibrium phases, show up in the mechanically stable regime of the active elastomer,
	$\tilde{B}_0>0$.


	\subsection*{Segregation Instability}

	As $\lambda_b$ increases beyond $0$,  $\lambda_{max}$ goes from being negative (stable) to positive, leading to a long-wavelength instability in $\phi$,
	with a fastest growing wave-vector $q_{seg}$
	which sets the characteristic width $q_{seg}^{-1}$ of the segregated pattern (Fig.\,\ref{fig:phasedia-disp}(d),  SI-Eq.\,(46)), provided $\tilde{B}_0$ is bounded between,
	\begin{equation}
		0	<\tilde{B}_0< 2\frac{\zeta_{\text{avg}}}{k_1}\bigg(\frac{C}{A}+\frac{k_3}{k_1}\bigg) + 2\frac{\zeta_{\text{rel}}}{k_1}\frac{k_4}{k_1}\,.
		\label{segrecond}
	\end{equation}
	This linear segregation regime is typically realised 	
	when the relative activity $\zeta_{\text{rel}}$ and relative strain dependent unbinding $k_4$ have the same sign, which since the stresslets are contractile, implies  $k^u_1(\epsilon)<k^u_2(\epsilon)$.  To drive segregation, the stresslet with stronger contractile activity must have a lower strain dependent unbinding rate.  Note that the density peaks of the individual stresslets colocalise (Fig.\,\ref{fig:phasedia-disp}(f) inset, also \href{https://drive.google.com/file/d/137avxjvNrZ2WnVWZMiJcew6dv_PjA2r2/view?usp=sharing}{SI-Movie\,S1}) unlike in conventional phase separation, which is reminiscent of the study in  \cite{Beachetal2014}. This occurs when both stresslets exhibit catch-bond behaviour ($k_3>0$); for slip-bond response ($k_3<0$), the individual density peaks separate as in usual segregation (\href{https://drive.google.com/file/d/1SGvCRVNLZQ_KcbroFJ4WCzA07wu7iXje/view?usp=sharing}{SI-Movie\,S2}). 
	It is worth emphasizing that even a small difference in contractility or strain dependent unbinding rate manifests as a large segregation width $q^{-1}_{seg}$ in the real space (SI-Fig.\,S2).
	
	
	
	
What is the driving force for this segregation in the linear theory? Since the stresslets do not directly interact with each other, the driving force must come from their indirect interaction through the elastomer strain. We find that to linear order, the power density
	\begin{equation}
		\dot{W}(t)=\frac{1}{2L}\int_{-L}^{L}\Bigg(\frac{\partial w}{\partial \epsilon}\dot{\epsilon}+\frac{\partial w}{\partial \rho}\dot{\rho}+\frac{\partial w}{\partial \phi}\dot{\phi}\Bigg)dx\nonumber
	\end{equation}
	associated with the effective elastic energy density	 $w:=\sigma_0(\rho,\phi)\,\epsilon+\frac{1}{2}\tilde{B}(\rho,\phi)\,\epsilon^2$ ($2L$ is the system size)
	is negative in the segregated phase (SI-Fig.\,S3), that is to say $W$ is a Lyapunov functional driving segregation of the stresslets. 	The appearance of $\rho_2$ micro-domains within the $\rho_1$ (more contractile) domain, is a consequence of the interplay between the strain dependent catch-bond turnover and this driving force. Note that the value of strain in the linearly segregated domains of high contractility is set by the minima of $w$,   $\epsilon_{\text{min}}=-\frac{\sigma_0}{B_{\textbf{pass}}-\frac{\chi'(\rho_a^0)}{\chi(\rho_a^0)}\frac{C}{A}\sigma_0}$, that depends directly on the active back pressure $\sigma_0$. Hence, the active back pressure is significant in keeping the segregated domains of stronger contractile stresslets well-separated, preventing them from clumping.

		\subsection*{Travelling Waves}	
		
		From the form of $\lambda_{max}$ (\eqref{egvalues}), we see that $\lambda_b< 0$
	characterizes the  various oscillatory phases (stable and unstable pulsations and/or waves), with frequency $\omega(q)= \vert \operatorname{Im}(\lambda_{max}) \vert$, and decay/growth rate $\tau_d(q)=\vert\operatorname{Re}(\lambda_{max}) \vert$. 
	For negative values of $\lambda_a$, the oscillations grow with a fastest growing mode at a wave vector $q^\star$ (SI-Eq.\,(S55)). However,
	as $\lambda_a(q^*)$ first touches $0$ (Fig.\,\ref{fig:phasedia-disp}(e)), we get travelling waves with wave vector $q^\star_c=\left({\frac{k_1}{D}}\right)^{1/4}$ and frequency $\omega(q^\star_c)$ (SI-Eq.\,(S59)), whose speed is set by $\sqrt{\zeta_{\text{rel}}k_4}$ (See Fig.\,\ref{fig:phasedia-disp}(g), \href{https://drive.google.com/file/d/1taNDBEVuwdG5PSU6NgP_QmAbmG9t508w/view?usp=sharing}{SI-Movie\,S3}).

	\subsection*{Swap} 
As we have seen, the rate of the  contractile instability is determined by the time scale  $Re[\lambda_{max}(q)]^{-1}$, while the time scale of unbinding of the stronger stresslet is ${k^u_1(\epsilon)}^{-1}$. Within the oscillatory phase, i.e., when $\zeta_{\text{rel}}k_4<0$, if the stronger stresslet unbinds before the contractile instability sets in, i.e., if $Re[\lambda_{max}(q)]^{-1}\le {k^u_1(\epsilon)}^{-1}$, then the contracting domain bounces back. This is the
	swap phase, a standing wave that breaks time translation symmetry \cite{Fruchartetal2021} (See Fig.\,\ref{fig:phasedia-disp}(h), \href{https://drive.google.com/file/d/1PYB37ATjee2QveQQ05l4hUCd1XBvqUmP/view?usp=sharing}{SI-Movie\,S4}). The swap phase does not appear as a distinct phase in the linear stability phase diagrams based on the dispersion curves. However, it appears in the full phase diagram, at the boundary between the damped oscillations and contractile instability phase,  as discussed below.


	\subsection*{Exceptional Points} 
	
	So far our discussion of the instabilities has been based on the behaviour of the maximum eigenvalue 
	$\lambda_{max}$. However, since the dynamical matrix $\mathbf{M}$ is non-Hermitian (see SI Eq.\,(S36)), the nature of the instabilities depends  crucially on the angle between the eigenvectors, in particular on {\it exceptional} points (EPs), where two eigenvalues coincide and the corresponding eigenvectors co-align \cite{Kato1984,Fruchartetal2021}. In general, eigenvalue based linear stability analysis gives robust predictions only about asymptotic phases, i.e., for  $t\to\infty$. In the vicinity of EPs, however, short time `transient growth' becomes  several orders of magnitude large so that linearity fails and the system `bootstraps' into nonlinear phases  \cite{TrefethenEmbree2005, Trefethenetal93}. 
	We defer a detailed analysis and classification of these exceptional points to a later study. Here we only mention that in the linear stability phase diagrams (Fig.\,\ref{fig:phasedia-disp}(a)), the only EPs present are on the boundary between stable and damped oscillation phases. Hence, the system goes into the oscillatory phase through an EP \cite{Youetal2020}.
	
		\subsection*{Numerical phase diagram} The swap and the travelling wave phases show up distinctly in the full phase diagram (Fig.\,\ref{fig:phasedia-disp}(c)) obtained from a numerical analysis of the scalar version of  \eqref{eqn:main} using our own code based on finite difference Euler scheme with a stencil adaptive algorithm, and the spectral methods based pde solver Dedalus \cite{dedalus2020}, with periodic boundary conditions (see SI Sec.\,6 for details).	
		For $\zeta_{\text{rel}}k_4<0$, we observe from numerical phase diagram Fig.\,\ref{fig:phasedia-disp}(c) that there is a region where travelling waves and swap are coexisting phases in time (\href{https://drive.google.com/file/d/1VXxqBFuo5Deq1-xESXc-qcML_YBWR6Dc/view?usp=sharing}{SI-Movie\,S5}). Starting with small random perturbations about the uniform, symmetric unstrained steady state in the parameter regime $\zeta_{\text{rel}}k_4<0$, sustained oscillations appear at the unstable oscillations and damped oscillations phase boundary, either as a travelling wave train, or as a standing wave (i.e., swap). Typically the system exhibits a long transient, where it first goes into a swap phase, then a coexistence between swap and travelling wave, and then finally transitions into a travelling wave \cite{Lehtimakietal2021}. The transient time decreases with an increase in the average contractility. We draw the phase diagrams by making note of the configuration at a fixed large time $t_{max}$ starting from statistically identical initial conditions.
The numerical code shows an eventual blowup at a very large run time in the travelling wave phase, due to the sharpness of the slopes of the travelling front. 
In a later publication, we will study in detail the instabilities through which swap phase transitions into the travelling wave phase. At the boundary between the oscillatory phases and the contractile instability, we see a travelling wave train with amplitude that grows indefinitely, giving rise to an array of moving tension lines.

	\begin{figure*}[t]
		\centering
		\includegraphics[scale=.83]{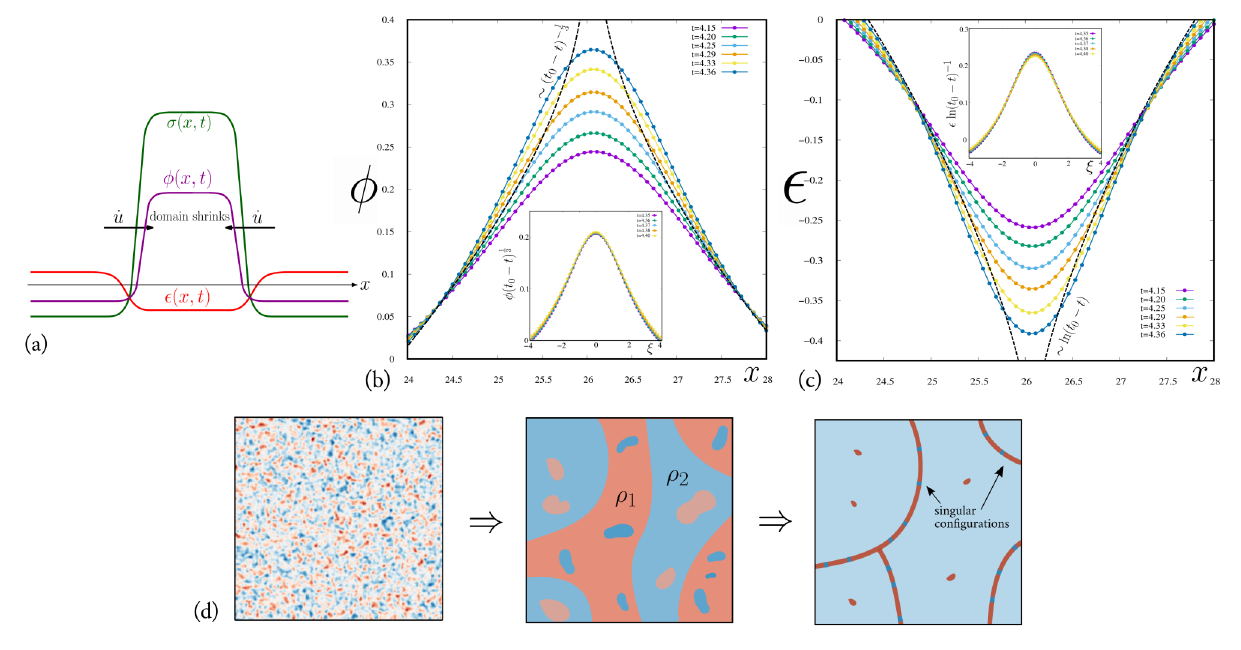}
		\caption{{\bf Emergence of singular tension chains.} (a) Schematic showing typical  profiles of the order parameter $\phi$ and strain $\epsilon$ immediately following the linear segregation. The stress $\sigma$  jumps of opposite signs across the right and left fronts cause them to move towards each other with speed ${\dot u}$, leading to a growing amplitude and an eventual singularity in the middle. (b,c) Numerical results verifying the formation of a singularity in $\phi$ and $\epsilon$ in finite time. Insets show scaling collapse of the $\phi$ and $\epsilon$ profiles near the singularity, as predicted from theory. Starting from a homogeneous unstrained symmetric state, a numerical solution of the scalar version of  \eqref{eqn:main}, gives a value  $t_0=4.55$ in units of $t^{\star}$  for the finite time blowup (see SI Sec.\,7B).  (d) Schematic shows the time evolution of a uniform distribution of a mixture of stresslets, to a linear segregation of the stresslets $\rho_1$ and $\rho_2$, to the eventual formation of tension chains and punctae in 2D  of the stronger stresslets in a sea of the weaker stresslets, with some weaker stresslets embedded within the tension chains (darker shade represents higher density).}
		\label{fig:singularformation}
	\end{figure*}
		\subsection*{Single stresslets} 
		Our results for a mixture of stresslets carry over to 
		the case of a single stresslet too, provided the dependence of the active stress on stresslet density ($\zeta(\rho)$) is steep (see SI Sec.\,4), resulting in a phase separation between regions of low and high stresslet density, akin to a gas-liquid phase separation (\href{https://drive.google.com/file/d/1KDbiGwtyOtWpiFQGyXurWQ44JwraGiT0/view?usp=sharing}{SI-Movie\,S6}).
		In the mixture of stresslets, the segregation depends on the profiles of both the average and relative contractility, and thus appears over a wider parameter range compared to the single stresslet.
		
		\subsection*{Stresslets in fluid}
		  In case of fluid mediated interaction between the stresslets,  {\it non-monotonic} dependence of the active stress on the stresslet density $\rho$ is necessary for segregation (here,  a monotonic $\zeta(\rho)$ with steep positive slope results in a clumping instability  instead \cite{GowrishankarRao2016,HusainRao2017}.  This non-monotonicity in $\rho$ naturally arises from the binding of contractile stresslets on finite patches of actin mesh with free boundaries embedded in a fluid, where the elastic response of the patches is taken to  be fast. The crucial role of the 
		  active back pressure in driving this segregation, is played by the negative slope branch
		  of the non-monotonic $\zeta(\rho)$, that separates the positive slope branches corresponding to  low and high stresslet density (SI Sec.\,5, \href{https://drive.google.com/file/d/1pp3cNh6lbD8cQk3Kwiox_jqrvffA9n8i/view?usp=sharing}{SI-Movie\,S7}).

	\section*{Nonlinear Effects: emergence of tension chains}
	
	The exponential growth of the linear segregation instability quickly leads to a stage where nonlinear effects become significant. However unlike usual segregation, where nonlinearities temper the exponential growth to a slower power-law~\cite{Bray2002}, here the effect of nonlinearities is to drive it to form singular structures in finite time~\cite{eggersfontelos-book2015}. This happens through a feedback mechanism where a contractile instability rides atop the segregation instability.
To see this, we note that the typical order parameter profile of a segregation after the linear instability regime would look like Fig.\,\ref{fig:singularformation}(a), with width 
$q^{-1}_{seg}$. In the region between the two fronts, $\phi>0$ and so the strain $\epsilon<0$ (i.e., the stress $\sigma$ is highly tensile). Outside this region $\phi<0$ and so $\epsilon>0$ and the stress is compressive, though of relatively low magnitude, Fig.\,\ref{fig:singularformation}(a).
Since $\dot{u}=\partial_{x}\sigma$, the  stress jumps across the fronts, cause them to move {\it towards} each other,
resulting in an ever increasing concentration of $\rho_1$ within the shrinking $\phi>0$ domain. Enhanced catch bond response accelerates this shrinking. Eventually, this shrinking domain enters the contractile instability regime $\tilde{B}\le0$ when $\zeta_{\text{rel}}\phi\gg 1$, where there is no escape from collapse, leading to the formation of singular structures in finite time! These tensile structures remain well-separated in space through the actively produced back pressure. This is very different from the algebraically growing domains in usual phase segregation~\cite{Bray2002}. 

To compute the scaling behaviour as one approaches the finite-time singularity, we find it convenient to turn off the contributions from stresslet turnover, thus making $\rho$ and $\phi$ conserved.  In this situation, the $\epsilon^3$ term in the effective strain energy density $w:=\sigma_0\epsilon+\frac{1}{2}\tilde{B}\epsilon^2+\frac{1}{3}B_2\epsilon^3+\frac{1}{4}B_3\epsilon^4$ is the dominant driver of  the  concentration of the stresslet densities towards a singularity in a finite time $t_0$ at spatial location $x_0$. Using the method of dominant balance~\cite{eggersfontelos-book2015} in the vicinity of the singularity, we find that $\rho$, $\phi$ and $\epsilon$ exhibit self-similar forms, 
$\rho(x,t)\sim\frac{1}{(t_0-t)^{\frac{1}{3}}}R\Big(\frac{x-x_0}{(t_0-t)^{\frac{1}{3}}}\Big)$, $\phi(x,t)\sim\frac{1}{(t_0-t)^{\frac{1}{3}}}\Phi\Big(\frac{x-x_0}{(t_0-t)^{\frac{1}{3}}}\Big)$, $\epsilon(x,t)\sim\ln(t_0-t)E\Big(\frac{x-x_0}{(t_0-t)^{\frac{1}{3}}}\Big)$ (see SI Sec.\,7A). For an initial segregating domain of width $l$, dimensional analysis suggests that the domain width goes to zero at time $t_0\sim \bar{t}(l/\bar{l})^3$, where $\bar{t}:=\Big(\frac{\Gamma}{{\bar{B}_2}^2}\Big)^{\frac{1}{3}}$ and  $\bar{l}:=\Big(\frac{\bar{B}_2}{\Gamma^2}\Big)^{\frac{1}{3}}$ are the characteristic time and length scales, with    $\bar{B}_2:=\chi''(\rho^0_a)\,C^2\,\zeta_{\text{rel}}$. We verify these self-similar forms in a careful numerical study
of the scalar version of \eqref{eqn:main} (Fig.\,\ref{fig:singularformation}(b,c)).
These singularities are  {\it physical} in that their resolution involves incorporation of additional
physical effects such as steric hindrance (represented by the $\epsilon^4$ term in $w$). 
Thus the singularity is never reached, resulting in highly concentrated tensile regions of finite width, $\sim 300$\,nm, the length of myosin-II bipolar filament~\cite{Bershadsky2017}.




Evidently the geometry of these singular structures depends on dimensionality -- in 1D the singular regions are punctae, in 2D the singular regions appear as tension chains and punctae, while in 3D, they would manifest as sheets, lines and punctae. In a finite system,
these singular structures would need to be stabilized by anchoring conditions at the boundary.

Note that, these actively generated chains and sheets of tension are  anisotropic structures that emerge through an unconventional spontaneous breaking of the underlying isotropic symmetry of the active elastomer. 
The anisotropy is a consequence of the local nature of the nonlinear effects, viz.~the entrapment of contractile instability within the segregated domains of high contractility, since the effective elastic moduli depend on $\rho$ and $\phi$. As $\phi$ increases in the linearly segregated domains,  $\tilde{B}$ crosses zero (where the linear elastic response loses positive definiteness) and quickly becomes negative. As soon as the local elasticity crosses its ellipticity threshold at $\tilde{B}=-\mu$, `characteristic' lines of the ensuing parabolic, and eventually hyperbolic response emerge; this characteristic direction sets the local tangent direction of the tension chain. This emergent anisotropy is encoded in the uniaxial nature of the singular stress field along the tension chain: $\boldsymbol{\sigma}=\gamma\,\mathbf{t}_S\otimes\mathbf{t}_S$ \cite{SimhaBhattacharya1998}, where $\gamma$ is the tension in the chain and $\mathbf{t}_S$ is the local tangent vector specifying the orientation of the anisotropy. The uniaxial form is equivalent to the existence of a local {\it fabric tensor} $\mathbf{P}:=\mathbf{t}_S\otimes\mathbf{t}_S-\frac{1}{d}\mathbf{I}$ \cite{Kanatani1984}, symmetric and traceless, such that $\mathbf{P}\cdot\boldsymbol{\sigma}=0$. Evolution of this fabric tensor, hence the anisotropy, depends on the local history of formation and motion of the singular structure that will come from solving the complete initial-boundary-value problem, as discussed below.


		\begin{figure*}[t]
		\centering
		\includegraphics[scale=0.7]{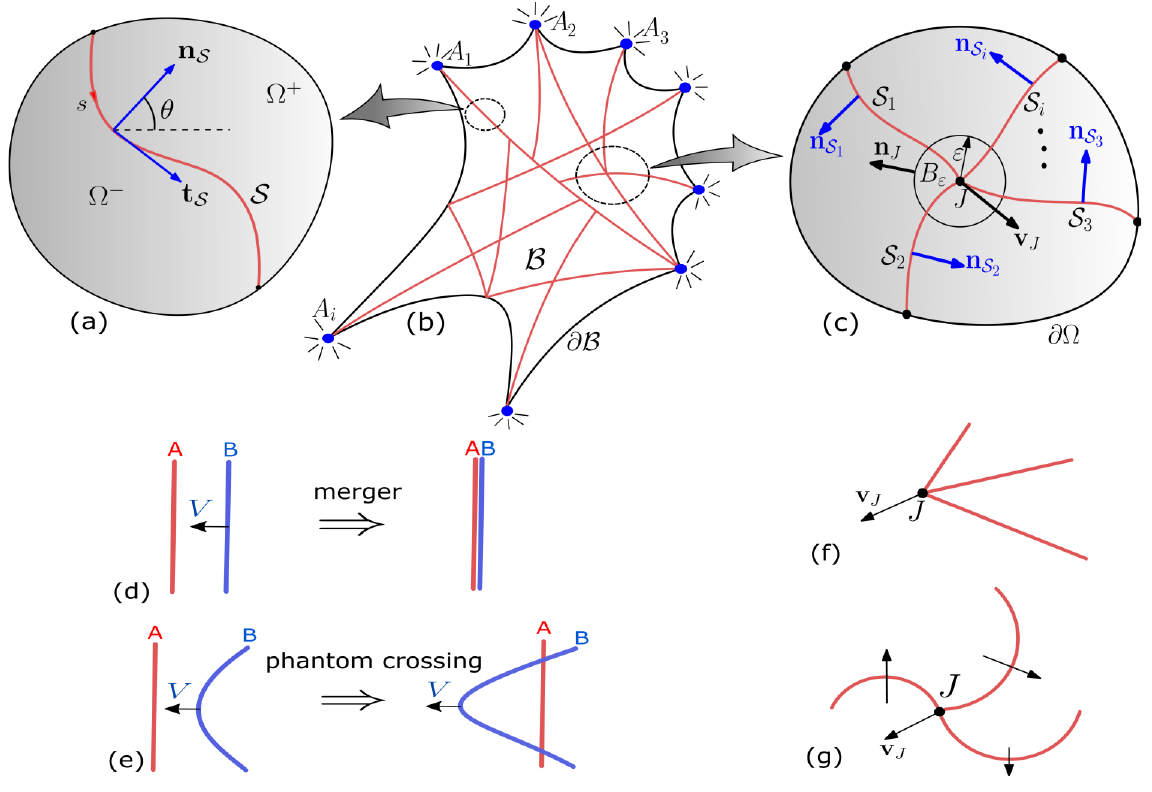}
		\caption{{\bf Mechanics of interacting and moving tension chains.} (a) Schematic showing an arbitrary region $\Omega$ containing a curved tension chain, carved from a 2D cell body $\mathcal{B}$ shown in (b), which is anchored at some of its boundary points $A_i\in\partial\mathcal{B}$. (c) Schematic showing a junction $J$ where $N$ tension chains meet.  (d,e) A merger and a phantom crossing of two tension chains moving relatively towards each other. (d) Merger of two straight tension chains moving relatively towards each other.  (e) Phantom crossings of curved tension chains moving relatively towards each other. (f) A moving polar junction of tension chains.   (g) A moving and rotating spiral junction of tension chains.}
		\label{fig:frontmove}
	\end{figure*}
	
	 \section*{Mechanics of Tension Chains}
	  The initial profile of the two fronts bounding the segregating domain in Fig.\,\ref{fig:singularformation} was taken to be symmetric, suggesting that there is no pressure difference on either side of the domain. In general this is not guaranteed, leading to the possibility of moving tension chains \cite{Shutovaetal2017}.  
	 A moving tension chain in a fixed 2D domain $\Omega$ is represented by a smooth evolving curve  $\mathcal{S}$, see Fig.\,\ref{fig:frontmove}(a). $\mathcal{S}$ is a material curve consisting predominantly of the stronger stresslet species, across which bulk fields of the predominant  weaker stresslet species suffer jump discontinuities.
	$\mathcal{S}$, in the reference configuration, has a local parametrization $\boldsymbol{r}(s,t)$, where $s$ is an arc length parameter. The unit tangent and normal fields on $\mathcal{S}$ are $\mathbf{t}_S$ and  $\mathbf{n}_S$, respectively, and the normal speed is $V:=\frac{\partial\boldsymbol{r}}{\partial t}\cdot\mathbf{n}_S$. We denote the jump in a bulk discontinuous field $\psi$  across $\mathcal{S}$ by $\llbracket\psi\rrbracket:=\psi^+-\psi^-$, where $\psi^\pm$ are the limiting values of $\psi$ as one approaches $\mathcal{S}$ from $\Omega^\pm$. We also define the average value of $\psi$ at $\mathcal{S}$ as $\langle\psi\rangle:=(\psi^++\psi^-)/2$. Then, using the divergence and transport theorems for fields with line singularities in a 2D domain (details in SI Sec.\,8), one can derive the governing system of equations for fields in the bulk and fields defined on the singular structure.
	The singular counterparts of the mass balance equations \eqref{eqn:main:b} and \eqref{eqn:main:c} relate the rate of change of the average and relative densities,  $\rho_S$ and $\phi_S$, of the two stresslets on the tension chain to the jump in the bulk mass flux of the stresslet species across the chain and their intrinsic flux along the chain, with  contributions due to its curvature and intrinsic turnover  (Eq.\,(S126) in SI).  
	On the other hand, the singular counterpart of the force balance \eqref{eqn:main:a} yields
	\begin{equation}
		\partial_s\gamma\,\mathbf{t}_S+\gamma H\,\mathbf{n}_S+\llbracket\boldsymbol{\sigma}\rrbracket\mathbf{n}_S=\Gamma_S\mathbf{v}_S,
	\end{equation}
	where, $\mathbf{v}_S:=\langle\dot{\boldsymbol{u}}\rangle+V\,\langle\mathbf{I}+\nabla\boldsymbol{u}\rangle\mathbf{n}_S$ is the intrinsic velocity field, $\gamma$ is the tension,  and $H$ is the curvature of the tension chain $\mathcal{S}$ \cite{SimhaBhattacharya1998}.
	If we assume, for simplicity, that both the passive and the active parts of the bulk stress are isotropic, i.e., $\boldsymbol{\sigma}^{\pm}=p^\pm\mathbf{I}$, then the normal and tangential components of the above equation along $\mathcal{S}$ gives
	\begin{subequations}
		\begin{align}
			& \gamma H+\llbracket p^e\rrbracket+\llbracket p^a\rrbracket=\Gamma_S\mathbf{v}_S\cdot\mathbf{n}_S,~~\text{and}\\
			&\partial_s\gamma=\Gamma_S\mathbf{v}_S\cdot\mathbf{t}_S.
		\end{align}
		\label{singmonbal}
	\end{subequations}

	The static tension chain, for which $\mathbf{v}_S=\mathbf{0}$, gives the active Young-Laplace law, with the following consequences: (1) the tension chain is straight ($H=0$) if the active pressure jump $		\llbracket p^a \rrbracket$ counter-balances the (passive) elastic pressure jump $	\llbracket p^e \rrbracket$ (i.e., $\llbracket p^e\rrbracket+\llbracket p^a\rrbracket=0$); (2) in the absence of a passive  pressure jump (i.e., $\llbracket p^e\rrbracket=0$), active pressure jump  gives rise to a curved tension chain ($H=-\llbracket p^a\rrbracket/\gamma$); and (3) tension $\gamma$ along the chain is constant.

	It can be readily shown (SI Sec.\,8G) that, if two tension chains are moving relatively towards each other 
	(Fig.\,\ref{fig:frontmove}(d,e)), they merge into a single chain if they are straight and parallel and they  scatter if they are curved towards each other.

	\subsection*{Finite geometry -- surface anchoring and wetting}

	In the finite geometry of the cell, one needs to specify appropriate boundary anchoring conditions or boundary interactions at the cell surface. A natural choice is to declare surface anchoring at the locations of integrin-based focal adhesions and cadherin-based adherens junctions. These molecular complexes embedded in the cell surface bind strongly to the actomyosin filaments via linker proteins on the intracellular side and to the substrate or adjoining cells on the extracellular side.
	
	The complete initial boundary value problem for a cell body $\mathcal{B}$ consists of the dynamical \eqref{eqn:main}, supplemented by the anchoring conditions $\boldsymbol{u}=0$ and $\nabla\rho_i\cdot\mathbf{n}=0$ at a finite number of boundary points $A_i\in\partial\mathcal{B}$, and no flux conditions $\boldsymbol{\sigma}\mathbf{n}=\mathbf{0}$ and $\nabla\rho_i\cdot\mathbf{n}=0$ on the rest of the boundary $\partial\mathcal{B}\setminus\cup_iA_i$, together with appropriate initial condition respecting the boundary data (Fig.\,\ref{fig:frontmove}(b)). Due to contractile activity, the stress fields naturally concentrate near $A_i$, and stable tension chains emerge from these anchoring points that span the whole system $\mathcal{B}$.

	One may deduce the force-balance condition at these rigid anchoring sites to be,
	\begin{equation}
		\mathbf{f}_{A_j}+\sum_{i=1}^{N_j}\gamma_i\mathbf{t}_{\mathcal{S}_i}\big|_{A_j}+\gamma_{\partial \mathcal{B}}\mathbf{t}_{\partial\mathcal{B}}=\mathbf{P}_{A_j},
	\end{equation}
	where $N_j$ tension chains meet at the anchoring point $A_j$. Here, $\mathbf{f}_{A_j}:=\lim_{\varepsilon\to 0}\int_{\partial H_{\varepsilon}}\boldsymbol{\sigma}\mathbf{n}\,dl$ is a singular force at  $A_{j}$ coming from the bulk stress field $\boldsymbol{\sigma}$, $H_{\varepsilon}$ is a small half-ball, centered at $A_{j}$, of radius $\varepsilon$ and unit normal $\mathbf{n}$ pointing into the bulk; $\gamma_{\partial \mathcal{B}}$ is the tension on the boundary at $A_j$ acting along the tangent $\mathbf{t}_{\partial\mathcal{B}}$ (coming from  surface elasticity of the membrane); and $\mathbf{P}_{A_j}$ is the reaction force due to the anchoring condition $\boldsymbol{u}(A_j)=\mathbf{0}$ (Fig.\,\ref{fig:frontmove}(c)). 
	Supported by these stable anchoring sites at the cell surface, the intracellular side may support a tension web, viz. a web of tension lines with multi-valent junctions.
	For $N$ tension chains meeting at a junction $J$, force balance at the junction gives
	\begin{equation}
	    \mathbf{f}_J+\sum_{i=1}^{N}\gamma_i\mathbf{t}_{\mathcal{S}_i}\big|_{J}=\Gamma_J\mathbf{v}_J,
	\end{equation}
	where $\mathbf{f}_J:=\lim_{\varepsilon\to 0}\int_{\partial B_{\varepsilon}}\boldsymbol{\sigma}\mathbf{n}_J\,dl$ is a singular force at $J$ coming from the bulk stress field $\boldsymbol{\sigma}$; here, $B_{\varepsilon}$ is a small ball of radius $\varepsilon$ and unit normal $\mathbf{n}_J$ containing the junction \cite{SimhaBhattacharya1998}, $\mathbf{v}_J$ is the velocity and $\Gamma_J$ is the friction coefficient at the junction (Fig.\,\ref{fig:frontmove}(b)). 
	If $\boldsymbol{\sigma}$ is isotropic, $\mathbf{f}_J=0$. Using the compatibility condition for velocity at the junction  \cite{Fischeretal2012}: 
$V_i\big|_{J}=\mathbf{v}_J\cdot\mathbf{n}_{\mathcal{S}_i}\big|_{J}$, $i=1,2,\ldots,N$, and 
	 the identities $(\mathbf{t}_{\mathcal{S}_j}\cdot\mathbf{n}_{\mathcal{S}_i})\big|_{J}=\sin(\theta_j-\theta_i)$ and $\mathbf{f}_J\cdot\mathbf{n}_{\mathcal{S}_i}\big|_{J}=f_J\sin(\theta_J-\theta_i)$, where $\theta_i$ and $\theta_J$ are the angles that $\mathbf{n}_{\mathcal{S}_j}\big|_{J}$ and $\mathbf{f}_J$, respectively, make with a fixed global axis (say, the $x$-axis), and $f_J:=|\mathbf{f}_J|$, the force balance equation gives
\begin{equation}
	\frac{\Gamma_J}{\Gamma_{S_i}}\gamma_iH_i(J)=f_J\sin(\theta_J-\theta_i)+\sum_{j=1}^{N}\gamma_j\sin(\theta_j-\theta_i),
	\label{junccond}
\end{equation}
	for $i=1,2,\ldots,N$; here, $(\theta_2-\theta_1)+(\theta_3-\theta_2)+\cdots+(\theta_1-\theta_N)=2\pi$. This is the generalized Young-Dupr\'{e} equation for junctions of force chains embedded in an elastic medium. 
	In striking contrast to the usual Young-Dupr\'{e} equations, these junctions can display polarity and 
	hence move, or  can support spiral junctions which will rotate owing to active stresses 
	(Fig.\,\ref{fig:frontmove}(f,g)). 
	
	Since the effective elasticity is hyperbolic when the tension chain formation takes place, the standard boundary conditions for conventional elliptic elasticity discussed above may become {\it incompatible} for the hyperbolic regimes. This is because  hyperbolic equations require half of the boundary conditions than its elliptic counterpart, as the rest are determined by `propagating' the boundary data along the characteristics 
	to the other half \cite{Bouchaud2002}. Such incompatible boundary conditions, or more interestingly, perturbations of compatible boundary conditions (thereby altering the geometry of focal adhesions), for a given set of steady state tension chains, would make the web unstable, resulting in dynamical remodelling of the web through binding-unbinding, so that the boundary conditions become compatible again for the remodelled network \cite{Bouchaud2002,RamaswamyRao2007}. The steady state configuration of a web of tension chains is, hence, {\it fragile} in this precise sense \cite{Catesetal1998}.

	Alternatively, one may also specify boundary interactions, either at the inner leaflet of the cell membrane or on the membranes of intracellular organelles, such as the Golgi, endosomes or nucleus.  A moving actomyosin web can then adhere to the inner surface of the cell membrane or wrap over organelles. Considering a mixture of stresslets, one species may have a  preferential wetting to a substrate, say the inner leaflet of the cell membrane.
	In SI Sec.\,9, we demonstrate that active bulk segregation together with preferential wetting to the substrate will ensure macroscopic segregation where one of the stresslets wets the substrate (\href{https://drive.google.com/file/d/1YTP1iSPCpkZuJG1GfY-_M6ORvpvc1n45/view?usp=sharing}{SI-Movie\,S8} and \href{https://drive.google.com/file/d/1NAcaYcbJCWnnJa0AjEAuaB6D47Ndi-QX/view?usp=sharing}{S9}), leading to a stratified layering of the two stresslets.
	
	

	


	
	\section*{Discussion}
	
To summarize, we have shown using a hydrodynamic approach, how active cytoskeletal stresslets that act as molecular force generators and sensors, give rise to striking singular patterns of nonequilibrium tension
chains within the cell. It is the
nonlinear coupling between the linear segregation and contractile instability that leads to the formation of these singular tension patterns. The associated anisotropy emerges via local departure from elliptic elasticity. 
In the finite geometry of the cell, these patterns are stabilised by cell surface anchors such as focal adhesions and adherens junctions. These fragile force patterns are sensitive to boundary conditions, and thus naturally shaped by the current geometry of the system \cite{Matthewsetal2006,banerjeeetal2020}. Thus, different FA and AJ
geometries would result in different networks with different mechanical properties \cite{GeigerSpatzBershadsky2009,GuptonWaterman-Storer2006}. Simultaneously, the system exhibits a variety of oscillatory force patterns such as travelling waves, swap, and their temporal coexistence, that are accessed 
through an exceptional point of this underlying non-reciprocal dynamics.




	An immediate extension of this work is to analyse the force patterning in 3D, where we expect both tension lines and sheets to emerge. Further, we will study the implications of nonequilibrium force patterning for force adaptation and strain homeostasis and analyse the frequency dependent rheological response of the stable steady state of tension chains with rigid anchors at the boundary.
 In an ongoing study, we find that segregation is substantially enhanced in a mixture of contractile and extensile stresslets due to presence of both attraction and repulsion mediated through the elastomer \cite{jimenezetal2021}, which, with nonlinear feedback, result in singular force patterns that include both tension and compression chains (the latter endogeneously stabilizes the former, together with boundary anchoring) -- with implications for an active hydrodynamic theory of cellular tensegrity~\cite{ingberetal2014} and the nonequilibrium assembly of active metamaterials \cite{PishvarHarne2020,Qietal2022}.
 




\vspace{10mm}
\noindent
{\bf Acknowledgement.} We thank T. van Zanten for a careful reading and pointing us to relevant experimental references, A. Nagilla for help in the initial numerical implementation and the members of Simons Centre for critical discussions. We acknowledge support from the Department of Atomic Energy (India), under project no.\,RTI4006, and the Simons Foundation (Grant No.\,287975), and computational facilities at NCBS. MR acknowledges a JC Bose Fellowship from DST-SERB (India)



\begin{thebibliography}{9}
	
	\bibitem{PollardGoldman2016}
	T. D. Pollard and R. D. Goldman (Editors), {\it The Cytoskeleton}, Cold Spring Harbor Laboratory Press (2016).
	
	
	\bibitem{Tanejaetal2020}
	N. Taneja, M. R. Bersi, S. M. Baillargeon, A. M. Fenix, J. A. Cooper, R. Ohi, V. Gama, W. D. Merryman and D. T. Burnette, {\it Precise Tuning of Cortical Contractility Regulates Cell Shape during Cytokinesis}, Cell Reports {\bf 31}, 107477 (2020).
	
	\bibitem{Matthewsetal2006}
	B. D. Matthews,  D. R. Overby,  R. Mannix and  D. E. Ingber, {\it Cellular adaptation to mechanical stress: role of integrins, Rho, cytoskeletal tension and mechanosensitive ion channels}, Journal of Cell Science {\bf 119}, 508--518 (2006).
	
	\bibitem{banerjeeetal2020} S. Banerjee, M. L. Gardel and U. S. Schwarz, {\it The actin cytoskeleton as an active adaptive material}, Annual Review of Condensed Matter Physics {\bf 11}, 421--439  (2020).
	
	
	
	
	\bibitem{Brangwynne2021}
	S. F. Shimobayashi, P. Ronceray, D. W. Sanders, M. P. Haataja and C. P. Brangwynne, {\it Nucleation landscape of biomolecular condensates}, Nature {\bf 599}, 503--506 (2021).
	
	

	\bibitem{Marshall2020}
	W. F. Marshall, {\it Pattern formation and complexity in single cells}, Current Biology {\bf 30}, R544--R552 (2020).
	
	\bibitem{jimenezetal2021} A. J. Jimenez, A. Schaeffer, C. D. Pascalis, G. Letort, B. Vianay, M. Bornens, M. Piel, L. Blanchoin and M. Th{\'e}ry, {\it Acto-myosin network geometry defines centrosome position}, Current Biology {\bf 31}, 1206--1220  (2021).
	
	
	
	\bibitem{Makhija2015}
	E. Makhija, D. S. Jokhun and G. V. Shivashankar, {\it Nuclear deformability and telomere dynamics are regulated by cell geometric constraints}, The Proceedings of the National Academy of Sciences {\bf 113}, E32--E40 (2015).
	
	\bibitem{Sunetal2020}
	X. Sun, D. Y. Z. Phua, L. Axiotakis Jr., M. A. Smith, E. Blankman, R. Gong, R. C. Cail, S. E. de los Reyes, M. C. Beckerle, C. M. Waterman and G. M.Alushin, {\it Mechanosensing through Direct Binding of Tensed F-Actin by LIM Domains}, Developmental Cell {\bf 55}, 468--482 (2020).
	
	
	
	\bibitem{Shutovaetal2017}
	M. S. Shutova, S. B. Asokan, S. Talwar, R. K. Assoian, J. E. Bear and T. M. Svitkina, {\it Self-sorting of nonmuscle myosins IIA and IIB polarizes the cytoskeleton and modulates cell motility}, Journal of Cell Biology {\bf 216}, 2877--2889 (2017).
	
	\bibitem{Kovacsetal2007}
	M. Kov\'{a}cs, K. Thirumurugan, P. J. Knight, and J. R. Sellers, {\it Load-dependent mechanism of nonmuscle myosin 2}, The Proceedings of the National Academy of Sciences {\bf 104}, 9994--9999 (2007).
	
	
	\bibitem{Fernandez-Gonzalezetal2009}
	R. Fernandez-Gonzalez, S. M. Simoes, J-C R\"{o}per, S. Eaton, and J. A. Zallen, {\it Myosin II dynamics are regulated by tension in intercalating cells}, Developmental Cell {\bf 17}, 736--743 (2009)
	
	\bibitem{Mullaetal2022}
	Y. Mulla, M. J. Avellaneda, A. Roland, L. Baldauf, S. J. Tans and G. H. Koenderink, {\it Weak catch bonds make strong networks}, bioRxiv preprint doi: https://doi.org/10.1101/2020.07.27.219618 (2022).
	
	
	\bibitem{Beachetal2014}
	J. R. Beach, L. Shao, K. Remmert, D. Li, E. Betzig and J. A. Hammer III, {\it Nonmuscle Myosin II Isoforms Coassemble in Living Cells}, Current Biology {\bf 24}, 1160--1166 (2014).
	
	\bibitem{HotulainenLappalainen2006}
	P. Hotulainen and P. Lappalainen, {\it Stress fibers are generated by two distinct actin assembly mechanisms in motile cells}, The Journal of Cell Biology {\bf 173}, 383--394 (2006).
	
	\bibitem{Bershadsky2017}	
	S. Hu, K. Dasbiswas, Z. Guo, Y-H. Tee, V. Thiagarajan, P. Hersen, T-L. Chew, S. A. Safran, R. Zaidel-Bar, and A. D. Bershadsky , {\it Long-range self-organization of cytoskeletal myosin II filament stacks}, Nature Cell Biology {\bf 19}, 133--141 (2017).
	
	
	\bibitem{Weissenbruchetal2021} K. Wei{\ss}enbruch, J. Grewe, M. Hippler, M. Fladung, M. Tremmel, K. Stricker, U. S. Schwarz and M. Bastmeyer, {\it Distinct roles of nonmuscle myosin {II} isoforms for establishing tension and elasticity during cell morphodynamics}, eLife {\bf 10}, e71888 (2021).
	
	
	
	    \bibitem{Vicente-Manzanaresetal2008}
	    M. Vicente-Manzanares,  M. A. Koach,  L. Whitmore,  M. L. Lamers and  A. F. Horwitz,
        {\it Segregation and activation of myosin IIB creates a rear in migrating cells}, Journal of Cell Biology {\bf 183}, 543--554 (2008).
	
		\bibitem{silvaetal2011} M. S. de Silva, M. Depken, B. Stuhrmann, M. Korsten, F. C. MacKintosh and G. H. Koenderink, {\it Active multistage coarsening of actin networks driven by myosin motors}, The Proceedings of the National Academy of Sciences {\bf 108}, 9408--9413  (2011).
		
		\bibitem{koenderink-paluch2018} G. H. Koenderink and E. K. Paluch, {\it Architecture shapes contractility in actomyosin networks}, Current Opinion in Cell Biology {\bf 50}, 79--85  (2018).
		
	    

		
		
		
		
		
		
		
		\bibitem{zhangetal2021} R. Zhang, S. A. Redford, P. V. Ruijgrok, N. Kumar, A. Mozaffari, S. Zemsky, A. R. Dinner, V. Vitelli, Z. Bryant, M. L. Gardel and J. J. de Pablo, {\it Spatiotemporal control of liquid crystal structure and dynamics through activity patterning}, Nature Materials {\bf 20}, 875--882  (2021).
		
		
			\bibitem{zallen2011}R. Fernandez-Gonzalez, and J.A. Zallen, {\it Oscillatory behaviors and hierarchical assembly of contractile structures in intercalating cells},
		Physical Biology {\bf 8}, 045005 (2011).
		
		
		\bibitem{munjal2015}
		A. Munjal, J-M. Philippe, E. Munro and T. Lecuit,
		{\it A self-organized biomechanical network drives shape changes during tissue morphogenesis},
		 Nature {\bf 524}, 351--355 (2015).
		
		
		\bibitem{debsankar2017} D. S. Banerjee, A. Munjal, T. Lecuit and M. Rao,
		{\it Actomyosin pulsation and flows in an active elastomer with turnover and network remodeling},
		Nature Communications {\bf 8}, 1121 (2017).
	
\bibitem{Lehtimakietal2021}
J. I. Lehtim\"{a}ki, E. K. Rajakyl\"{a}, S. Tojkander and Pekka Lappalainen, {\it Generation of stress fibers through myosin-driven re-organization of the actin cortex}, eLife {\bf 10}, e60710 (2021).
			\bibitem{Marchettietal2013}M. C. Marchetti, J.-F. Joanny, S. Ramaswamy, T. B. Liverpool, J. Prost, M. Rao and R.A. Simha, {\it Hydrodynamics of Soft Active Matter}, Review of Modern Physics {\bf 85}, 1143--1189 (2013).
	
		
	
	
	
	
	
		\bibitem{BanerjeeLiverpoolMarchetti2011}  S. Banerjee, T.B. Liverpool and M.C. Marchetti,
		{\it Generic phases of cross-linked active gels: Relaxation, oscillation and contractility}, 
		Europhysics Letters {\bf 96}, 58004 (2011).
		
		
		\bibitem{BanerjeeMarchetti2011} S. Banerjee and M.C. Marchetti, {\it Instabilities and oscillations in isotropic active gels},
		Soft Matter {\bf 7}, 463 (2011).
		
		
				\bibitem{Bray2002}
		A. J. Bray, {\it Theory of phase ordering kinetics}, Advances in Physics {\bf 51}, 481--587 (2002).
		
				\bibitem{Fruchartetal2021}
		M. Fruchart, R. Hanai, P. B. Littlewood, and V. Vitelli, {\it Non-reciprocal phase transitions}, Nature {\bf 592}, 363--369 (2021).
		
		\bibitem{Kato1984}
		T. Kato, {\it Perturbation Theory for Linear Operators}, 2nd edn, Springer (1984).	
		
		
		
		
				\bibitem{Youetal2020} Z. You, A. Baskaran and M. C. Marchetti, {\it Nonreciprocity as a generic route to traveling states}, The Proceedings of the National Academy of Sciences {\bf 117}, 19767--19772 (2020).
	
		
		
		
			\bibitem{Roncerayetal2016}P. Ronceray, C. P. Boedersz and M. Lenz, {\it Fiber networks amplify active stress}, The Proceedings of the National Academy of Sciences {\bf 113}, 2827--2832 (2016).
		
		\bibitem{Roncerayetal2019}P. Ronceray, C. P. Boedersz and M. Lenz, {\it Stress-dependent amplification of active forces in nonlinear elastic media}, Soft Matter {\bf 15}, 331--338 (2019).
		
		\bibitem{GeigerSpatzBershadsky2009}
	B. Geiger, J. P. Spatz and A. D. Bershadsky, {\it Environmental sensing through focal adhesions},  Nature Reviews Molecular Cell Biology {\bf 10}, 21--33 (2009).
		
		
	\bibitem{GuptonWaterman-Storer2006} S. L. Gupton and C. M. Waterman-Storer, {\it Spatiotemporal Feedback between Actomyosin and Focal-Adhesion Systems Optimizes Rapid Cell Migration}, Cell {\bf 125}, 1361 (2006).	


	
	\bibitem{Rosakisetal2015}
		P. Rosakis, J. Notbohm, and G. Ravichandran, {\it A model for compression-weakening materials and the elastic fields due to contractile cells}, Journal of The  Mechanics and Physics of Solids, {\bf 85}, 16--32 (2015).
		
		
		\bibitem{Grekasetal2021}
		G. Grekas, M. Proestaki, P. Rosakis, J. Notbohm, C. Makridakis and G. Ravichandran, {\it Cells exploit a phase transition to mechanically remodel the fibrous extracellular matrix}, Journal of The Royal Society Interface, {\bf 18}, 20200823 (2021).
		
		\bibitem{Keelingetal2017}
		M. C. Keeling, L. R. Flores, A. H. Dodhy, E. R. Murray and N. Gavara, {\it Actomyosin and vimentin cytoskeletal networks regulate nuclear shape, mechanics and chromatin organization},Scientific Reports {\bf 7}, 5219 (2017). 
		
		
		
		
		\bibitem{KnowlesSternberg1978}
		J. K. Knowles and E. Sternberg, {\it On the failure of ellipticity and the emergence of discontinuous deformation gradients in plane finite elastostatics}, Journal of Elasticity, {\bf 8}, 329--379 (1978).
		
		\bibitem{SimpsonSpector1985}
		H. C. Simpson and S. J. Spector, {\it On failure of the complementing condition and nonuniqueness in linear elastostatics}, Journal of Elasticity, {\bf 15}, 229--231 (1985).
		
		\bibitem{EgorovPalencia2011}
		Yu. V. Egorov and E. Sanchez-Palencia, {\it On ill-posedness of free-boundary problems for highly compressible two-dimensional elastic bodies}, St. Petersburg Math. J. {\bf 22}, 913--926 (2011).
		
		\bibitem{Blumenfeld2006}
		R. Blumenfeld, {\it Isostaticity and controlled force transmission in the cytoskeleton: A model awaiting experimental evidence}, Biophysical Journal {\bf 91}, 1970--1983 (2006).
		
		\bibitem{Catesetal1998}
		M. E. Cates, J. P. Wittmer, J. P. Bouchaud, and P. Claudin, {\it Jamming, Force Chains, and Fragile Matter}, Physical Review Letters {\bf 81}, 1841 (1998). 
		
		
		\bibitem{Bouchaud2002}
		 J. P. Bouchaud, {\it Granular Media: Some ideas from Statistical Physics}, in Slow Relaxations and Nonequilibrium Dynamics in Condensed Matter, NATO Advanced Study Institute, Les Houches, Session LXXVII, eds. J.L. Barrat, M. Feigelman, J. Kurchan and J. Dalibard. (2002).
	
		
	\bibitem{Otto2003} M. Otto, J.P. Bouchaud, P. Claudin and J.E.S. Socolar, {\it Anisotropy in granular media: Classical elasticity and directed-force chain network}, Physical Review E {\bf 67}, 031302 (2003).
	


    
		
		
		
		
		
		
		
	
		
		
		

		
		
		
		
	
		
		
		
		
		
		
		
		
		
		

		
		
		
		
		
		
		
		
		

		
		
	
		

		
		
		
		
		\bibitem{rct22}
		A. Roychowdhury and L. Truskinovsky, {\it Towards Marginal Elascity}, to be submitted.
		
		\bibitem{Trefethenetal93} L. N. Trefethen, A. E. Trefethen, S. C. Reddy and T A Driscoll, {\it Hydrodynamic Stability Without Eigenvalues}, Science {\bf 261}, 578--584  (1993).
		
	
		
		\bibitem{TrefethenEmbree2005} L. N. Trefethen and M. Embree, {\it Spectra and Pseudospectra}, Princeton University Press  (2005).
		
		\bibitem{dedalus2020}
		K. J. Burns, G. M. Vasil, J. S. Oishi, D. Lecoanet and B. P. Brown, {\it Dedalus: A Flexible Framework for Numerical Simulations with Spectral Methods}, Physical Review Research {\bf 2}, 023068 (2020).
		
		\bibitem{GowrishankarRao2016}
		K. Gowrishankar and M. Rao,
		{\it Nonequilibrium phase transitions, fluctuations and correlations in an active contractile polar fluid}, Soft matter {\bf 12}, 2040--2046 (2016).
		
		\bibitem{HusainRao2017}
		K. Husain and M. Rao,
		{\it Emergent structures in an active polar fluid: Dynamics of shape, scattering, and merger}, Physical Review Letters {\bf 118}, 078104 (2017).
		
		
		
		\bibitem{eggersfontelos-book2015} J. Eggers and M. A. Fontelos, {\it Singularities: {F}ormation, {S}tructure, and {P}ropagation}, Cambridge University Press, Cambridge, UK  (2015).
		
	
		
		
		

		
		\bibitem{Kanatani1984}
		K-I. Kanatani, {\it Distribution of directional data and fabric tensors}, International Journal of Engineering Science {\bf 22}, 149--164 (1984).
		
		
		
		
		
		
		
		
		
		
		
		\bibitem{SimhaBhattacharya1998}  N. K. Simha and K. Bhattacharya, 
	    {\it Kinetics of Phase Boundaries with Edges and Junctions}, 
	    Journal of The Mechanics and Physics of Solids {\bf 46}, 2323--2359 (1998).
		
	    \bibitem{Fischeretal2012}  F. D. Fischer, J. Svoboda and K. Hackl, 
	    {\it Modelling the kinetics of a triple junction}, 
	    Acta Materiala {\bf 60}, 4704--4711 (2012).
	    
	    \bibitem{RamaswamyRao2007}
	    S. Ramaswamy and M. Rao, {\it Active-filament hydrodynamics: instabilities, boundary conditions and rheology}, New Journal of Physics {\bf 9}, 423 (2007).
	    
	    \bibitem{ingberetal2014} D. E. Ingber, N. Wang and D. Stamenovi{\'c}, {\it Tensegrity, cellular biophysics, and the mechanics of living systems}, Reports on Progress in Physics {\bf 77}, 046603  (2014).
	    
	    
	    
	    \bibitem{PishvarHarne2020}
	    M. Pishvar and R. L. Harne, {\it Foundations for Soft, Smart Matter by Active Mechanical Metamaterials}, Advanced Science {\bf 7}, 2001384 (2020).
	    
	    \bibitem{Qietal2022}
	    J. Qi, Z. Chen, P. Jiang, W. Hu, Y. Wang, Z. Zhao, X. Cao, S. Zhang, R. Tao, Y. Li and D. Fang,
	    {\it Recent Progress in Active Mechanical Metamaterials and Construction Principles}, Advanced Science {\bf 9}, 2102662 (2022).

	\end{thebibliography}


\end{document}


	\title{Supplementary Information\\
	Emergence of Tension Chains and Active Force Patterning}
	\author{Ayan Roychowdhury}
	\thanks{Joint first author}
	\affiliation{Simons Centre for the Study of Living Machines, National Centre for Biological Sciences-TIFR, Bengaluru, India  560065.}
	\author{Saptarshi Dasgupta}
        \thanks{Joint first author}
	\affiliation{Simons Centre for the Study of Living Machines, National Centre for Biological Sciences-TIFR, Bengaluru, India  560065.}
	\author{Madan Rao}
	\thanks{rao.madan@gmail.com}
	\affiliation{Simons Centre for the Study of Living Machines, National Centre for Biological Sciences-TIFR, Bengaluru, India  560065.}
	

\maketitle



	
	



	

\tableofcontents
\newpage
\section{Hydrodynamic Equations for a Mixture of Contractile Stresslets on an Elastomer}

We describe the dynamics of active stress propagation in the active medium of the cell using hydrodynamic equations for the crosslinked actin mesh and the density of different species of myosin 
filaments, embedded in the viscous cytosol.

\subsection{Equations for the Crosslinked Actin Meshwork embedded in the Cytosol}
We start with a  passive  $d$-dimensional elastomeric meshwork of mass density $\rho_a$, whose displacement with respect to an unstrained reference state is $\boldsymbol{u}$. The meshwork moves in the fluidic cytosol whose velocity is $\boldsymbol{v}$. The hydrodynamic equations for its linear momentum balance and mass balance are, therefore,
\begin{subequations}
	\begin{align}
		&\rho_a\ddot{\boldsymbol{u}}+\Gamma\,(\dot{\boldsymbol{u}}-\boldsymbol{v})=\nabla\cdot\boldsymbol{\sigma},~~\text{and}\label{actin-dyn1}\\
		& \dot{\rho_a}+\nabla\cdot(\rho_a\,\dot{\boldsymbol{u}})=M\nabla^2\frac{\delta F}{\delta \rho_a}+\mathcal{S}_a.\label{actin-dyn2}
	\end{align}
	\label{actin-dyn}
\end{subequations}
Here, $\Gamma>0$ is the friction coefficient of the elastomer with respect to the fluidic cytosol, and $\boldsymbol{\sigma}$ is the total stress in the elastomer; $M$ represents the mobility of permeation of the meshwork, and $\mathcal{S}_a$ represents turnover of the actin meshwork. $F$ is the free energy functional for the passive meshwork: 
\begin{equation}
	F(\boldsymbol{\epsilon},\rho_a)=\int_{\Omega}f_B\,d^{d}r;
\end{equation}
where the free energy density $f_B(\boldsymbol{\epsilon},\rho_a)$ depends on the linearized strain $\boldsymbol{\epsilon}:=(\nabla\boldsymbol{u}+\nabla\boldsymbol{u}^T)/2$ of the elastomer, and the mass density $\rho_a$.

The hydrodynamic equations of the fluidic cytosol are given by

\begin{equation}
	\rho_f\Big(\dot{\boldsymbol{v}}+\boldsymbol{v}\cdot\nabla\boldsymbol{v}\Big)=\eta^s_f\nabla^2\boldsymbol{v}+\Big(\eta^b_f+\frac{\eta^s_f}{d}\Big)\nabla(\nabla\cdot\boldsymbol{v})-\nabla p+\Gamma\,(\dot{\boldsymbol{u}}-\boldsymbol{v});
\end{equation}
here, $\rho_f$ is the density of the fluid, and $\eta^s_f$ and $\eta^b_f$ are the shear and bulk viscosities of the fluid. The fluid pressure $p$ appears due to the total incompressibility of the meshwork-fluid system:
\begin{equation}
	\nabla\cdot\Big(c_a\,\dot{\boldsymbol{u}}+(1-c_a)\boldsymbol{v}\Big)=0;
\end{equation} 
here, $c_a$ is volume fraction of the meshwork. 
Here, for convenience, we ignore the hydrodynamics of the fluid,  
permissible in the limit when $c_a \approx 1$.

\subsection{Equations for the Stresslets}

Consider a mixture of active contractile stresslets with different contractilities undergoing turnover onto this elastomer. We assume that the stresslets binding onto the elastomer are recruited from an infinite pool of stresslets unbound to the elastomer, and, hence, disregard the dynamics of these unbound stresslets.  Restricting to a binary mixture, let $\rho_i$, $i=1,2$,  be the density fields of the two species of bound stresslets. The bound stresslets get advected by the local velocity $\dot{\boldsymbol{u}}$ of the elastomer, and diffuse on it with the same diffusion coefficient $D$. Let the stresslets bind onto the elastomer with rates $k_i^{b}>0$, and unbind from the elastomer with rates $k_i^{u}(\boldsymbol{\epsilon})>0$ that in principle depends on the strain $\boldsymbol{\epsilon}$ of the elastomer. The dynamics of these bound stresslets is, hence, governed by
\begin{equation}
	\dot\rho_i+\nabla\cdot(\rho_i\,\dot{\boldsymbol{u}})=\nabla\cdot(D\,\nabla\rho_i)+\mathcal{S}_i,
	\label{stresslets-dyn}
\end{equation}
here, $\mathcal{S}_i:=k_i^{b}\,\rho_a - k_i^{u}(\boldsymbol{\epsilon})\,\rho_i$ represent turnover of the bound stresslets.

We will assume the Hill  form for the  unbinding rates:
\begin{equation}
	k_i^{u}(\boldsymbol{\epsilon})=k_{i0}^{u}\, e^{\alpha_i\,\epsilon},~~~~~i=1,2
	\label{unbindingrates}
\end{equation}
where $k_{i0}^{u}>0$, $i=1,2$,  are the strain independent parts of the respective rates, $\epsilon:=\text{tr}\,\boldsymbol{\epsilon}$ is the isotropic strain, and the  dimensionless numbers $\alpha_i$ capture whether the bond is catch or slip type: $\alpha_i>0$  ensures  that local  contraction (extension) will decrease (increase) the unbinding of the stresslets, i.e., the bond is of catch type, while $\alpha_i<0$  ensures  that local  contraction (extension) will increase (decrease) the unbinding of the stresslets, i.e., the bond is of slip type.

In the small timescale of the turnover processes of the stresslets comparatively larger than the tunrover of the crosslinked actin meshwork, the elastomer can be considered as long lived, i.e., right hand side of \eqref{actin-dyn2} is zero. This implies that elastomer density $\rho_a$ is enslaved to the isotropic strain of the elastomer: $\delta\rho_a\propto-\epsilon$ (obtained from a variation of \eqref{enrg} below); here,  $\delta\rho_a:=\rho_a-\rho_a^0$ is the deviation of the elastomer density from its state value  $\rho_a^0$

\subsection{Constitutive Equations}

The total stress $\boldsymbol{\sigma}=\boldsymbol{\sigma}^p+\boldsymbol{\sigma}^a$ is the summation of
passive stress $\boldsymbol{\sigma}^p$ and the active stress $\boldsymbol{\sigma}^a$.

\subsubsection{Passive stress}
  The elastic part $\boldsymbol{\sigma}^e := \frac{\delta F}{\delta \,\boldsymbol{\epsilon}}$ comes from the  free-energy functional   $F(\boldsymbol{\epsilon},\rho_a)
=\int d^2r f_B$, where 
\begin{eqnarray}
	f_B&=&\frac{1}{2}\mathbb{C}[\boldsymbol{\epsilon}]\cdot\boldsymbol{\epsilon}+C\,\delta\rho_a\,\epsilon+\frac{A}{2}\delta\rho_a^2
	\label{enrg}
\end{eqnarray}
is the free energy density; 
the elastic stiffness tensor $\mathbb{C}$ is positive definite, and $C>0$, $A>0$ from thermodynamic stability. Assuming $\mathbb{C}$ to be isotropic, we obtain
\begin{equation}
	\boldsymbol{\sigma}^e=\frac{\delta F}{\delta \boldsymbol{\epsilon}}\nonumber\\
	=B\,\epsilon\,\mathbf{I}+2\mu\,\tilde{\boldsymbol{\epsilon}}+C\,\delta\rho_a\,\mathbf{I}
	=\Big(B-\frac{C^2}{A}\Big)\,\epsilon\,\mathbf{I}+2\mu\,\tilde{\boldsymbol{\epsilon}},
\end{equation}
noting that $\delta\rho_a=-\frac{C}{A}\epsilon$. Here, $\tilde{\boldsymbol{\epsilon}}:=\boldsymbol{\epsilon}-(1/d)\epsilon\,\mathbf{I}$ is the deviatoric strain tensor. The elastic moduli of the actin mesh is set by the crosslinker density.

The viscous stress is 
\begin{equation}
	{\boldsymbol{\sigma}}^d =\boldsymbol{\eta}\,\dot{\boldsymbol{\epsilon}}
\end{equation}
where $\boldsymbol{\eta}$ is the positive definite viscosity tensor.

\subsubsection{Active stress}


At the macroscopic/coarse-grained scale, the isotropic active stress  $\boldsymbol{\sigma}^a$ is of the form  $\boldsymbol{\sigma}^a=\Delta\mu\,\chi(\rho_a)\zeta(\{\rho_i\})\mathbf{I}$, where $\Delta\mu$ is the chemical potential change due to ATP hydrolysis, and $\chi(\rho_a)$ is a sigmoidal function that encodes the dependence of the active stress on the meshwork density. We will take $\Delta\mu=1$ for simplicity.

For a single stresslet system, Taylor expanding the function $\chi(\rho_a)$ about the state value $\rho_a=\rho_a^0$ upto cubic order leads to

\begin{eqnarray}
	\boldsymbol{\sigma}^a&=&\chi(\rho_a)\, \zeta(\rho)\,\mathbf{I}\nonumber\\
	&=&\Bigg(\chi(\rho_a^0)+\chi'(\rho_a^0)\,\delta\rho_a+\frac{1}{2!}\chi''(\rho_a^0)\,\delta\rho_a^2+\frac{1}{3!}\chi''(\rho_a^0)\,\delta\rho_a^3+o(\delta\rho_a^3)\Bigg)\,\zeta(\rho)\,\mathbf{I}\nonumber\\
	&=&\Bigg(\chi(\rho_a^0)-\chi'(\rho_a^0)\frac{C}{A}\,\epsilon+\frac{1}{2!}\chi''(\rho_a^0)\,\Big(\frac{C}{A}\,\epsilon\Big)^2-\frac{1}{3!}\chi''(\rho_a^0)\,\Big(\frac{C}{A}\,\epsilon\Big)^3+o(\epsilon^3)\Bigg)\,\zeta(\rho)\,\mathbf{I}.
	\label{activestress-sigle}
\end{eqnarray}

In case of mixtures, we assume that the stresslets do not interact directly but only via the strain of the elastomer. Hence, the function $\zeta(\{\rho_i\})$ can be additively decomposed into individual contributions coming from each stresslet species, i.e., $\zeta(\{\rho_i\})=\sum_i\zeta_i(\rho_i)$. We assume the functions $\zeta_i(\rho_i)$ to be linear in $\rho_i$, i.e., $\zeta_i(\rho_i)=\zeta_i\rho_i$ (no sum over $i$) where the contractilities $\zeta_i>0$ are different for different species $i$. 
For a binary mixture, we obtain
\begin{eqnarray}
	\boldsymbol{\sigma}^a&=&\chi(\rho_a) \Big(\zeta_{\text{avg}}\,\rho_1+\zeta_{\text{rel}}\,\rho_2\Big)\mathbf{I}\nonumber\\
	&=&\Bigg(\chi(\rho_a^0)-\chi'(\rho_a^0)\frac{C}{A}\,\epsilon+\frac{1}{2!}\chi''(\rho_a^0)\,\Big(\frac{C}{A}\,\epsilon\Big)^2-\frac{1}{3!}\chi''(\rho_a^0)\,\Big(\frac{C}{A}\,\epsilon\Big)^3+o(\epsilon^3)\Bigg)\times\Big(\zeta_{\text{avg}}\,\rho_1+\zeta_{\text{rel}}\,\rho_2\Big)\mathbf{I}.
	\label{activestress3}
\end{eqnarray}

The dependence of the active stress on the stresslet density has a  similar for the single species and for mixtures.

\subsubsection{Total stress}
The expression for the total stress, assuming isotropic form of the viscosity tensor $\boldsymbol{\eta}$, becomes

\begin{equation}
	\boldsymbol{\sigma}
	=\Big(\sigma_0
	+\tilde{B}\,\epsilon
	+B_2\,\epsilon^2
	+B_3\,\epsilon^3\Big)\mathbf{I}+2{\mu}\,\tilde{\boldsymbol{\epsilon}}+\eta^b\text{tr}\,\dot{\boldsymbol{\epsilon}}+2\eta^s\dot{\tilde{\boldsymbol{\epsilon}}},
	\label{totalstress}
\end{equation}

where

\begin{eqnarray}
	\sigma_0&:=& \chi(\rho_a^0)\Big(\zeta_{\text{avg}}\,\rho_1+\zeta_{\text{rel}}\,\rho_2\Big),\nonumber\\
	\tilde{B}&:=&B-\frac{C^2}{A}-\chi'(\rho_a^0)\frac{C}{A}\Big(\zeta_{\text{avg}}\,\rho_1+\zeta_{\text{rel}}\,\rho_2\Big),\nonumber\\
	B_2&:=&\frac{1}{2!}\chi''(\rho_a^0)\,\Big(\frac{C}{A}\Big)^2\Big(\zeta_{\text{avg}}\,\rho_1+\zeta_{\text{rel}}\,\rho_2\Big),\nonumber\\
	B_3&:=&-\frac{1}{3!}\chi'''(\rho_a^0)\,\Big(\frac{C}{A}\Big)^3\Big(\zeta_{\text{avg}}\,\rho_1+\zeta_{\text{rel}}\,\rho_2\Big).
\end{eqnarray}

\begin{figure}[H]
	\centering
	\includegraphics[scale=.5]{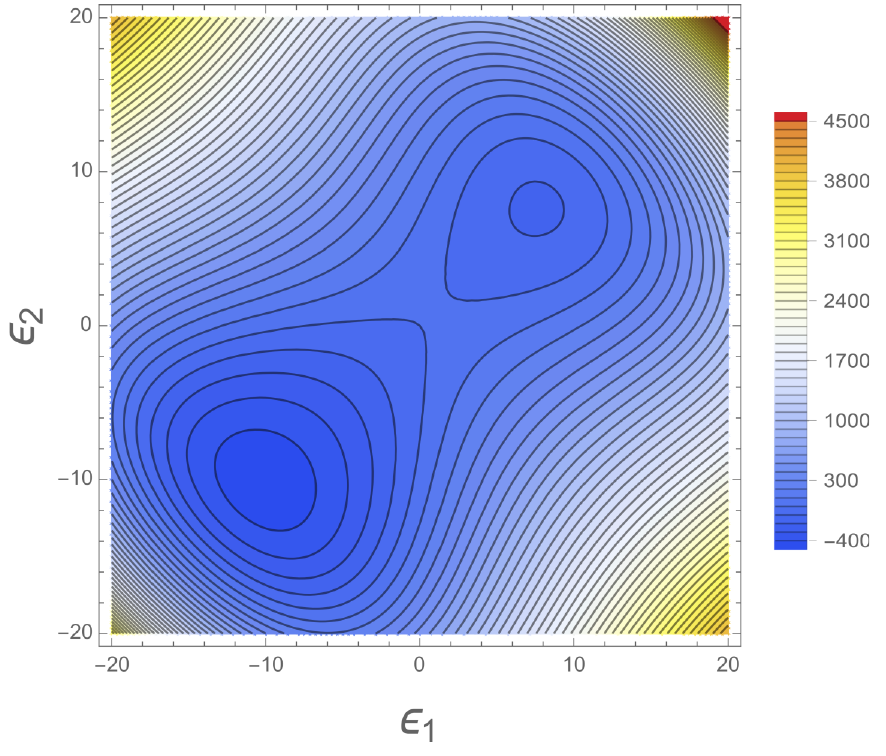}
	\caption{Active elastomer with double minima of the effective strain energy density function $w=\sigma_0\epsilon+\frac{1}{2}\tilde{B}(\rho,\phi)\epsilon^2+\frac{1}{3}B_2(\rho,\phi)\epsilon^3+\frac{1}{4}B_3(\rho,\phi)\epsilon^4+\mu\,\text{tr}\tilde{\boldsymbol{\epsilon}}^2$ in the principal strain space $\epsilon_1$-$\epsilon_2$, for $B=4$, $\mu=5$, $C=1$, $A=1$,  $\chi(\rho_a^0)=0.1$, $\chi'(\rho_a^0)=0.1$, $\chi''(\rho_a^0)=0.001$, $\chi'''(\rho_a^0)=-0.001$, and $\zeta_{\text{avg}}=2$, $\zeta_{\text{rel}}=5$, $\rho=1$, $\phi=6$}
	\label{doubleminima}
\end{figure}

Here, $\sigma_0$ is the purely active back pressure, $\tilde{B}$ is the activity renormalized bulk modulus of linear elasticity,  $B_2$ and $B_3$ are purely active nonlinear bulk moduli, which depend on
$\rho_1$ and $\rho_2$. For the effective material to show contractile response, $\chi''(\rho_a^0)$ and hence $B_2$ needs to be positive (Fig.\,\ref{doubleminima}). For stability of the nonlinear elastic material,
$\chi'''(\rho_a^0)$  must be negative (rendering $B_3$ positive).


\subsection{Hydrodynamic equations in terms of average and relative densities of the stresslets}

Define the average and the relative densities of the bound stresslets,
\begin{equation}
	\rho:=\frac{\rho_1+\rho_2}{2}~~\text{and}~~ \phi:=\frac{\rho_1-\rho_2}{2},
\end{equation}
 respectively; note that $\phi$ is the order parameter for segregation. 

We will further assume the overdamped limit of the elastomer, i.e., $|\rho_a\,\ddot{\boldsymbol{u}}|\ll |\Gamma\,\dot{\boldsymbol{u}}|$.  With this, \eqref{actin-dyn1} and \eqref{stresslets-dyn} reduce to the following system of equations
\begin{subequations}
	\begin{align}
		&\Gamma\,\dot{\boldsymbol{u}}= \nabla\cdot\boldsymbol{\sigma}, \label{dyn-u-f12}\\
		& \dot{\rho} +\nabla\cdot(\rho\,\dot{\boldsymbol{u}}) = {D}\,\nabla^2\rho+k^b_{\text{avg}}\,\bigg(\rho_a^0-\frac{C}{A}\nabla\cdot\boldsymbol{u}\bigg)-k^{u}_{\text{avg}}(\boldsymbol{\epsilon})\,\rho-k^{u}_{\text{rel}}(\boldsymbol{\epsilon})\,\phi,\label{dyn-e-f12}\\
		&\dot{\phi} +\nabla\cdot(\phi\,\dot{\boldsymbol{u}}) = {D}\,\nabla^2\phi+k^b_{\text{rel}}\,\bigg(\rho_a^0-\frac{C}{A}\nabla\cdot\boldsymbol{u}\bigg)-k^{u}_{\text{avg}}(\boldsymbol{\epsilon})\,\phi-k^{u}_{\text{rel}}(\boldsymbol{\epsilon})\,\rho;\label{dyn-c-f12}
	\end{align}
	\label{dyn-f12}
\end{subequations}
where $\boldsymbol{\sigma}$ is defined in \eqref{totalstress}.

Here,
\begin{equation}
	k^{u}_{\text{avg}}:=\frac{k^u_1+k^u_2}{2},~~~k^{u}_{\text{rel}}:=\frac{k^u_1-k^u_2}{2}
\end{equation}
are the average and relative unbinding rates, respectively;
\begin{equation}
	k^{b}_{\text{avg}}:=\frac{k^{b}_1+k^{b}_2}{2}>0~~\text{and}~~k^{b}_{\text{rel}}:=\frac{k^{b}_1-k^{b}_2}{2}
\end{equation}
are the average and relative binding rates, respectively; and

\begin{equation}
	\zeta_{\text{avg}}:=\frac{\zeta_{\text{avg}}+\zeta_{\text{rel}}}{2}>0,~\zeta_{\text{rel}}:=\frac{\zeta_{\text{avg}}-\zeta_{\text{rel}}}{2}
\end{equation}
are the average and relative contractility, respectively.

Using the Hill form \eqref{unbindingrates}, we write
\begin{subequations}
	\begin{align}
		& k^{u}_{\text{avg}}(\boldsymbol{\epsilon})=k_1+k_3\,\epsilon+o(|\boldsymbol{\epsilon}|),\\
		& k^{u}_{\text{rel}}(\boldsymbol{\epsilon})=k_2+k_4\,\epsilon+o(|\boldsymbol{\epsilon}|);
	\end{align}
\end{subequations}
where
\begin{equation}
	k_1:=\frac{k^{u}_{10}+k^{u}_{20}}{2}>0,~~\text{and}~~k_2:=\frac{k^{u}_{10}-k^{u}_{20}}{2}
\end{equation}
are the bare (i.e., strain independent) average and relative unbinding rates, respectively, and
\begin{equation}
	k_3:=\frac{k^{u}_{10}\alpha_1+k^{u}_{20}\alpha_2}{2},~~\text{and}~~k_4:=\frac{k^{u}_{10}\alpha_1-k^{u}_{20}\alpha_2}{2}
\end{equation}
are the coefficients of the linear strain dependent parts of the  relative and average unbinding rates, respectively. 

Note that if $\zeta_{\text{rel}}=0$, $k_2=0$ and $k_4=0$, the distinction between the two contractile species disappears and the system becomes effectly one species. Hence, these three parameters in our model cannot be made zero simultaneously.

\subsection{Non-dimensionalizing the equations}

Let the characteristic time scale be $t^\star:=1/k^{b}_{\text{avg}}$, the characteristic length scale $l^\star:=\sqrt{\eta_b/\Gamma}$, and the characteristic density  $\rho_a^0$. We non-dimensionalize all the variables with the following redefinitions:
\begin{subequations}
	\begin{align}
		&\frac{t}{t^\star}\to t,~~\frac{{x}}{l^\star}\to{x},~~l^\star\partial_{x}\to \partial_{x},~~{l^\star}^2\partial^2_{xx}\to \partial^2_{xx}\\
		&\frac{{u}}{l^\star}\to {u},~~\frac{\rho}{\rho_a^0}\to \rho,~\frac{\phi}{\rho_a^0}\to\phi,~~\frac{D\,t^\star}{{l^{\star}}^2}\to D,\\
		&t^\star\,k^{b}_{\text{rel}}\to k^{b}_{\text{rel}},~~t^\star\,k^{u}_{\text{avg}}\to k^{u}_{\text{avg}},~~t^\star\,k^{u}_{\text{rel}}\to k^{u}_{\text{rel}},\\
		&\frac{B\, t^\star}{\Gamma {l^\star}^2}\to B,~~\frac{C \,t^\star \rho_a^0}{\Gamma {l^\star}^2}\to C,~~A\to A,~~\frac{\eta_s}{\eta_b}\to\eta_s\\
		&\frac{\zeta_{\text{avg}} \,t^\star \rho_a^0}{\Gamma {l^\star}^2}\to\zeta_{\text{avg}},~~\frac{\zeta_{\text{rel}}\, t^\star \rho_a^0}{\Gamma {l^\star}^2}\to \zeta_{\text{rel}}.
	\end{align}
\end{subequations}

With this, the non-dimensional form of the governing equations become

\begin{subequations}
	\begin{align}
		&\dot{\boldsymbol{u}}= \nabla\cdot\boldsymbol{\sigma}, \label{dyn-d-u-f12}\\
		& \dot{\rho} +\nabla\cdot(\rho\,\dot{\boldsymbol{u}}) = {D}\,\nabla^2\rho+\bigg(\rho_a^0-\frac{C}{A}\nabla\cdot\boldsymbol{u}\bigg)-k^{u}_{\text{avg}}(\boldsymbol{\epsilon})\,\rho-k^{u}_{\text{rel}}(\boldsymbol{\epsilon})\,\phi,\label{dyn-nd-e-f12}\\
		&\dot{\phi} +\nabla\cdot(\phi\,\dot{\boldsymbol{u}}) = {D}\,\nabla^2\phi+k^b_{\text{rel}}\,\bigg(\rho_a^0-\frac{C}{A}\nabla\cdot\boldsymbol{u}\bigg)-k^{u}_{\text{avg}}(\boldsymbol{\epsilon})\,\phi-k^{u}_{\text{rel}}(\boldsymbol{\epsilon})\,\rho.\label{dyn-nd-c-f12}
	\end{align}
	\label{dyn-nd-f12}
\end{subequations}
with


\begin{eqnarray}
	\boldsymbol{\sigma}&=&\Big(\sigma_0+\tilde{B}\, \epsilon +B_2\,\epsilon^2+B_3\,\epsilon^3+\dot{\epsilon} \Big)\mathbf{I}+2\mu\tilde{\boldsymbol{\epsilon}}+2\eta_s\,\dot{\tilde{\boldsymbol{\epsilon}}}
\end{eqnarray}
as the non-dimensional stress, where

\begin{eqnarray}
	\sigma_0&:=&2\,\chi(\rho^0_a)\big(\zeta_{\text{avg}}\,\rho+\zeta_{\text{rel}}\,\phi\big),\\
	\tilde{B}&:=&B-\frac{C^2}{A}-2\,\chi'(\rho^0_a)\,\frac{C}{A}\big(\zeta_{\text{avg}}\,\rho+\zeta_{\text{rel}}\,\phi\big),\\
	B_2&:=&\chi''(\rho^0_a)\,\bigg(\frac{C}{A}\bigg)^2\,\big(\zeta_{\text{avg}}\,\rho +\zeta_{\text{rel}}\,\phi\big),\\
	B_3&:=&-\frac{\chi'''(\rho^0_a)}{3}\,\bigg(\frac{C}{A}\bigg)^3\,\big(\zeta_{\text{avg}}\,\rho +\zeta_{\text{rel}}\,\phi\big).
\end{eqnarray}

\subsection{Equations in terms of strain}

 The non-dimensional form of the $d$-dimensional equation \eqref{dyn-nd-f12} can be conveniently written in terms of $\epsilon_{ij}$ as
	
		\begin{subequations}
		\begin{align}
			& (1/d)\dot{\epsilon}\,\delta_{ik}+\dot{\tilde{\epsilon}}_{ik}=(1/d)\sigma_{,ik}+(\tilde{\sigma}_{ij,jk}+\tilde{\sigma}_{kj,ji})/2, \label{u-nondim-1d141}\\
			& \dot{\rho}+(1/d)(\rho\,\sigma_{,i})_{,i}+(\rho\,\tilde{\sigma}_{ij,j})_{,i}=D\, \rho_{,ii} +1 -\frac{C}{A}\,\epsilon -(k_1+k_3\,\epsilon+o(\epsilon))\,\rho -(k_2+k_4\,\epsilon+o(\epsilon))\,\phi, \label{rho-nondim-1d141}\\
			& \dot{\phi}+(1/d)(\phi\,\sigma_{,i})_{,i}+(\phi\,\tilde{\sigma}_{ij,j})_{,i}=D\, \phi_{,ii} + k^b_{\text{rel}}\bigg(1-\frac{C}{A}\,\epsilon\bigg) - (k_1+k_3\,\epsilon+o(\epsilon))\,\phi - (k_2+k_4\,\epsilon+o(\epsilon))\,\rho. \label{phi-nondim-1d141}
		\end{align}
		\label{eqn-strain}
	\end{subequations}
	where
	\begin{subequations}
		\begin{align}
			&\sigma=\sigma_0+\tilde{B}\, \epsilon +B_2\,\epsilon^2+B_3\,\epsilon^3 +\dot{\epsilon},\\
			&\tilde{\sigma}_{ij}=2\mu\tilde{\epsilon}_{ij}+2\eta_s\,\dot{\tilde{\epsilon}}_{ij},
		\end{align}
	\end{subequations}
are the isotropic and deviatoric parts of the non-dimensional stress, respectively, and $(\cdot)_{,i}:=\frac{\partial}{\partial x^i}(\cdot)$.

\section{Linear Stability Analysis}

\subsection{Homogeneous unstrained steady state}

For the homogeneous unstrained steady states of the system \eqref{eqn-strain}, we have $\boldsymbol{\epsilon}=\mathbf{0}$, and $\nabla\rho=\nabla\phi=\mathbf{0}$. The dynamical equations reduce to
\begin{subequations}
	\begin{align}
		&k_1\,\rho+k_2\,\phi = 1,\\
		& k_2\rho + k_1\,\phi= k^{b}_{\text{rel}},
	\end{align}
\end{subequations}
which yields
\begin{equation}
	\rho=\rho_0:=\frac{k_1-k_2\,k^{b}_{\text{rel}}}{(k_1)^2-(k_2)^2},~~\text{and}~~	\phi=\phi_0:=\frac{k_1\,k^{b}_{\text{rel}}-k_2}{(k_1)^2-(k_2)^2}.
	\label{hom-unstrained-ss}
\end{equation}
For $\rho_0 >0$, we require either  $k_1>k_2 k^{b}_{\text{rel}}$ and $(k_1)^2>(k_2)^2$, or $k_1<k_2 k^{b}_{\text{rel}}$ and $(k_1)^2<(k_2)^2$.

\subsection{Linear stability of the homogeneous unstrained  steady state}

To study the linearized dynamics of \eqref{eqn-strain} around the homogeneous unstrained steady state, we substitute $\boldsymbol{\epsilon}=\delta\epsilon\,\mathbf{I}$ (i.e., considering the perturbation in the strain direction to be purely isotropic), $\rho=\rho_0+\delta\rho$ and $\phi=\phi_0+\delta\phi$ in \eqref{eqn-strain}, neglect all the terms containing higher powers of ${\delta \epsilon}$, ${\delta \rho}$ and ${\delta \phi}$, and obtain
\begin{subequations}
	\begin{align}
		& \dot{\delta \epsilon} = \tilde{B}_0\,\nabla^2 \delta \epsilon+2\,\chi(\rho_a^0)\big(\,\zeta_{\text{avg}}\,\nabla^2 \delta\rho+\zeta_{\text{rel}}\,\nabla^2 \delta\phi \big) 
		+\nabla^2 \dot{\delta \epsilon} , \label{u-lin1-1d}\\
		& \dot{\delta\rho}+\rho_0\,\dot{\delta \epsilon}=\big(D\,\nabla^2 -k_1\big)\delta\rho  -\bigg(\frac{C}{A} +k_3\,\rho_0 + k_4\,\phi_0\bigg) \delta\epsilon -k_2\,\delta\phi, \label{rho-lin1-1d}\\
		& \dot{\delta\phi}+\phi_0\,\dot{\delta \epsilon}=\big(D\,\nabla^2  - k_1\big)\delta\phi - \bigg(k^{b}_{\text{rel}}\frac{C}{A}  +k_3\,\phi_0 + k_4\,\rho_0\bigg)\,\delta\epsilon- k_2\,\delta\rho, \label{phi-lin1-1d}
	\end{align}
	\label{dyn-lin1-1d}
\end{subequations}
where
\begin{equation}
	\tilde{B}_0:=B-\frac{C^2}{A}-2\chi'(\rho_a^0)\frac{C}{A}\big(\zeta_{\text{avg}}\,\rho_0
	+\zeta_{\text{rel}}\,\phi_0\big)
\end{equation}
is  the (spatiotermporally constant) renormalized linear elastic modulus at the homogeneous steady state.

\subsection{Linear stability of a symmetric mixture of stresslets} 
\label{phi0case}
We focus on the special case  $\phi_0=0$, i.e., symmetric binary mixture. From \eqref{hom-unstrained-ss}, we obtain the necessary condition to maintain this, namely,  $k^{b}_{\text{rel}}=k_2/k_1$. As a consequence we see that      $\rho_0=1/k_1$, and $\tilde{B}_0=B-\frac{C^2}{A}-2\,\chi'(\rho_a^0)\,\frac{C}{A}\,\frac{\zeta_{\text{avg}}}{k_1}$. 


The Fourier transform of \eqref{dyn-lin1-1d} with respect to $\mathbf{x}$ is obtained by substituting the anstatz $\delta A(\mathbf{x},t)=\frac{1}{(2\pi)^d}\int{\delta \hat{A}(t)}e^{i\mathbf{q}\cdot\mathbf{x}}\,d\mathbf{q}$, where $A$ stands for $\epsilon$, $\rho$, $\phi$, and ${\bf q}$ is the wave vector. The resulting system can be written as

\begin{equation}
	\dot{\mathbf{w}}=\mathbf{M}\mathbf{w}
\end{equation}

where $\mathbf{w}({\bf q},t)=\Big(\hat{\delta \epsilon}({\bf q},t)\,\,\hat{\delta \rho}({\bf q},t)\,\,\hat{\delta \phi}({\bf q},t)\Big)^T$, and

\begin{equation}
	\mathbf{M}=\left[\begin{array}{ccc}
		-\frac{\tilde{B}_0\,q^2}{1+q^2} & -\frac{2\chi(\rho_a^0)\,\zeta_{\text{avg}}\,q^2}{1+q^2} & -\frac{2\chi(\rho_a^0)\,\zeta_{\text{rel}}\,q^2}{1+q^2} \\
		-\bigg(\frac{C}{A}+\frac{k_3}{k_1}-\frac{\tilde{B}_0q^2}{(1+q^2)k_1}\bigg) & -D q^2-k_1+\frac{2\chi(\rho_a^0)\,\zeta_{\text{avg}}\,q^2}{(1+q^2)k_1} & -k_2+\frac{2\chi(\rho_a^0)\,\zeta_{\text{rel}}\,q^2}{(1+q^2)k_1} \\
		-\bigg(\frac{k_2}{k_1}\frac{C}{A}+\frac{k_4}{k_1}\bigg) & -k_2 & -D q^2- k_1
	\end{array}\right],
	\label{simplified-lin-system}
\end{equation}
with $q:=|{\bf q}|$.

If $\lambda_i(q)$ are distinct eigenvalues of  $\mathbf{M}(q)$ and $\mathbf{v}_i(q)$ are the corresponding (linearly independent) eigenvectors, then we can write the general solution as
\begin{equation}
	\mathbf{w}(q,t)=\sum_{i=1}^{3}c_i(q)e^{\lambda_i(q)\,t}\,\mathbf{v}_i(q),
\end{equation}
where the coefficients $c_i(q)$ are the projections of the initial data $\mathbf{w}(q,0)$ along the respective eigenvectors:\\
$\Big(c_1(q)\,c_2(q)\,c_3(q)\Big)^T={\mathbf{V}(q)}^{-1}\mathbf{w}(q,0)$, where  $\mathbf{V}(q):=\Big[\mathbf{v}_1(q)\,\mathbf{v}_2(q)\,\mathbf{v}_3(q)\Big]$ is the matrix containing the eigenvectors as columns.

Since the matrix $\mathbf{M}$ is non-Hermitian (due to presence of activity and turnover), its eigenvalues $\lambda_i$ are not real and the eigenvectors $\mathbf{v}_i$ are not orthogonal in general.
We make one further simplifying assumption: $k_2=0$, hence, $k^{b}_{\text{rel}}=0$, meaning that the bare (strain independent) part of the unbinding rates are idential (i.e., $k^{u}_{10}=k^{u}_{20}$), and the binding rates are identical as well (i.e., $k^{b}_{1}=k^{b}_{2}$). It follows that $k_4=\frac{k^{u}_{10}}{2}(\alpha_1-\alpha_2)$.

Note that, (a) the coupling $M_{12}$ between $\epsilon$ and $\rho$ in the $\epsilon$-equation  and the coupling $M_{21}$ between $\epsilon$ and $\rho$ in the $\rho$-equation can be  of opposite signs depending upon the relative signs and magnitudes of $k_3$ and $\tilde{B}_0$; (b)  the coupling $M_{13}$ between $\epsilon$ and $\phi$ in the $\epsilon$-equation  and the coupling $M_{31}$ between $\epsilon$ and $\phi$ in the $\phi$-equation are  of opposite signs when $\zeta_{\text{rel}}$ and $k_4$ have the opposite signs; and (c) the coupling $M_{23}$ between $\rho$ and $\phi$ in the $\rho$-equation can have any sign while the coupling $M_{32}$ between $\phi$ and $\rho$ in the $\phi$-equation is zero. Hence, $\rho$-$\phi$ interaction is always non-reciprocal, while the $\epsilon$-$\phi$ interaction is  non-reciprocal when the stronger stresslet  unbinds faster; $\epsilon$-$\rho$ interaction can be  non-reciprocal depending upon the relative signs and magnitudes of $k_3$ and $\tilde{B}_0$.

The three eigenvalues, assuming $A=1$ and $\chi(\rho_a^0)=\chi'(\rho_a^0)=1$,  are
\begin{equation}
	\lambda_1=-k_1-D\,q^2,~~~~
	\lambda_2=-\frac{\lambda_a+\sqrt{\lambda_b}}{2(1+q^2)},~~~~
	\lambda_3=-\frac{\lambda_a-\sqrt{\lambda_b}}{2(1+q^2)};
	\label{eigenvalues}
\end{equation}
where 
\begin{subequations}
	\begin{align}
		&\lambda_a:=k_1+\Big(\tilde{B}_0-\frac{2\zeta_{\text{avg}}}{k_1}+D+k_1\Big)\,q^2+D\,q^4,\label{lambda_a}\\
		&\lambda_b:= \lambda_a^2-4\,q^2(1+q^2)\bigg[\tilde{B}_0\,D\,q^2+k_1\bigg(\tilde{B}_0-2\frac{\zeta_{\text{avg}}}{k_1}\bigg(\frac{C}{A}+\frac{k_3}{k_1}\bigg)-2\frac{\zeta_{\text{rel}}}{k_1}\frac{k_4}{k_1}\bigg)\bigg];\label{lambda_b}
	\end{align}
\end{subequations}
and the corresponding eigenvectors are

\begin{equation}
	\mathbf{v}_1=\left[\begin{array}{c}
		0\\
		-\frac{\zeta_{\text{rel}}}{\zeta_{\text{avg}}}\\
		1
	\end{array}
	\right],~\mathbf{v}_{2}=\left[\begin{array}{c}
		\frac{1}{2(1+q^2)}\frac{k_1}{k_4}\Big(v_a-\sqrt{v_b}\Big)\\
		\frac{1}{2(1+q^2)}\frac{1}{k_4}\Big(v_c-\frac{1}{k_1}\sqrt{v_b}\Big)\\
		1
	\end{array}
	\right],~\mathbf{v}_{3}=\left[\begin{array}{c}
		\frac{1}{2(1+q^2)}\frac{k_1}{k_4}\Big(v_a+\sqrt{v_b}\Big)\\
		\frac{1}{2(1+q^2)}\frac{1}{k_4}\Big(v_c+\frac{1}{k_1}\sqrt{v_b}\Big)\\
		1
	\end{array}
	\right],
\end{equation}

where

\begin{eqnarray}
	v_a&:=&-k_1+\Big(\tilde{B}_0-\frac{2\zeta_{\text{avg}}}{k_1}-D-k_1\Big)q^2-Dq^4;\\
	v_b&:=&2q^2(1+q^2)(-B+C^2+D+Dq^2)k_1^3+(1+q^2)^2k_1^4+4\zeta^2_{\text{avg}}(1+C)^2q^4    \nonumber   \\
	&&+q^2k_1^2\Bigg[q^2\Big(-B+C^2+D+Dq^2\Big)^2+4(3C-1)(1+q^2)\zeta_{\text{avg}}\Bigg]\nonumber\\
	&&+4q^2k_1\Big[\zeta_{\text{avg}}\Big(\Big(-B(1+C)-D+C(C+C^2+D)+(C-1)Dq^2\Big)+2(1+q^2)k_3\Big)+2(1+q^2)k_4\zeta_{\text{rel}}\Big];\\
	v_c&:=&-(1-2C)k_1+2k_3-\Bigg(\tilde{B}_0-\frac{2\zeta_\text{avg}}{k_1}+D+(1-2C)k_1-2k_3\Bigg)q^2-Dq^4.
\end{eqnarray}


	\section{Linear Instabilities and Phases}
	Our first observation is that $\lambda_1$ is real and  $\lambda_1< 0$ for all $q$, since $k_1>0$ and $D>0$ by definition. Hence, all modes corresponding to the first eigenvalue $\lambda_1$ are asymptotically  stable (i.e., monotonically decaying). Secondly, we observe that  $Re[\lambda_3]\ge Re[\lambda_2]$ for all $q$, i.e., $\lambda_{max}=\lambda_3$. Hence, the largest eigenvalue  $\lambda_{max}=\frac{1}{2(1+q^2)}(-\lambda_a+\sqrt{\lambda_b})$ determines the asymptotic stability of the linearized system, i.e., the `linear' phases. We give the definitions of various phases in Table \ref{table:phases}.
	
	\begin{table}[H]
		\centering
		\begin{tabular}{|l|c|}
			\hline
			{\it Phase} & {\it Definition} \\ 
			\hline
			\hline
			Mechanical stability of the elastomer & $\tilde{B}_0> 0$ \\
			\hline
			Mechanical instability of the elastomer & $\tilde{B}_0\le0$  \\
			\hline
			\hline
			\makecell{Segregation instability}  & $\tilde{B}_0>0$, $\lambda_b>0$, $\lambda\ge 0$  \\
			\hline
			\makecell{Monotonic stability} & $\tilde{B}_0> 0$, $\lambda_b> 0$, $\lambda< 0$  \\
			\hline
			\hline
			\makecell{Damped oscillations} & $\tilde{B}_0> 0$, $\lambda_b<0$, $\lambda_a< 0$  \\
			\hline
			\makecell{Growing oscillations (unstable)} & $\tilde{B}_0> 0$, $\lambda_b<0$, $\lambda_a>0$  \\
			\hline
			\makecell{Sustained oscillations\\(swap, travelling waves)} & $\tilde{B}_0> 0$, $\lambda_b<0$, $\lambda_a= 0$  \\
			\hline
		\end{tabular}
		\caption{Definition of the instabilities and phases in the linear analysis}
		\label{table:phases}
	\end{table}
	
	Based on the definitions given in Table \ref{table:phases}, we can construct various linear stability phase diagrams, which, in principle, depend on the wave vector magnitude $q$. The phase diagrams presented in the main text are constructed assuming $q$ to be small, i.e., in the long wavelength limit.  We Taylor expand $\lambda_{max}(q)$ and $\lambda_b(q)$: $\lambda_{max}(q)=P_0+P_2\,q^2+P_4\,q^4+\cdots$, $\lambda_{b}(q)=Q_0+Q_2\,q^2+Q_4\,q^4+\cdots$; then, for small $q$, sign of $\lambda_{max}(q)$ and $\lambda_b(q)$ will be determined by the signs of the lower order coefficients $P_{0,2,4}$, $Q_{0,2,4}$ etc.~which are dependent on various parameters of our system. Based on the signs of these coefficients, we construct $q$-independent phase diagrams.

	\subsection{Ellipticity of Linear Elasticity}

	In standard elasticity, the linear elastic constitutive law $\sigma_{ij}=C_{ijkl}\epsilon_{ij}$, where $C_{ijkl}$ are the components of the spatially homogeneous elastic stiffness tensor (with the standard major and minor symmetries $C_{ijkl}=C_{jikl}=C_{klij}$), and linear strain displacement relation $\epsilon_{ij}=(u_{i,j}+u_{j,i})/2$. Substituting this in the static force balance equation $\sigma_{ij,j}=0$, one obtains the pde for the displacement field $u_i$ as
	\begin{equation}
	    C_{ijkl}u_{k,lj}=0.\label{statelas}
	\end{equation}
	
	Taking Fourier transform of this pde yields 
	\begin{equation}
	    A_{ik}\, \hat{u}_{k}=0,
	\end{equation}
	where $A_{ik}:=C_{ijkl}q_jq_l$ is the elastic acoustic tensor.
	
	The pde \eqref{statelas} is {\it elliptic}, if $\text{det}(A_{ik})>0$, {\it parabolic}, if $\text{det}(A_{ik})=0$, and {\it hyperbolic}, if $\text{det}(A_{ik})<0$.
	
	The active material we have considered is described by isotropic elasticity, $C_{ijkl}=\lambda\delta_{ij}\delta_{kl}+\mu(\delta_{ik}\delta_{jl}+\delta_{il}\delta_{jk})$ where $\lambda$ and $\mu$ are the Lam\'{e} and shear moduli, respectively. In this case, the acoustic tensor is $A_{ik}=(\lambda+\mu)q_iq_k+\mu\,q_jq_j\,\delta_{ik}$. In 2D, we can readily calculate that $\text{det}(A_{ik})=(B+\mu)\mu\,(q_jq_j)^2$, where $B:=\lambda+\mu$ is the  bulk modulus. Assuming $\mu>0$, we see that the static pde for 2D isotropic elasticity is elliptic if $B+\mu>0$, parabolic if $B+\mu=0$, and hyperbolic if $B+\mu<0$ \cite{KnowlesSternberg1978}. The Fourier wave vector $q_i$ determines the (real) characteristic directions for the non-elliptic cases. For the parabolic and hyperbolic cases, there exist, respectively, one and two (real) characteristic directions along which solution to the Cauchy problem (specified data $u_i$, $u_{i,j}$ on a curve in the 2D domain of interest) `propagates', i.e., remains constant.
	
		In a finite material with a boundary, the static elasticity pde is said to satisfy the {\it complementing condition}, if there are no `surface wave' solutions when the boundary is traction free \cite{silhavy-book97,EgorovPalencia2011}. For isotropic elasticity, it can be readily shown that, if $B=0$, there exist surface wave solutions of arbitrary frequency and amplitude for a traction free boundary, meaning that $B=0$ yields surface instabilities.

	\subsection{Segregation} 
	
	When $\lambda_b>0$, then $\lambda_{2,3}$ are real, hence, we have the monotonic stable/unstable phases.  The  onset of segregation instability is dictated by the sign change of $\lambda_{max}$ from negative to positive.
	
	The fastest growing mode $q_{seg}$ is where $\partial_q\lambda_{max}=0$. With $\chi(\rho_a^0)=\chi'(\rho_a^0)=1$, $C=1$, $A=1$, $D=1$ and $k_1=1$, we find
	\begin{equation}
	    q_{seg}=\frac{1}{\sqrt{2}}\sqrt{\frac{B-1-2\zeta_{\text{avg}}(2+k_3)-2\zeta_{\text{rel}}k_4}{B-1-6\zeta_{\text{avg}}(1+k_3)+2Bk_3\zeta_{\text{avg}}-4k_3\zeta_{\text{avg}}^2(2+k_3)-2\zeta_{\text{rel}}k_4(3-B)-8\zeta_{\text{avg}}\zeta_{\text{rel}}k_4(1+k_3)-4\zeta_{\text{rel}}^2k_4^2}}.
	\end{equation}
	
	From $\eqref{eigenvalues}$, we observe that $\lambda_{max}=0$ when $\lambda_a=\sqrt{\lambda_b}$, which, using \eqref{lambda_b}, implies that
	
	\begin{equation}
		\tilde{B}_0\,D\,q^2+k_1\bigg(\tilde{B}_0-2\frac{\zeta_{\text{avg}}}{k_1}\bigg(\frac{C}{A}+\frac{k_3}{k_1}\bigg)-2\frac{\zeta_{\text{rel}}}{k_1}\frac{k_4}{k_1}\bigg)=0.
	\end{equation}
	The (positive) solution to this equation
	\begin{equation}
		q=\sqrt{\frac{k_1}{D\,\tilde{B}_0}}\sqrt{-\Bigg(	\tilde{B}_0-2\frac{\zeta_{\text{avg}}}{k_1}\bigg(\frac{C}{A}+\frac{k_3}{k_1}\bigg)-2\frac{\zeta_{\text{rel}}}{k_1}\frac{k_4}{k_1}\Bigg)}
	\end{equation}
	is real and non-zero for $\tilde{B}_0>0$ if and only if
	\begin{equation}
		\tilde{B}_0-2\frac{\zeta_{\text{avg}}}{k_1}\bigg(\frac{C}{A}+\frac{k_3}{k_1}\bigg)-2\frac{\zeta_{\text{rel}}}{k_1}\frac{k_4}{k_1}< 0.
		\label{segrecond}
	\end{equation}

	Hence, if the condition \eqref{segrecond} is met, $\lambda_3$ becomes non-negative in the long wavelength limit, thus, triggering segregation instability. The characteristic width of the segregated regime is $\sim\frac{1}{q_{seg}}$.

	
	\subsubsection{Force spectroscopy}

	\begin{figure}[H]
		\centering
		\includegraphics[scale=1]{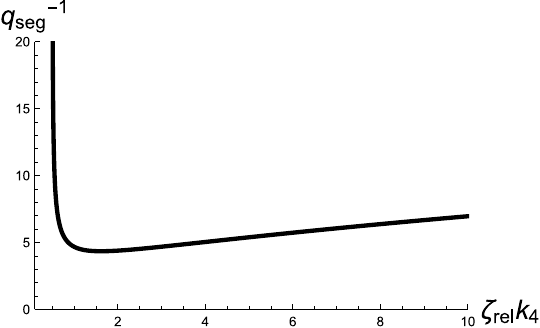}
		\caption{Plot of the width of the segregated domain $q^{-1}_{seg}$ vs $\zeta_{\text{rel}} k_4$. Rest of the parameters are fixed at $C=1$, $A=1$, $D=1$, $k_1=1$, $\chi(\rho_a^0)=1$, $\chi'(\rho_a^0)=1$, $B=8$, $\zeta_{\text{avg}}=1$, $k_3=1$.}
		\label{qsegcont}
	\end{figure}
	
	Small variations in the relative activity and turnover in chemical space get mapped onto large changes in the widths of the segregated regions in real space. To see this, we plot $q^{-1}_{seg}$ as a function of $\zeta_{\text{rel}}k_4$ in Fig.\,\ref{qsegcont}, keeping other parameters fixed.
	
	
	
	\subsubsection{Lyapunov functional for Linear Segregation}

\begin{figure}[H]
	\centering
	\subfigure[]{\includegraphics[scale=.6]{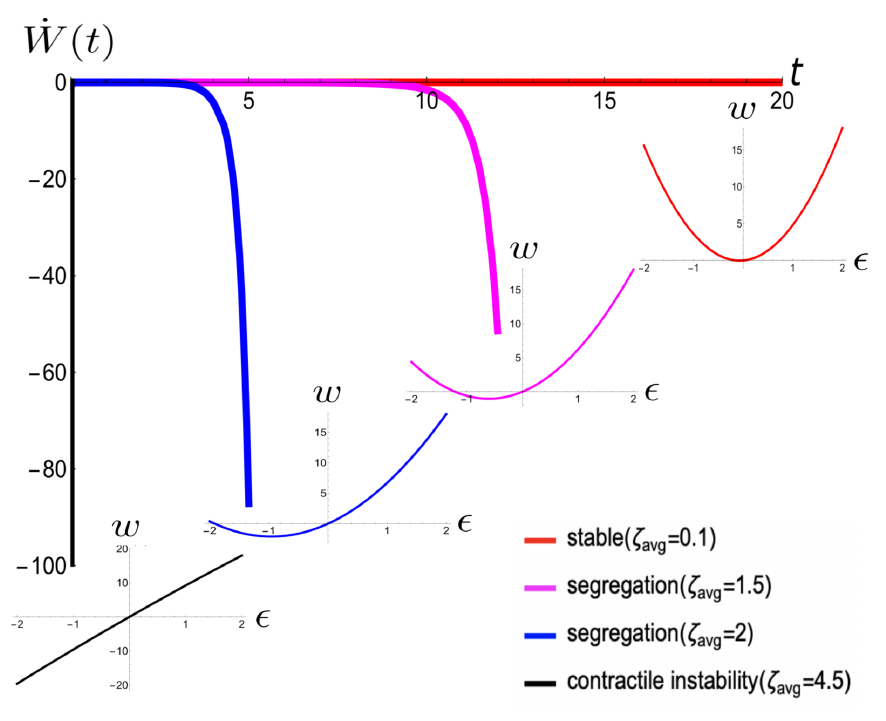}}
	\subfigure[]{\includegraphics[scale=.6]{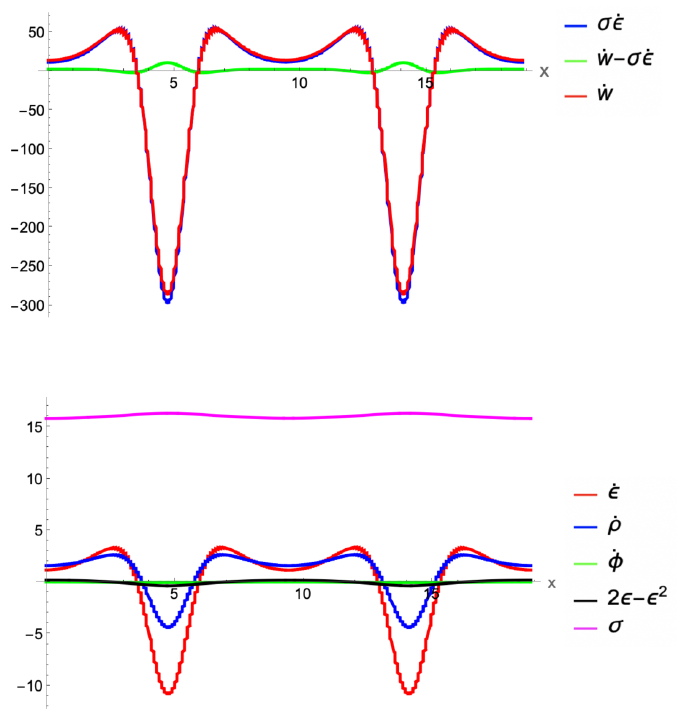}}
	\caption{(a) A plot of the power dissipation $\dot{W}(t)$ for $B=10$, $C=1$, $k_1=1$, $\zeta_{\text{rel}}=2$, $k_4=2$, with initial condition $\delta\phi(x,0)=0.01\sin(8x)$, $\delta\rho(x,0)=0$, $\delta\epsilon(x,0)=0$, shows that it is always negative and monotonic in this regime. The insets show plots of the effective strain energy density function for the linear  elastomer $w(\epsilon)$, for $B=10$, $C=1$, $k_1=1$, $\zeta_{\text{rel}}=2$, $\rho=1$, $\phi=0.1$. The minima $\epsilon_{min}$ shifts towards more contractile strain with increasing average contractility, and eventually goes to $-\infty$ at the contractile instability. (b) The top figure shows plots of the individual terms in $\dot{w}$ \eqref{eqn:spow} at a later stage of linear segregation. We see that the term $\sigma\dot{\epsilon}$ is the main contributor to $\dot{w}$. In the bottom figure, we see that the dominance of the $\sigma\dot{\epsilon}$ term is due to a large value of the stress field $\sigma$ in comparison to $2\epsilon-\epsilon^2$; the dominant part in $\sigma$ being the active back pressure $\sigma_0$.}
	\label{stresspower}
\end{figure}

Towards finding a Lyapunov functional for the segregation instability, we consider the ``effective strain energy density function'' for the linear  elastomer as
\begin{equation}
	w:=\sigma_0(\rho,\phi)\,\epsilon+\frac{1}{2}\tilde{B}(\rho,\phi)\,\epsilon^2.
\end{equation}

The minima of $w$ occurs at the strain
\begin{equation}
	\epsilon_{min}=-\frac{\sigma_0}{\tilde{B}}=-\frac{2\,\chi(\rho^0_a)\big(\zeta_{\text{avg}}\,\rho+\zeta_{\text{rel}}\,\phi\big)}{B-\frac{C^2}{A}-2\,\chi'(\rho^0_a)\,\frac{C}{A}\big(\zeta_{\text{avg}}\,\rho+\zeta_{\text{rel}}\,\phi\big)};
\end{equation}

the energy at this minima is

\begin{equation}
	w_{min}=w(\epsilon_{min})=-\frac{\sigma^2_0}{2\,\tilde{B}}=-\frac{1}{2}\Bigg(\frac{\Big(2\,\chi(\rho^0_a)\big(\zeta_{\text{avg}}\,\rho+\zeta_{\text{rel}}\,\phi\big)\Big)^2}{B-\frac{C^2}{A}-2\,\chi'(\rho^0_a)\,\frac{C}{A}\big(\zeta_{\text{avg}}\,\rho+\zeta_{\text{rel}}\,\phi\big)}\Bigg).
\end{equation}

In absence of activity, $\epsilon_{min}=0$ and $w_{min}=0$. In presence of activity with  large $\zeta_{\text{rel}}$ (implying segregation), $\sigma_0$ is large and $\tilde{B}$ is a small positive number; hence, $\epsilon_{min}$ acquires a large non-zero value, and, consequently $w_{min}$  decreases from zero (see Fig.\,\ref{stresspower}).

We calculate the power density

\begin{eqnarray}
	\dot{W}(t)&:=&\frac{1}{2L}\int_{-L}^{L}\dot{w}\,dx\nonumber\\
	&=&\frac{1}{2L}\int_{-L}^{L}\Bigg(\frac{\partial {w}}{\partial \epsilon}\dot{\epsilon}+\frac{\partial {w}}{\partial \rho}\dot{\rho}+\frac{\partial {w}}{\partial \phi}\dot{\phi}\Bigg)dx\nonumber\\
	&=&\frac{1}{2L}\int_{-L}^{L}\Big(\sigma\,\dot{\epsilon}+\zeta_{\text{avg}}(2\epsilon-\epsilon^2)\dot{\rho}+\zeta_{\text{rel}}(2\epsilon-\epsilon^2)\dot{\phi}\Big)dx
	\label{eqn:spow}
\end{eqnarray}
to linear order, where $2L$ is the system size. We show numerically in Fig.\,\ref{stresspower}(a) that $\dot{W}(t)<0$ for all $t$ in the linear segregation regime.
That is to say, $W$ is a Lyapunov functional driving segregation of the stresslets.

Indeed, the first term of the integral in \eqref{eqn:spow} is always negative, since $\int\sigma\dot{\epsilon}dx=-\int q^2\Big\vert\frac{\partial w}{\partial\epsilon_q}\Big\vert^2dq$. We also see from (Fig.\,\ref{stresspower}(b)-top) that
$|\zeta_{\text{avg}}(2\epsilon-\epsilon^2)\dot{\rho}+\zeta_{\text{rel}}(2\epsilon-\epsilon^2)\dot{\phi}|\ll|\sigma\dot{\epsilon}|$, i.e., contribution of the rest of the terms is negligible in comparison to this first term $\sigma\dot{\epsilon}$, as $|2\epsilon-\epsilon^2|\ll|\sigma|$ (see Fig.\,\ref{stresspower}(b)-bottom). As a consequence of the dominance of the first negative term, the whole integral in \eqref{eqn:spow} is negative. We note that the comparatively large magnitude of $\sigma$ is due to a large active back stress $\sigma_0$ (which linearly depends on $\rho$).
Once again, we see that the driving force for segregation depends on a large active back pressure.

	\subsection{Travelling Wave} 
	
	When $\lambda_b< 0$, we have $Im[\lambda_{max}]=\frac{\sqrt{|\lambda_b|}}{2(1+q^2)}\ne0$, and  $Re[\lambda_{max}]=-\frac{\lambda_a}{2(1+q^2)}$. Hence, the condition  $\lambda_b< 0$, characterizes the  various oscillatory phases (stable and unstable pulsations and/or waves), with frequency $\omega(q)=| Im[\lambda_{max}] |$, and decay/growth rate $\tau_d(q)=|Re[\lambda_{max}] |$. 
	
	The oscillations are stable (damped oscillations) for $Re[\lambda_{max}]<0$, i.e., when $\lambda_a>0$, and unstable (growing oscillations) for $Re[\lambda_{mas}]>0$, i.e., when $\lambda_a<0$.

	For small $q$, we have
	
	\begin{equation}
		Re[\lambda_{max}(q)]=-\frac{k_1}{2}-\frac{1}{2}\Bigg(\tilde{B}_0-\frac{\zeta_{\text{avg}}}{k_1}+D+k_1\Bigg)q^2+o(q^2)
	\end{equation}

	The fastest growing mode $q=q^\star$ satisfies $\partial_q Re[\lambda_{max}]=0$, yielding
	\begin{equation}
		q^\star=\sqrt{-1+\sqrt{\frac{-\tilde{B}_0+\frac{2\zeta_{\text{avg}}}{k_1}}{D}}}.
		\label{fgm}
	\end{equation}
	
	Note that $q^\star$ is real iff
	\begin{equation}
		-\tilde{B}_0+\frac{2\zeta_{\text{avg}}}{k_1}\ge0~~\text{and}~~-1+\sqrt{\frac{-\tilde{B}_0+\frac{2\zeta_{\text{avg}}}{k_1}}{D}}\ge 0.
	\end{equation}
	
	The critical  mode $q_c$ at the transition from stability to instability satisfies  $\lambda_a(q_c)=0$. Hence, we must have $q_c=q^\star$; this condition provides a mode independent characterization of marginal stability/criticality in the oscillatory phase as
	\begin{equation}
		\bigg(\sqrt{-\tilde{B}_0+\frac{2\zeta_{\text{avg}}}{k_1}}-\sqrt{D}\bigg)^2=k_1
		\label{osc-ms}
	\end{equation}
	
	Using this relation in \eqref{fgm}, we find the wave vector for travelling waves is
	\begin{equation}
		q^\star_c=\sqrt[4]{\frac{k_1}{D}}.
	\end{equation}

	The mode independent characterization of the emergence of  instability is obtained by replacing ``$=$'' with ``$<$'' in \eqref{osc-ms}.
	
	At $q=q^\star_c$, the frequency is given by (recall that $\lambda_a(q^\star_c)=0$)
	
	\begin{equation}
		\omega(q^\star_c)
		=\frac{\sqrt{\frac{k_1}{D}}}{\sqrt{1+\sqrt{\frac{k_1}{D}}}} \sqrt{\tilde{B}_0\,\sqrt{Dk_1}+k_1\bigg(\tilde{B}_0-2\frac{\zeta_{\text{avg}}}{k_1}\bigg(\frac{C}{A}+\frac{k_3}{k_1}\bigg)-2\frac{\zeta_{\text{rel}}}{k_1}\frac{k_4}{k_1}\bigg)}.
	\end{equation}
	
	Hence, for fixed values of $\tilde{B}_0>0$ and $\tilde{B}_0-2\frac{\zeta_{\text{avg}}}{k_1}\bigg(\frac{C}{A}+\frac{k_3}{k_1}\bigg)>0$, we must have $\zeta_{\text{rel}}k_4<0$ for real $\omega(q^\star_c)$, i.e, $\zeta_{\text{rel}}$ and $k_4$ should be of opposite signs.  If $\zeta_{\text{avg}}>\zeta_{\text{rel}}$ then one should have $\alpha_1<\alpha_2$ to trigger travelling wave phase. Since the stresslets are contractile, i.e., $\epsilon<0$, this implies that $k^u_1(\epsilon)>k^u_2(\epsilon)$.  Hence, in order to have travelling waves, the species with larger activity must unbind more than the species with lower activity.

	\subsection{Swap phase}

	The Swap phase is charaterized by standing wave solutions of the form $\phi(x,t)=\sin(qx)\sin(\omega t)$.
	
	In the linear regime, the onset of the  contractile instability is dictated by the time scale  $Re[\lambda_{max}(q)]^{-1}$, while the time scale of unbinding of the stronger stresslet (assuming it to be $\rho_1$) is ${k^u_1(\epsilon)}^{-1}$. Within the oscillatory phase, i.e., when $\zeta_{\text{rel}}k_4<0$, if the stronger stresslet unbinds before the contractile instability kicks in, i.e., if
	\begin{equation}
		Re[\lambda_{max}(q)]^{-1}\le {k^u_1(\epsilon)}^{-1},
	\end{equation}
	then the contracting domain expands again before ever reaching the contractile instability.

	The swap phase occurs at the boundary between the damped wave phase and mechanical instability phase.



\begin{figure}[H]
	\centering
	\includegraphics[scale=1.15]{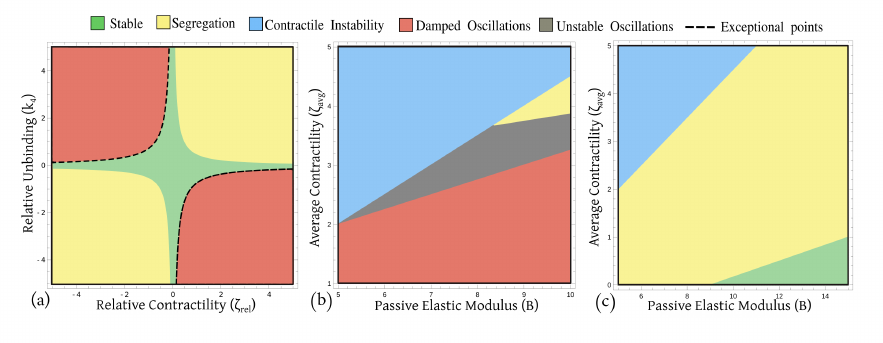}
	\caption{Phase diagram from Linear Stability Analysis, with $C=1$, $A=1$, $D=1$, $k_1=1$, $k_2=0$, $\chi(\rho_a^0)=1$, $\chi'(\rho_a^0)=1$. (a) For $B=8$, $\zeta_{\text{avg}}=1$, $k_3=1$; (b) For $k_3=1$, $\zeta_{\text{rel}}=-2$, $k_4=3$; (c) For $k_3=1$, $\zeta_{\text{rel}}=2$, $k_4=2$. The different phases are shown in the colour legend and the dashed line denotes the exceptional points.}
	\label{phasediagram}
\end{figure}

\begin{figure}[H]
	\centering
\includegraphics[scale=1.1]{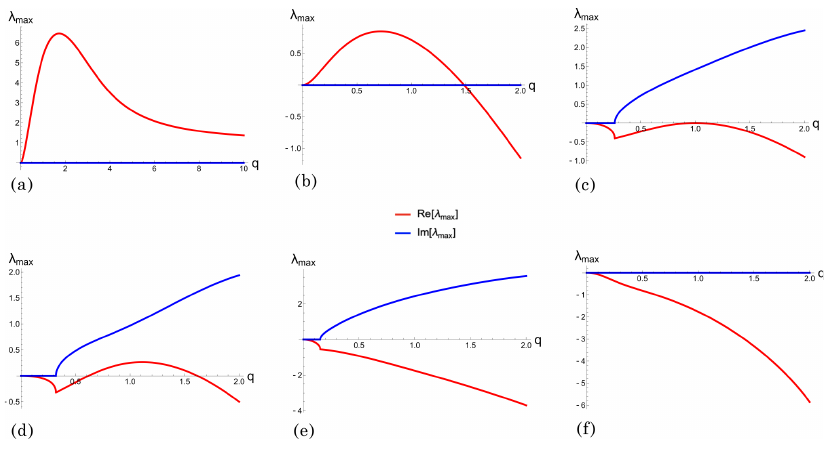}
\caption{Linear stability dispersion curves, with $C=1$, $A=1$, $D=1$, $k_1=1$, $k_2=0$, $\chi(\rho_a^0)=1$, $\chi'(\rho_a^0)=1$, showing (a) Contractile instability, for $B=8$, $\zeta_{\text{avg}}=4$, $k_3=1$, $\zeta_{\text{rel}}=-2$,  $k_4=3$; (b) Segregation instability, for $B=8$, $\zeta_{\text{avg}}=1$, $k_3=1$, $\zeta_{\text{rel}}=2$, $k_4=3$; (c) Travelling wave, for $B=8$, $\zeta_{\text{avg}}=2.75$, $k_3=1$, $\zeta_{\text{rel}}=-2$, $k_4=3$; (d) Unstable oscillations, for $B=8$, $\zeta_{\text{avg}}=3$, $k_3=1$, $\zeta_{\text{rel}}=-2$, $k_4=3$; (e) Damped oscillations (stable), for $B=8$, $\zeta_{\text{avg}}=1$, $k_3=1$, $\zeta_{\text{rel}}=-2$, $k_4=3$; (f) (Monotonically) Stable, for $B=17$, $\zeta_{\text{avg}}=1$, $k_3=1$, $k_4=1$, $\zeta_{\text{rel}}=1$. Red curves
denote $Re[\lambda_{max}]$ and blue curves denote $Im[\lambda_{max}]$ versus wave-vector $q$.}
\label{dispersion}
\end{figure}


\subsection{Exceptional Points}
Conclusions about various phases drawn from the above  linear stability analysis are robust only in regions far from the Exceptional Points (EPs) of the non-Hermitian dynamical matrix $\mathbf{M}$, where eigenvalues coalesce and eigenvectors co-align.  Near EPs, the system undergoes strong transient growth and `bootstraps' into nonlinear phases. We will analyse the effect of EPs in the mechanics of active matter in a later publication. We only mention here that in the above linear stability phase diagrams, the only EPs present are on the boundary between monotonically stable and damped wave phases. Hence, the system goes into the oscillatory phase through an EP.



\section{Segregation in a Single Species of Contractile Stresslets on an Elastomer}

So far our analysis of segregation has been for a mixture of stresslets on an elastomer. We now show that even a single species of stresslets on an elastomer can exhibit segregation into low and high density regions, akin to a gas-liquid segregation. The non-dimensional form of the governing {\it scalar}  equations in 1-dim  for the single species of stresslets are 
\begin{subequations}
	\begin{align}
		& \dot{\epsilon} = \partial^2_{xx}\,\sigma, \label{epsilon-single}\\
		& \dot{\rho}+\partial_x(\rho\,\dot{u})=D\,\partial^2_{xx} \rho +1-C\epsilon -k^u(1+\alpha\,\epsilon)\,\rho. \label{rho-single}
	\end{align}
	\label{dyn-single}
\end{subequations}

where $\rho$ is the density of the bound stresslet, and the total stress is given by
\begin{equation}
	\sigma
	=\sigma_0
	+\tilde{B}\,\epsilon
	+B_2\,\epsilon^2
	+B_3\,\epsilon^3+\dot{\epsilon}
	\label{totalstress-singless}
\end{equation}

where (assuming $A=1$)

\begin{eqnarray}
	\sigma_0&:=& \chi(\rho_a^0)\,\zeta(\rho),\nonumber\\
	\tilde{B}&:=&B-C^2-\chi'(\rho_a^0)C\,\zeta(\rho),\nonumber\\
	B_2&:=&\frac{1}{2!}\chi''(\rho_a^0)\,C^2\,\zeta(\rho),\nonumber\\
	B_3&:=&-\frac{1}{3!}\chi'''(\rho_a^0)\,C^3\,\zeta(\rho),
\end{eqnarray}
with
\begin{equation}
	\zeta(\rho)=\zeta_{\text{avg}}\rho+\zeta_{\text{rel}}\rho^2+\zeta_3\rho^3,
\end{equation}
a general nonlinear form that enters the active stress.

The homogeneous unstrained steady state is $\epsilon=\epsilon_0=0$, $\rho=\rho_0=(k^u)^{-1}$.

The linearised equations for perturbations $\delta\epsilon$ and $\delta\rho$ about this state are

\begin{subequations}
\begin{align}
& \dot{\delta\epsilon}=\tilde{B}_0\,\partial^2_{xx}\delta\epsilon+\chi(\rho_a^0)\,\zeta'(\rho_0)\,\partial^2_{xx}\delta\rho+\partial^2_{xx}\dot{\delta\epsilon},\\
&\dot{\delta\rho}+(k^u)^{-1}\Big[\chi(\rho_a^0)\,\zeta'(\rho_0)\,\partial^2_{xx}\delta\rho+\tilde{B}_0\partial^2_{xx}\delta\epsilon+\partial^2_{xx}\dot{\delta\epsilon}\Big]=D\partial^2_{xx}\delta\rho-(C+\alpha)\,\delta\epsilon-k^u\,\delta\rho,
\end{align}
\end{subequations}
where
\begin{equation}
	\tilde{B}_0:=B-C^2-\chi'(\rho_a^0)C\,\zeta(\rho_0).
\end{equation}

After Fourier transformation w.r.t.~$x$, we get the dynamical system
\begin{equation}
 \dot{\mathbf{w}}=
 	\left[\begin{array}{cc}
 		-\frac{\tilde{B}_0\,q^2}{1+q^2}&-\frac{\chi(\rho_a^0)\,\zeta'(\rho_0)\,q^2}{1+q^2}\\
 		-(C+\alpha)+\frac{\tilde{B}_0}{k^u}q^2-\frac{\tilde{B}_0\,q^4}{k^u(1+q^2)}&-D\,q^2-k^u+\frac{\chi(\rho_a^0)\,\zeta'(\rho_0)}{k^u}q^2-\frac{\chi(\rho_a^0)\zeta'(\rho_0)\,q^4}{k^u(1+q^2)}
 	\end{array}
 	\right]\mathbf{w}
\end{equation}
where $\mathbf{w}(q,t):=\Big(\delta\hat{\epsilon}(q,t)\,\,\delta\hat{\rho}(q,t)\Big)^T$.

The two eigenvalues are of the form $\lambda_{1,2}=-\frac{\lambda_a\pm\sqrt{\lambda_b}}{1+q^2}$.
For small $q$, assuming that $\chi(\rho_a^0)=\chi'(\rho_a^0)=1$, the two eigenvalues are

\begin{subequations}
	\begin{align}
		&\lambda_1=-k^u-\Bigg(D+\frac{(C+\alpha-1)\zeta'(\rho_0)}{k^u}\Bigg)\,q^2+O(q^3)\\
		&\lambda_2=-\Big(\tilde{B}_0-\frac{(C+\alpha)\zeta'(\rho_0)}{k^u}\Big)q^2+O(q^3).
	\end{align}
\end{subequations}

\begin{figure}[H]
	\centering
	\includegraphics[scale=1]{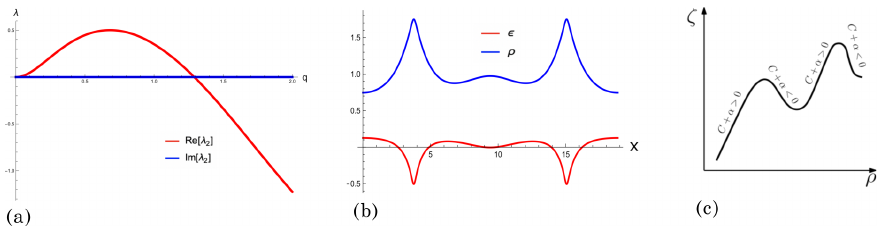}
	\caption{(a) Linear stability dispersion curves for segregation of a single stresslet species on an elastomer, with $B=6$, $C=1$, $D=1$, $\zeta=2$, $k^u=1$, $\alpha=1$. (b) The corresponding profiles of the strain $\epsilon$ and density $\rho$ showing segregation of low and high density regions. (c) Nonlinear dependence of the active stress $\zeta(\rho)$ on density for a single stresslet species in a fluid showing  catch-bond ($C+\alpha>0$) on the positive slope branches and slip-bond ($C+\alpha<0$) on the negative slope branches. The negative slope branch produces the `active back pressure' needed for the segregation instability to take place. }
	\label{singdisp}
\end{figure}

The dispersion curves corresponding to the maximum eigenvalue $\lambda_2$ shows a segregation instability (Fig.\,\ref{singdisp}(a)); the segregation condition for the single stresslet system  is

\begin{equation}
	0<\tilde{B}_0<\frac{\zeta'(\rho_0)}{k^u}(C+\alpha).
\end{equation}

Recall that for the binary mixture of stresslets, the segregation condition was

\begin{equation}
	0<\tilde{B}_0<2\frac{\zeta_{\text{avg}}}{k_1}\bigg(C+\frac{k_3}{k_1}\bigg)+2\frac{\zeta_{\text{rel}}}{k_1}\frac{k_4}{k_1},
\end{equation}
with
\begin{equation}
	k_3:=\frac{k^{u}_{10}\alpha_1+k^{u}_{20}\alpha_2}{2},~~\text{and}~~k_4:=\frac{k^{u}_{10}\alpha_1-k^{u}_{20}\alpha_2}{2}.
\end{equation}

For the binary mixture, the $\zeta_{\text{rel}}k_4$ term expands the window over which we see segregation.

The segregation condition for a binary mixture with non-monotonic functions  $\zeta_i(\rho_i)$ is

\begin{equation}
	0<\tilde{B}_0<2\frac{\zeta_{\text{avg}}'(\rho_0,\phi_0)}{k_1}\bigg(C+\frac{k_3}{k_1}\bigg)+2\frac{\zeta'_{\text{rel}}(\rho_0,\phi_0)}{k_1}\frac{k_4}{k_1},
\end{equation}

where
\begin{equation}
	\tilde{B}_0:=B-C^2-\chi'(\rho_a^0)C\,\Big(\zeta_{\text{avg}}(\rho_{10})+\zeta_{\text{rel}}(\rho_{20})\Big)=B-C^2-2\chi'(\rho_a^0)C\,\Big(\zeta_{\text{avg}}(\rho_0,\phi_0)+\zeta_{\text{rel}}(\rho_0,\phi_0)\Big).
\end{equation}

The corresponding $\epsilon$ and $\rho$ profiles are shown in Fig.\,\ref{singdisp}(b), and show segregation between low density and high density regions, similar to a gas-liquid separation.


\section{Segregation in a Single Species of Contractile Stresslets in Fluid}
\label{sect:singlefluid}

Instead of the elastomer medium discussed thus far, let us now embed the stresslets in a viscous fluid.
  In case of fluid mediated interaction between the stresslets,
the non-dimensional form of the {\it scalar} governing equation for a single species of stresslets  is
\begin{subequations}
	\begin{align}
		& v = \partial_{x}\,\sigma, \label{v-single}\\
		& \dot{\rho}+\partial_x(\rho\,v)=D\,\partial^2_{xx} \rho +1 -k\,\rho. \label{rho-single-f}
	\end{align}
	\label{dyn-single-f}
\end{subequations}
where $\rho$ is the density of the bound stresslet, $v$ is the fluid velocity, and the total stress is given by
\begin{equation}
	\sigma
	=\partial_x v+\zeta(\rho)
	\label{totalstress-singless-f}
\end{equation}
with
\begin{equation}
   \zeta(\rho):= \zeta_{\text{avg}}\rho+\zeta_{\text{rel}}\rho^2+\zeta_3\rho^3,
\end{equation}
the active osmotic pressure. Unlike in the previous cases, we see here that a   {\it non-monotonic} dependence of the active stress on the stresslet density $\rho$ is {\it necessary} for segregation (Fig.\,\ref{singdisp}(c));  a monotonic $\zeta(\rho)$ with steep positive slope results in a clumping instability  instead.  This non-monotonicity in $\rho$ naturally arises from the binding of contractile stresslets on finite patches of actin mesh with free boundaries embedded in a fluid, where the elastic response of the patches is taken to  be fast.

Homogeneous stress-free steady state is $\rho_0=1/k$, $v_0=0$.
Linearised dynamics of the perturbation $(\delta v, \delta\rho)$ about this state is
\begin{subequations}
	\begin{align}
		& \delta v = \partial^2_{xx}\delta v+\zeta'(\rho_0)\partial_{x}\delta\rho, \label{v-lin-single}\\
		& \dot{\delta\rho}+\rho_0\partial_x\delta v=D\,\partial^2_{xx} \delta\rho  -k\,\delta\rho. \label{rho-lin-single}
	\end{align}
	\label{dyn-lin-single}
\end{subequations}

Taking Fourier transform w.r.t.~$x$, we obtain
\begin{subequations}
	\begin{align}
		& \hat{\delta v} = -q^2\,\hat{\delta v}+iq\,\zeta'(\rho_0)\hat{\delta\rho}, \label{v-linft-single}\\
		& \dot{\hat{\delta\rho}}+iq\,\rho_0\hat{\delta v}=-Dq^2\, \hat{\delta\rho}  -k\,\hat{\delta\rho}, \label{rho-linft-single}
	\end{align}
	\label{dyn-linft-single}
\end{subequations}
which yields
\begin{equation}
    \dot{\hat{\delta\rho}}=-\Bigg(D-\frac{\rho_0\zeta'(\rho_0)}{1+q^2}\Bigg)q^2\, \hat{\delta\rho}  -k\,\hat{\delta\rho}.
\end{equation}

Hence, there is a long wave length instability when
\begin{equation}
   \rho_0\zeta'(\rho_0)>D. 
\end{equation}
The effective negative diffusion leads to finite-time singularity.

Note that the crucial role of the 
		  active back pressure in driving this segregation, is played by the negative slope branch
		  of the non-monotonic $\zeta(\rho)$, that separates the positive slope branches corresponding to  low and high stresslet density (Fig.\,\ref{singdisp}(c)).  The bond is of `catch'-type ($C+\alpha>0$) on the positive slope branches and of `slip'-type ($C+\alpha<0$) on the negative slope branches.

\newpage

\section{Numerical Solution of governing equations}

We numerically solve the scalar system of pdes written in terms of strain:
\begin{subequations}
	\begin{align}
		& \dot{\epsilon} = \partial^2_{xx}\,\sigma, \label{strain-eqn}\\
		& \dot{\rho}+\partial_x(\rho\,\partial_x	\sigma)=D\,\partial^2_{xx} \rho +1 -\frac{C}{A}\,\epsilon -(k_1+k_3\,\epsilon)\,\rho -(k_2+k_4\,\epsilon)\,\phi, \label{rho-eqn}\\
		& \dot{\phi}+\partial_x(\phi\,\partial_x\sigma)=D\,\partial^2_{xx} \phi + k^{b}_{\text{rel}}\bigg(1-\frac{C}{A}\,\epsilon\bigg) - (k_1+k_3\,\epsilon)\,\phi - (k_2+k_4\,\epsilon)\,\rho, \label{phi-eqn}
			\end{align}
			\label{num:eqns}
	\end{subequations}
where
		\begin{subequations}
			\begin{align}
		&\sigma:=2\,\chi(\rho^0_a)\bigg(\zeta_{\text{avg}}\,\rho+\zeta_{\text{rel}}\,\phi\bigg)+\Bigg[B-\frac{C^2}{A}-2\,\chi'(\rho^0_a)\,\frac{C}{A}\big(\zeta_{\text{avg}}\,\rho+\zeta_{\text{rel}}\,\phi\big)\Bigg] \epsilon \\
		&\hspace{8mm}+\Bigg[\chi''(\rho^0_a)\,\bigg(\frac{C}{A}\bigg)^2\,\big(\zeta_{\text{avg}}\,\rho +\zeta_{\text{rel}}\,\phi\big)\Bigg]\,\epsilon^2-\Bigg[\frac{\chi'''(\rho^0_a)}{3}\,\bigg(\frac{C}{A}\bigg)^3\,\big(\zeta_{\text{avg}}\,\rho +\zeta_{\text{rel}}\,\phi\big)\Bigg]\,\epsilon^3+\eta\,\dot{\epsilon},
	\end{align}
\end{subequations}
in a spatial domain $x\in[a,b]$  with periodic boundary conditions (w.r.t.~$x$) for $\epsilon(x,t)$, $\rho(x,t)$ and $\phi(x,t)$.

\subsection{Numerical Scheme}

\subsubsection{Finite Difference Euler Scheme}

A finite difference Euler scheme with a stencil adaptive algorithm was implemented via a code written in the language MATLAB. All fields and differential operators appearing in the partial differential equation \eqref{num:eqns} are discretized and the advection terms are calculated using an upwinding scheme. Periodic boundary conditions have been implemented, and the viscous contribution to the total stress (the $\dot{\epsilon}$ term) has been neglected.

Conventional stability schemes like the Crank-Nicholson scheme do very poorly at advecting waveforms with sharp leading or trailing edges. In order to suppress spurious oscillations at the leading and trailing edges of a sharp waveform, one can use upwinding schemes. In such a scheme, the spatial differences are skewed in the upwind direction, i.e., the direction from which the advecting flow emanates. Stability of the upwinding scheme is ensured by checking that the Courant–Friedrichs–Lewy (CFL) condition is satisfied. To study segregation and eventually the formation of a finite time singularity in the profiles of the relative density ($\phi$), average density ($\rho$) and strain ($\epsilon$), a constant time step ($dt$) has been used. We discuss the dependence of the blow-up time ($t_0$) on the integration step size ($dt$) below.  To study the moving phases (travelling waves, damped oscillations and swap), an adaptive time-step is used.

A three point stencil for the central difference scheme is used for functions $\verb|du-dx-central()|$ and $\verb|d2u-dx2-central()|$; whereas a five  point stencil is used for the function $\verb|du-dx-higher()|$ to compute the spatial derivatives. The advection terms have been calculated using the upwinding scheme with the function $\verb|du-dx-upwind-flux()|$. We start with white noise perturbation in $\phi$ and $\rho$ about a homogeneous, unstrained, uniform state. The choice of mesh size ($dx$) depends on the parameters and the phases under study. Thus, while studying segregation and the formation of the singular structures, we use a small mesh size of $dx=0.02$, and  while studying the oscillatory phases (travelling waves, swap and damped oscillations), we use a mesh size of $dx=0.2$.

\subsubsection{Spectral Methods}

We use Dedalus pseudospectral solver \cite{dedalus2020} to solve the system of partial differential equations. While analysing the sharp singular structures formed in the segregation regime, the finite difference solver written in the Euler scheme has severe limitations. Solving the set of pde's using Spectral schemes help with obtaining sharper singular profiles without the numerical scheme breaking down. To study segregation, the variables are represented on a periodic Fourier basis. Periodic basis functions like the Fourier basis provide exponentially converging approximations to smooth functions. The fast Fourier transform (implemented using sciPy and FFTW libraries) 
enables computations requiring both the series coefficients and grid values to be performed efficiently. For time evolving the system of equations, we use a SBDF2 time-stepper which is a 2nd order semi-implicit Backward Differentiation formula scheme.  As initial condition, we add random noise perturbations to $\phi$ with a magnitude of $0.2$ about its homogeneous steady state value. We use $500$ modes ($N_x=500$) in a domain length of $L_x=35$ and a time-step ($dt$) of $10^{-4}$ to solve the pde \eqref{num:eqns}.
\begin{figure}[H]
	\centering
	\subfigure[]{\includegraphics[scale=.6]{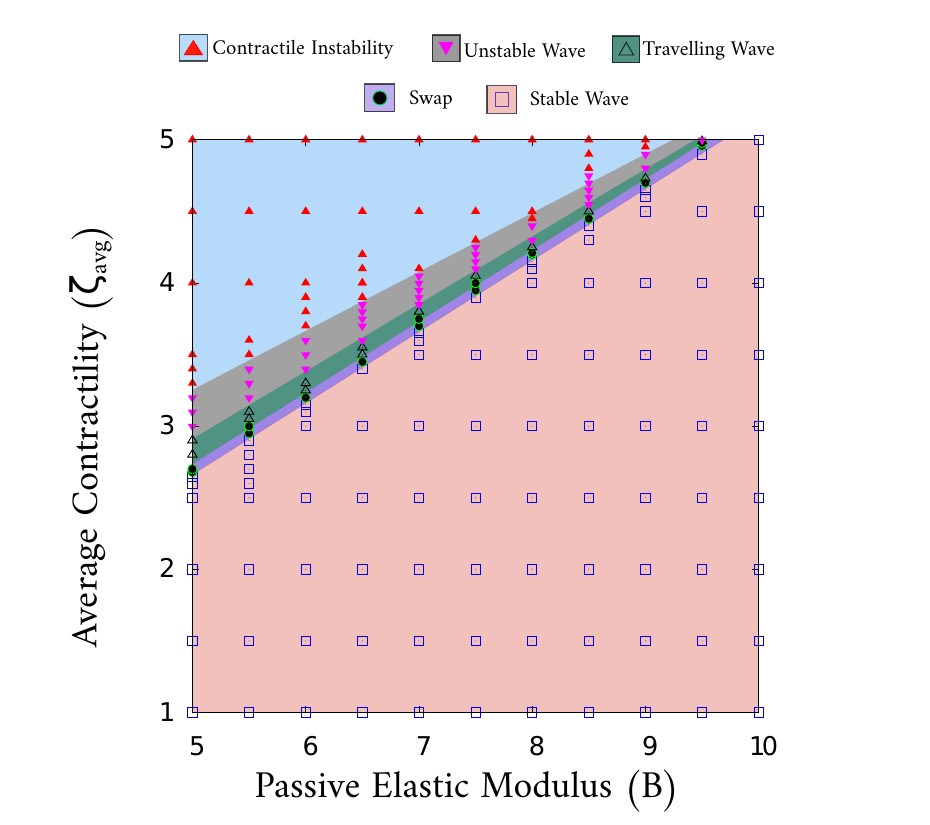}}
	\hspace{1mm}
	\subfigure[]{\includegraphics[scale=.6]{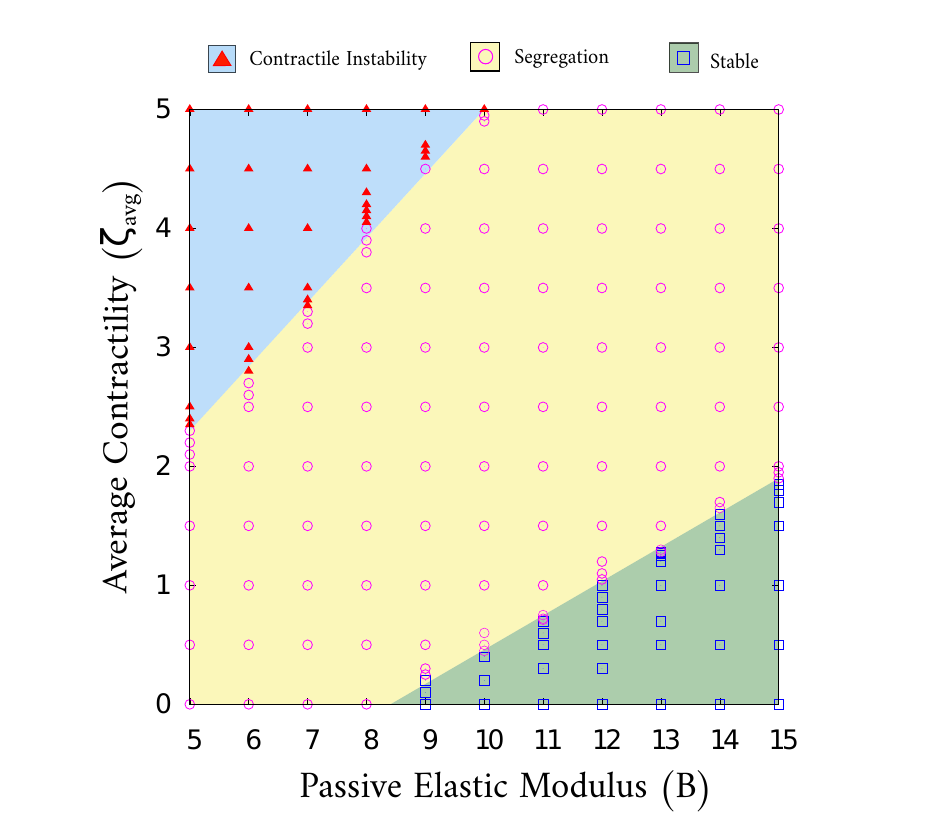}}
	\caption{Complete phase diagrams obtained from numerically solving the scalar version of the governing system of nonlinear pdes. Various phases are denoted in the colour legend. Symbols represent state points at which the numerical solution was obtained. (a) Average contractility versus passive elastic modulus when $\zeta_{\text{rel}}k_4<0$ shows the oscillatory phases (note the region of coexistence) and (b) Average contractility versus passive elastic modulus when $\zeta_{\text{rel}}k_4>0$ shows segregation.  }
	\label{numpd}
\end{figure}

\subsection{Complete Phase Diagrams}

The complete phase diagrams, obtained from numerically integrating the pde \eqref{num:eqns}, are qualitatively similar to the linear stability phase diagrams, apart from quantitative differences and the appearance of sustained oscillatory phases such as travelling waves and swap.  
\\

For $\zeta_{\text{rel}}k_4<0$, we observe from numerical phase diagram Fig.\,\ref{numpd}(a) that there is a region where travelling waves and swap are coexisting phases in time. Starting with random perturbations about the uniform, symmetric unstrained steady state in the parameter regime $\zeta_{\text{rel}}k_4<0$, sustained oscillations appear at the unstable wave and stable wave phase boundary, either as a travelling wave train, or as a standing wave (i.e., swap). Typically the system exhibits a long transient, where it first goes into a swap phase, then a coexistence between swap and travelling wave, and then finally transitions into a travelling wave. The transient time decreases with an increase in the average contractility. We draw the phase diagrams by making note of the configuration at a fixed large time $t_{max}$ starting from statistically identical initial conditions (Fig.\,\ref{numpd}). 

The numerical code shows an eventual blowup at a very large run time in the travelling wave phase, due to the sharpness of the slopes of the travelling front.

On the other hand, for $\zeta_{\text{rel}}k_4>0$, we observe from the numerical phase diagram 
Fig.\,\ref{numpd}(b), the appearance of the segregated phase that eventually evolves into a collection of static and moving singular lines as discussed in Sect.\,\ref{sect:nonlinear}. The numerical code for the governing
equations experiences a similar blowup at large times due to the formation of singular configurations.

At the boundary between the oscillatory phases and the contractile instability, we see a travelling wave train with amplitude that grows indefinitely, giving rise to an array of moving tension lines.




\section{Nonlinear Analysis of governing equations}
\label{sect:nonlinear}

We have seen that in the regime where $\zeta_{\text{rel}}k_4>0$, the uniform $\phi_0=0$ state is linearly
unstable and initiates the segregation of the two stresslets, giving rise to a bicontinuous segregated configuration whose width is given by $q_{seg}^{-1}$. How does this configuration further evolve under the influence of significant nonlinear effects? Here we show that boundaries between linearly segregated domains, proceed to move in the direction of stress jump across the boundary.

\begin{figure}[H]
	\centering
	\includegraphics[scale=1.5]{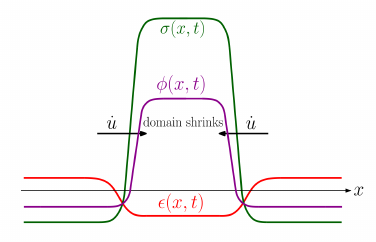}
	\caption{Profiles of $\phi$, $\epsilon$ and the stress $\sigma$ at the end of linear segregation instability. At this stage nonlinear effects become significant and drive the domain fronts to move towards each other with speed $\dot u$, shrinking the domain to  eventually form a singular tension field in a finite time.}
	\label{frontsmove}
\end{figure}

To see this quickly, take $\zeta_{\text{rel}}>0$, i.e., the stresslet $\rho_1$ is more contractile than $\rho_2$, and for the moment ignore the subsequent dynamics of the average density $\rho$.
Consider the dynamics of strain $\epsilon$ and relative density $\phi$ starting from an initial condition of the form shown in Fig.\,\ref{frontsmove}, that consists of two fronts continuously connecting constant positive and negative values of $\epsilon$ and $\phi$. This is the typical configuration achieved soon after the linear segregation instability starting from a homogeneous, symmetric unstrained steady state (i.e., $\rho_0=1/k_1$,  $\phi_0=0$, $\epsilon_0=0$).

 The central `bulk' region with $\phi>0$ consists predominantly of $\rho_1$, while the bulk regions on either side of it with $\phi<0$ have predominantly $\rho_2$.  The strain field $\epsilon$ in the central  bulk region 
is negative, hence, contractile relative to the extensile regimes on both sides.  This initial state can be taken to be of the form $\phi(x,0)=\tanh(x+10)-\tanh(x-10)-(\tanh(10)-\tanh(-10))/2$, $\epsilon(x,0)=-\phi(x,0)$, together with 
the corresponding total stress, i.e. elastic + active (schematised in Fig.\,\ref{frontsmove}).  We see that in the centre bulk region, the stress is highly tensile, while on either sides it is compressive of relatively low magnitude. Since $\dot{u}=\partial_{x}\sigma$, the stress jumps across the fronts cause the left front to move towards the right and the right front to move towards the left, resulting in high  concentration of $\rho_1$ within the shrinking central domain. This ever growing concentration of $\rho_1$ eventually makes the effective linear bulk modulus $\tilde{B}(\rho,\phi)$ negative within the shrinking domain, thus, triggering a contractile instability there. Hence, this localized nonlinear coupling of contractile instability and segregation leads to the formation of spatially well-separated singular structures in finite time.

If we  now turn off  the binding-unbinding of the stresslets, then $\phi$ will be conserved; in which case, the shrinkage of the central bulk region stops at the largest negative strain minima set by the activity induced $\epsilon^4$ term in the effective elastic strain energy density, arising from steric effects that are inevitably present. The singularity is never reached due to this physical constraint, and, one ends up with highly concentrated tensile regions of finite width set by the steric hindrance. In presence of binding-unbinding of catch bond type, with the catch bond for $\rho_1$ stronger than $\rho_2$ (i.e., $k_4>0$, same sign as $\zeta_{\text{rel}}$), this width will be modified due to violation of local detailed balance.

\subsection{Self-similar Solution for Finite Time Singularity}

Following the above discussion, we analyse the approach to the finite-time singularity by dropping the steric stabilising cubic term in elastic stress and the turnover of stresslets. We also neglect the  viscous stresses for convenience, though this does not pose any difficulties. The
dynamical equations in 1D then reduce to the conservative system
\begin{subequations}
	\begin{align}
		& \dot{\rho}=\partial_{x}\big(D\partial_{x}\rho-\partial_x\sigma\,\rho\big)\label{rho-1d-simple}\\
		& \dot{\phi}=\partial_{x}\big(D\partial_{x}\phi-\partial_x\sigma\,\phi\big)\label{phi-1d-simple}\\
		&\dot{\epsilon}=\partial^2_{xx}\sigma,\label{epsilon-1d-simple}
	\end{align}
	\label{dyn-1d-simple}
\end{subequations} 
where
\begin{eqnarray}
	\sigma &=& 2\,\chi(\rho^0_a)\,(\zeta_{\text{avg}}\,\rho+\zeta_{\text{rel}}\,\phi)
	+\bigg(B-C^2-2\,\chi'(\rho^0_a)\,C\big(\zeta_{\text{avg}}\,\rho+\zeta_{\text{rel}}\,\phi\big)\bigg)\,\epsilon +\chi''(\rho^0_a)\,C^2\,\big(\zeta_{\text{avg}}\,\rho +\zeta_{\text{rel}}\,\phi\big)\,\epsilon^2.
	\label{stress:eqn}
\end{eqnarray}

Starting from a white noise perturbation of the homogeneous unstrained symmetric mixture, the approach to singularity happens through stages.
Within the initial linear segregation regime where $\rho$, $\phi$ and $\epsilon$ are relatively small, the first two terms (active back pressure and linear elasticity) in the expression \eqref{stress:eqn} of the stress field $\sigma$ dominate. However, as $\rho$, $\phi$ and $\epsilon$ grow exponentially due to linear instability, they would  eventually become large inside the segregated domains where local density is high, so that the system crosses over to the nonlinear regime where the  $\epsilon^2$ term in \eqref{stress:eqn} starts dominating at locations where the density is high. In this final approach to the singularity at these locations,  this $\epsilon^2$ nonlinearity in elasticity is the dominant physical mechanism. This dominance of a particular power of the driving mechanism is the origin of the scale invariance of the equation, and hence the solution, near the singularity 
\cite{eggersfontelos-book2015}. Note that, at the relatively low density regions that separate the high density domains, the dominant mechanism is the active back pressure as the effective bulk modulus in the second term in \eqref{stress:eqn} is small. Thus the interface between the high and low density regions, is {\it pushed} by the active back pressure and {\it pulled} by the contractile instability.

From the physics of the problem discussed above, we expect that there exists a finite temporal location $t_0$ and a spatial location $x_0$ for the emergence of the singularity. The scale invariance of the solution near this singularity implies that,
starting from a typical segregated domain at time $t<t_0$ as shown in Fig.\,\ref{frontsmove}, the height of the profiles $\rho$, $\phi$ and $\epsilon$ must grow, and simultaneously the domain width must shrink as $t\to t_0$, as power laws with appropriate universal exponents. This universality of the scaling exponents is a consequence of the fact that they arise solely from the structure of the pde, and do not depend on the initial condition.
This motivates us to write the following self-similar form of the solution:
\begin{subequations}
	\begin{align}
		&\rho(x,t)=\frac{C_0}{(t_0-t)^{s}}R(\xi),\label{simtrans-rho}\\
		&\phi(x,t)=\frac{A_0}{(t_0-t)^{p}}\Phi(\xi),\label{simtrans-phi}\\
		&\epsilon(x,t)=\frac{B_0}{(t_0-t)^{q}}E(\xi),\label{simtrans-e}
	\end{align}
	\label{simtrans}
\end{subequations}
		where the scaling variable, 
	$$\xi:=\frac{x-x_0}{(t_0-t)^r}.$$
In the above,		
	$A_0$, $B_0$ and $C_0$ are constants; $s>0$, $p>0$, $q>0$ and $r>0$ are the scaling exponents; and $R$, $\Phi$ and $E$ are analytic scaling functions. 
These scaling forms imply the following for the derivatives,
\begin{subequations}
	\begin{align}
		& \dot \rho=\frac{C_0}{(t_0-t)^{1+s}}\bigg[s\,R(\xi)+r\,\xi\,R'(\xi)\bigg],~~\dot \phi=\frac{A_0}{(t_0-t)^{1+p}}\bigg[p\,\Phi(\xi)+r\,\xi\,\Phi'(\xi)\bigg],\nonumber\\
		&\dot \epsilon=\frac{B_0}{(t_0-t)^{1+q}}\bigg[q\,E(\xi)+r\,\xi\,E'(\xi)\bigg],\\
		&\partial_x \rho=\frac{C_0}{(t_0-t)^{s+r}}R'(\xi),~~\partial^2_{xx} \rho=\frac{C_0}{(t_0-t)^{s+2r}}R''(\xi),\\
		&\partial_x \phi=\frac{A_0}{(t_0-t)^{p+r}}\Phi'(\xi),~~\partial^2_{xx} \phi=\frac{A_0}{(t_0-t)^{p+2r}}\Phi''(\xi),\\
		&\partial_x \epsilon=\frac{B_0}{(t_0-t)^{q+r}}E'(\xi),~~\partial^2_{xx} \epsilon=\frac{B_0}{(t_0-t)^{q+2r}}E''(\xi),
	\end{align}
	\label{simtrans-der}
\end{subequations}
where, $(\cdot)'$ denotes differentiation with respect to $\xi$.

With this scaling form, we are left with the following  unknown quantities: the blowup time $t_0$ and location $x_0$, the exponents $s$, $p$, $q$ and  $r$, and the  similarity profiles $R(\xi)$, $\Phi(\xi)$ and $E(\xi)$. These are obtained by substituting the similarity forms \eqref{simtrans} and $\eqref{simtrans-der}$ into the system \eqref{dyn-1d-simple}. Also, since diffusion cannot produce any finite time singularity, we will ignore the diffusion term in the $\rho$ and $\phi$-equations.

The $\epsilon$-equation gives,

\begin{eqnarray}
	\frac{B_0}{(t_0-t)^{1+q}}\bigg(q\,E+r\,\xi\,E'\bigg) &=& 2\,\chi(\rho^0_a)\,\bigg(\zeta_{\text{avg}}\,\frac{C_0}{(t_0-t)^{s+2r}}R''+\zeta_{\text{rel}}\,\frac{A_0}{(t_0-t)^{p+2r}}\Phi''\bigg)\nonumber\\
&&+\big(B-C^2\big)\frac{B_0}{(t_0-t)^{q+2r}}E''-2\,\chi'(\rho^0_a)\,C\bigg(\zeta_{\text{avg}}\,\frac{C_0}{(t_0-t)^{s}}R+\zeta_{\text{rel}}\,\frac{A_0}{(t_0-t)^{p}}\Phi\bigg)\,\frac{B_0}{(t_0-t)^{q+2r}}E''\nonumber\\
&&-2\,\chi'(\rho^0_a)\,C\bigg(\zeta_{\text{avg}}\,\frac{C_0}{(t_0-t)^{s+r}}R'+\zeta_{\text{rel}}\,\frac{A_0}{(t_0-t)^{p+r}}\Phi'\bigg)\,\frac{B_0}{(t_0-t)^{q+r}}E'\nonumber\\
&&-2\,\chi'(\rho^0_a)\,C\bigg(\zeta_{\text{avg}}\,\frac{C_0}{(t_0-t)^{s+r}}R'+\zeta_{\text{rel}}\,\frac{A_0}{(t_0-t)^{p+r}}\Phi'\bigg)\,\frac{B_0}{(t_0-t)^{q+r}}E'\nonumber\\
&&-2\,\chi'(\rho^0_a)\,C\bigg(\zeta_{\text{avg}}\,\frac{C_0}{(t_0-t)^{s+2r}}R''+\zeta_{\text{rel}}\,\frac{A_0}{(t_0-t)^{p+2r}}\Phi''\bigg)\,\frac{B_0}{(t_0-t)^{q}}E\nonumber\\ &&+2\chi''(\rho^0_a)\,C^2\,\bigg(\zeta_{\text{avg}}\,\frac{C_0}{(t_0-t)^{s+r}}R' +\zeta_{\text{rel}}\,\frac{A_0}{(t_0-t)^{p+r}}\Phi'\bigg)\,\frac{B_0^2}{(t_0-t)^{2q+r}}EE'\nonumber\\
&&+\chi''(\rho^0_a)\,C^2\,\bigg(\zeta_{\text{avg}}\,\frac{C_0}{(t_0-t)^{s+2r}}R'' +\zeta_{\text{rel}}\,\frac{A_0}{(t_0-t)^{p+2r}}\Phi''\bigg)\,\frac{B_0^2}{(t_0-t)^{2q}}E^2\nonumber\\
&&+2\chi''(\rho^0_a)\,C^2\,\bigg(\zeta_{\text{avg}}\,\frac{C_0}{(t_0-t)^{s}}R +\zeta_{\text{rel}}\,\frac{A_0}{(t_0-t)^{p}}\Phi\bigg)\,\frac{B_0^2}{(t_0-t)^{2q+2r}}\Big(EE''+E'^2\Big)\nonumber\\
&&+2\chi''(\rho^0_a)\,C^2\,\bigg(\zeta_{\text{avg}}\,\frac{C_0}{(t_0-t)^{s+r}}R' +\zeta_{\text{rel}}\,\frac{A_0}{(t_0-t)^{p+r}}\Phi'\bigg)\,\frac{B_0^2}{(t_0-t)^{2q+r}}EE'.\label{epsilon-ode}
\end{eqnarray}

\newpage

The $\rho$-equation gives,
\begin{eqnarray}
	\frac{C_0}{(t_0-t)^{1+s}}\bigg(s\,R+r\,\xi\,R'\bigg)&=&-\frac{C_0}{(t_0-t)^{s+r}}R'\times\nonumber\\
	&&\Bigg[2\,\chi(\rho^0_a)\,\bigg(\zeta_{\text{avg}}\,\frac{C_0}{(t_0-t)^{s+r}}R'+\zeta_{\text{rel}}\,\frac{A_0}{(t_0-t)^{p+r}}\Phi'\bigg)\nonumber\\
	&&+\big(B-C^2\big)\frac{B_0}{(t_0-t)^{q+r}}E'-2\,\chi'(\rho^0_a)\,C\bigg(\zeta_{\text{avg}}\,\frac{C_0}{(t_0-t)^{s}}R+\zeta_{\text{rel}}\,\frac{A_0}{(t_0-t)^{p}}\Phi\bigg)\,\frac{B_0}{(t_0-t)^{q+r}}E'\nonumber\\
	&&-2\,\chi'(\rho^0_a)\,C\bigg(\zeta_{\text{avg}}\,\frac{C_0}{(t_0-t)^{s+r}}R'+\zeta_{\text{rel}}\,\frac{A_0}{(t_0-t)^{p+r}}\Phi'\bigg)\,\frac{B_0}{(t_0-t)^{q}}E\nonumber\\ &&+2\chi''(\rho^0_a)\,C^2\,\bigg(\zeta_{\text{avg}}\,\frac{C_0}{(t_0-t)^{s}}R +\zeta_{\text{rel}}\,\frac{A_0}{(t_0-t)^{p}}\Phi\bigg)\,\frac{B_0^2}{(t_0-t)^{2q+r}}EE'\nonumber\nonumber\\
	&&+\chi''(\rho^0_a)\,C^2\,\bigg(\zeta_{\text{avg}}\,\frac{C_0}{(t_0-t)^{s+r}}R' +\zeta_{\text{rel}}\,\frac{A_0}{(t_0-t)^{p+r}}\Phi'\bigg)\,\frac{B_0^2}{(t_0-t)^{2q}}E^2
	\Bigg]\nonumber\\
	&&-\frac{C_0}{(t_0-t)^{s}}R\times\nonumber\\
	&&\Bigg[2\,\chi(\rho^0_a)\,\bigg(\zeta_{\text{avg}}\,\frac{C_0}{(t_0-t)^{s+2r}}R''+\zeta_{\text{rel}}\,\frac{A_0}{(t_0-t)^{p+2r}}\Phi''\bigg)\nonumber\\
	&&+\big(B-C^2\big)\frac{B_0}{(t_0-t)^{q+2r}}E''-2\,\chi'(\rho^0_a)\,C\bigg(\zeta_{\text{avg}}\,\frac{C_0}{(t_0-t)^{s}}R+\zeta_{\text{rel}}\,\frac{A_0}{(t_0-t)^{p}}\Phi\bigg)\,\frac{B_0}{(t_0-t)^{q+2r}}E''\nonumber\\
	&&-2\,\chi'(\rho^0_a)\,C\bigg(\zeta_{\text{avg}}\,\frac{C_0}{(t_0-t)^{s+r}}R'+\zeta_{\text{rel}}\,\frac{A_0}{(t_0-t)^{p+r}}\Phi'\bigg)\,\frac{B_0}{(t_0-t)^{q+r}}E'\nonumber\\
	&&-2\,\chi'(\rho^0_a)\,C\bigg(\zeta_{\text{avg}}\,\frac{C_0}{(t_0-t)^{s+r}}R'+\zeta_{\text{rel}}\,\frac{A_0}{(t_0-t)^{p+r}}\Phi'\bigg)\,\frac{B_0}{(t_0-t)^{q+r}}E'\nonumber\\
	&&-2\,\chi'(\rho^0_a)\,C\bigg(\zeta_{\text{avg}}\,\frac{C_0}{(t_0-t)^{s+2r}}R''+\zeta_{\text{rel}}\,\frac{A_0}{(t_0-t)^{p+2r}}\Phi''\bigg)\,\frac{B_0}{(t_0-t)^{q}}E\nonumber\\ &&+2\chi''(\rho^0_a)\,C^2\,\bigg(\zeta_{\text{avg}}\,\frac{C_0}{(t_0-t)^{s+r}}R' +\zeta_{\text{rel}}\,\frac{A_0}{(t_0-t)^{p+r}}\Phi'\bigg)\,\frac{B_0^2}{(t_0-t)^{2q+r}}EE'\nonumber\\
	&&+\chi''(\rho^0_a)\,C^2\,\bigg(\zeta_{\text{avg}}\,\frac{C_0}{(t_0-t)^{s+2r}}R'' +\zeta_{\text{rel}}\,\frac{A_0}{(t_0-t)^{p+2r}}\Phi''\bigg)\,\frac{B_0^2}{(t_0-t)^{2q}}E^2\nonumber\\
	&&+2\chi''(\rho^0_a)\,C^2\,\bigg(\zeta_{\text{avg}}\,\frac{C_0}{(t_0-t)^{s}}R +\zeta_{\text{rel}}\,\frac{A_0}{(t_0-t)^{p}}\Phi\bigg)\,\frac{B_0^2}{(t_0-t)^{2q+2r}}\Big(EE''+E'^2\Big)\nonumber\\
	&&+2\chi''(\rho^0_a)\,C^2\,\bigg(\zeta_{\text{avg}}\,\frac{C_0}{(t_0-t)^{s+r}}R' +\zeta_{\text{rel}}\,\frac{A_0}{(t_0-t)^{p+r}}\Phi'\bigg)\,\frac{B_0^2}{(t_0-t)^{2q+r}}EE'
	\Bigg].\label{rho-ode}
\end{eqnarray}

\newpage

The $\phi$-equation gives,
\begin{eqnarray}
 \frac{A_0}{(t_0-t)^{1+p}}\bigg(p\,\Phi+r\,\xi\,\Phi'\bigg)&=&-\frac{A_0}{(t_0-t)^{p+r}}\Phi'\times\nonumber\\
 &&\Bigg[2\,\chi(\rho^0_a)\,\bigg(\zeta_{\text{avg}}\,\frac{C_0}{(t_0-t)^{s+r}}R'+\zeta_{\text{rel}}\,\frac{A_0}{(t_0-t)^{p+r}}\Phi'\bigg)\nonumber\\
 &&+\big(B-C^2\big)\frac{B_0}{(t_0-t)^{q+r}}E'-2\,\chi'(\rho^0_a)\,C\bigg(\zeta_{\text{avg}}\,\frac{C_0}{(t_0-t)^{s}}R+\zeta_{\text{rel}}\,\frac{A_0}{(t_0-t)^{p}}\Phi\bigg)\,\frac{B_0}{(t_0-t)^{q+r}}E'\nonumber\\
 &&-2\,\chi'(\rho^0_a)\,C\bigg(\zeta_{\text{avg}}\,\frac{C_0}{(t_0-t)^{s+r}}R'+\zeta_{\text{rel}}\,\frac{A_0}{(t_0-t)^{p+r}}\Phi'\bigg)\,\frac{B_0}{(t_0-t)^{q}}E\nonumber\\ &&+2\chi''(\rho^0_a)\,C^2\,\bigg(\zeta_{\text{avg}}\,\frac{C_0}{(t_0-t)^{s}}R +\zeta_{\text{rel}}\,\frac{A_0}{(t_0-t)^{p}}\Phi\bigg)\,\frac{B_0^2}{(t_0-t)^{2q+r}}EE'\nonumber\nonumber\\
 &&+\chi''(\rho^0_a)\,C^2\,\bigg(\zeta_{\text{avg}}\,\frac{C_0}{(t_0-t)^{s+r}}R' +\zeta_{\text{rel}}\,\frac{A_0}{(t_0-t)^{p+r}}\Phi'\bigg)\,\frac{B_0^2}{(t_0-t)^{2q}}E^2
 \Bigg]\nonumber\\
 &&-\frac{A_0}{(t_0-t)^{p}}\Phi\times\nonumber\\
 &&\Bigg[2\,\chi(\rho^0_a)\,\bigg(\zeta_{\text{avg}}\,\frac{C_0}{(t_0-t)^{s+2r}}R''+\zeta_{\text{rel}}\,\frac{A_0}{(t_0-t)^{p+2r}}\Phi''\bigg)\nonumber\\
 &&+\big(B-C^2\big)\frac{B_0}{(t_0-t)^{q+2r}}E''-2\,\chi'(\rho^0_a)\,C\bigg(\zeta_{\text{avg}}\,\frac{C_0}{(t_0-t)^{s}}R+\zeta_{\text{rel}}\,\frac{A_0}{(t_0-t)^{p}}\Phi\bigg)\,\frac{B_0}{(t_0-t)^{q+2r}}E''\nonumber\\
 &&-2\,\chi'(\rho^0_a)\,C\bigg(\zeta_{\text{avg}}\,\frac{C_0}{(t_0-t)^{s+r}}R'+\zeta_{\text{rel}}\,\frac{A_0}{(t_0-t)^{p+r}}\Phi'\bigg)\,\frac{B_0}{(t_0-t)^{q+r}}E'\nonumber\\
 &&-2\,\chi'(\rho^0_a)\,C\bigg(\zeta_{\text{avg}}\,\frac{C_0}{(t_0-t)^{s+r}}R'+\zeta_{\text{rel}}\,\frac{A_0}{(t_0-t)^{p+r}}\Phi'\bigg)\,\frac{B_0}{(t_0-t)^{q+r}}E'\nonumber\\
 &&-2\,\chi'(\rho^0_a)\,C\bigg(\zeta_{\text{avg}}\,\frac{C_0}{(t_0-t)^{s+2r}}R''+\zeta_{\text{rel}}\,\frac{A_0}{(t_0-t)^{p+2r}}\Phi''\bigg)\,\frac{B_0}{(t_0-t)^{q}}E\nonumber\\ &&+2\chi''(\rho^0_a)\,C^2\,\bigg(\zeta_{\text{avg}}\,\frac{C_0}{(t_0-t)^{s+r}}R' +\zeta_{\text{rel}}\,\frac{A_0}{(t_0-t)^{p+r}}\Phi'\bigg)\,\frac{B_0^2}{(t_0-t)^{2q+r}}EE'\nonumber\\
 &&+\chi''(\rho^0_a)\,C^2\,\bigg(\zeta_{\text{avg}}\,\frac{C_0}{(t_0-t)^{s+2r}}R'' +\zeta_{\text{rel}}\,\frac{A_0}{(t_0-t)^{p+2r}}\Phi''\bigg)\,\frac{B_0^2}{(t_0-t)^{2q}}E^2\nonumber\\
 &&+2\chi''(\rho^0_a)\,C^2\,\bigg(\zeta_{\text{avg}}\,\frac{C_0}{(t_0-t)^{s}}R +\zeta_{\text{rel}}\,\frac{A_0}{(t_0-t)^{p}}\Phi\bigg)\,\frac{B_0^2}{(t_0-t)^{2q+2r}}\Big(EE''+E'^2\Big)\nonumber\\
 &&+2\chi''(\rho^0_a)\,C^2\,\bigg(\zeta_{\text{avg}}\,\frac{C_0}{(t_0-t)^{s+r}}R' +\zeta_{\text{rel}}\,\frac{A_0}{(t_0-t)^{p+r}}\Phi'\bigg)\,\frac{B_0^2}{(t_0-t)^{2q+r}}EE'\Bigg].\label{phi-ode}
\end{eqnarray}

Now, since we have ignored binding-unbinding,   $\rho(x,t)$ and $\phi(x,t)$ are conserved, thus
	\begin{subequations}
		\begin{align}
			&\int_{-1}^{1}\rho(x,t)\,dx=\int\frac{C_0}{(t_0-t)^{s-r}}R\,d\xi=const.,~~\text{and}\label{consvrho}\\
			&\int_{-1}^{1}\phi(x,t)\,dx=\int\frac{A_0}{(t_0-t)^{p-r}}\Phi\,d\xi=const.\label{consvphi}
		\end{align}
	\label{consvcod}
	\end{subequations}
	
	Condition \eqref{consvcod} implies $s=p=r$.

		To get the asymptotic form of the singularity, we analyse the coupled nonlinear odes  \eqref{epsilon-ode},  \eqref{rho-ode} and \eqref{phi-ode} using the method of dominant balance~\cite{eggersfontelos-book2015}. The largest order nonlinearity should dominate near the singularity. From this balance, we obtain 
	
	\begin{equation}
		1+q=s+r+2q+r,~~1+s=s+s+r+2q+r,~~1+p=p+s+r+2q+r ~~~\Rightarrow~~~s=p=r=\frac{1}{3},~~~ q=0.	\end{equation}

Hence,

\begin{subequations}
	\begin{align}
	& 	\rho\sim\frac{1}{(t_0-t)^{\frac{1}{3}}} R\Big(\frac{x-x_0}{(t_0-t)^{\frac{1}{3}}}\Big),~~~~\dot{\rho}\sim\frac{1}{(t_0-t)^{\frac{4}{3}}} R_1\Big(\frac{x-x_0}{(t_0-t)^{\frac{1}{3}}}\Big),\\	
	&	\phi\sim\frac{1}{(t_0-t)^{\frac{1}{3}}} \Phi\Big(\frac{x-x_0}{(t_0-t)^{\frac{1}{3}}}\Big),~~~~\dot{\phi}\sim\frac{1}{(t_0-t)^{\frac{4}{3}}} \Phi_1\Big(\frac{x-x_0}{(t_0-t)^{\frac{1}{3}}}\Big),\\
	&	\dot{\epsilon} \sim \frac{1}{t_0-t} E_1\Big(\frac{x-x_0}{(t_0-t)^{\frac{1}{3}}}\Big)~~\Rightarrow~~\epsilon\sim\ln(t_0-t)E\Big(\frac{x-x_0}{(t_0-t)^{\frac{1}{3}}}\Big).
	\end{align}
	\label{scalingforms}
\end{subequations}

	The dominant balance gives the following leading order nonlinear advection equation as the singularity is approached:
	\begin{subequations}
		\begin{align}
			&\dot{\rho}+\partial_{x}\big(\rho\,\partial_{x}\sigma_2\big)=0,\label{rho-1d-simple-lo}\\
			& \dot{\phi}+\partial_{x}\big(\phi\,\partial_{x}\sigma_2\big)=0,\label{phi-1d-simple-lo}\\
			&\dot{\epsilon}=\partial^2_{xx}\sigma_2,\label{epsilon-1d-simple-lo}\\
			&\sigma_2:=\chi''(\rho^0_a)\,C^2\,\big(\zeta_{\text{avg}}\rho+\zeta_{\text{rel}}\phi\big)\,\epsilon^2.
		\end{align}
		\label{dyn-1d-simple-lo}
	\end{subequations}

The initial data in $\phi$ is of the form $\phi(x,0)\sim\phi_0\Big(\frac{x-x_0}{l}\Big)$, parameterized by an initial width $l$. We obtain the blowup time $t_0$ using dimensional analysis: $t_0\sim t^\star(l/l^\star)^3$, where $t^\star:=\Big(\frac{\Gamma}{{B_2}^2}\Big)^{\frac{1}{3}}$,  $l^\star:=\Big(\frac{B_2}{\Gamma^2}\Big)^{\frac{1}{3}}$, and  $B_2:=\chi''(\rho^0_a)\,C^2\,\zeta_{\text{rel}}$. 

Thus the spatial ($x_0$) and temporal ($t_0$) location of the singularity depend on the initial condition.


    \subsection{Numerical Analysis of the Finite Time Singularity}

\begin{figure}[H]
		\centering
		\subfigure[]{\includegraphics[scale=1.1]{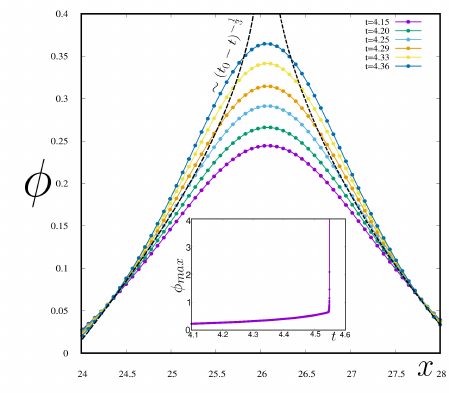}}
		\hspace{5mm}
		\subfigure[]{\includegraphics[scale=1.8]{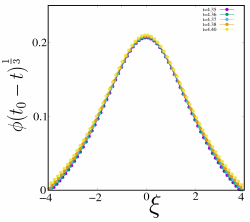}}
		\caption{(a) The profiles of $\phi$ at different times as it approaches the finite time singularity $t_0$. Inset shows our estimation of the blowup time $t_0=4.55$ from the divergence of $\phi_{max}(t)$.  (b) The profiles of $\phi$ at different times collapse to the scaling form \eqref{scalingforms} where $\xi:=\frac{x-x_0}{(t_0-t)^{1/3}}$ is the scaling variable.}
		\label{numscaling:phi}
	\end{figure}

\begin{figure}[H]
		\centering
		\subfigure[]{\includegraphics[scale=1.1]{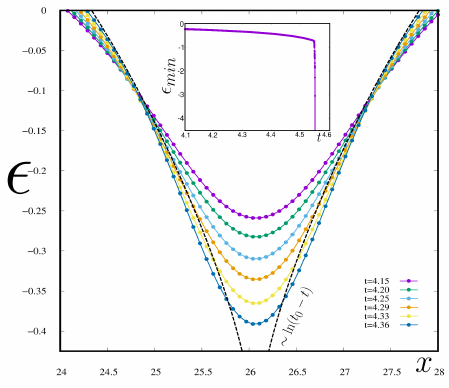}}
		\hspace{5mm}
		\subfigure[]{\includegraphics[scale=1.7]{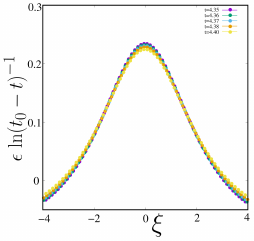}}
		\caption{(a) The profiles of $\epsilon$ at different times as it approaches the finite time singularity $t_0$. Inset shows our estimation of the blowup time $t_0=4.55$ from the divergence of $\epsilon_{max}(t)$.  (b) The profiles of $\epsilon$ at different times collapse to the scaling form \eqref{scalingforms} where $\xi:=\frac{x-x_0}{(t_0-t)^{1/3}}$ is the scaling variable. The scaling collapse here is not as good as that of the $\phi$-profile, since the growth of $\epsilon$ towards the finite time singularity is slower. }
		\label{numscaling:epsilon}
	\end{figure}

We determine the blowup time $t_0$ and the blowup location $x_0$ from the numerical solution of the pde \eqref{num:eqns} in the segregation regime with a fixed initial condition. For this part of the analysis, the pdes were solved using the spectral methods based scheme Dedalus~\cite{dedalus2020}. The parameters are set at $B=10$, $C=1$, $D=1$, $k_1=1$, $k_2=0$, $k^b_{\text{rel}}=0$, $k_3=1$, $\zeta_{\text{avg}}=3$, $\zeta_{\text{rel}}=2$, $k_4=2$, $\chi(\rho_a^0)=1$, $\chi'(\rho_a^0)=0.01$,  $\chi''(\rho_a^0)=0.001$, $\chi'''(\rho_a^0)=0.001$. We use  SBDF2 time-stepper with a time step size of $dt=10^{-4}$, taking 500 modes (Nx=500) in a domain length $Lx=35$.
The initial condition is set at a white noise perturbation of magnitude $0.2$ about the homogeneous unstrained symmetric mixture $\phi=0$.
We find that, eventually, singularities in $\rho$, $\phi$ and $\epsilon$ develop in the high density regimes. The profiles of $\phi$ and $\epsilon$ near the singularity, at different time points, are shown in Fig.\,\ref{numscaling:phi}(a) and Fig.\,\ref{numscaling:epsilon}(a), respectively.

We determine the numerical blowup time $t_0$ and location $x_0$, from the divergence of the maximum (minimum) value of $\phi$ ($\epsilon$), as seen in the insets of  Fig.\,\ref{numscaling:phi}(a) (Fig.\,\ref{numscaling:epsilon}(a)).
We find that at the numerical blowup location $x_0=26.06$ and  blowup time $t_0=4.55$, the reasonable collapse is achieved for both the scaled $\phi$ and $\epsilon$ profiles, see  Fig.\,\ref{numscaling:phi}(b) and Fig.\,\ref{numscaling:epsilon}(b).

		\begin{figure}[H]
		\centering
		\includegraphics[scale=0.72]{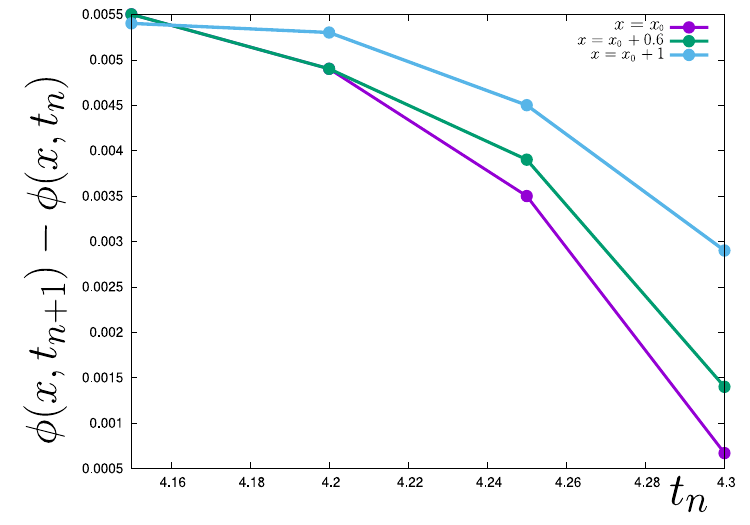}
		\caption{Cauchy convergence of scaled values of $\phi(x, t_{n+1})- \phi(x, t_n)$ as we approach singularity at $x=x_0$. }
		\label{fig:convergence}
	\end{figure}

	Since numerically we cannot approach the singular point  arbitrarily closely, we test the asymptotic convergence of the profiles approaching the singularity, using the Cauchy-convergence criterion for the sequence $\phi(x,t_n)$, for  fixed values of $x$ near $x_0$. In Fig.\,\ref{fig:convergence}, we have plotted 
	$\phi(x,t_{n+1})-\phi(x,t_{n})$ as a function of $t_n$, at three spatial locations $x-x_0=0$, $x-x_0=0.6$, and $x-x_0=1$. We see that $\phi(x,t_{n+1})-\phi(x,t_{n})\to 0$ faster for points closer to the centre of the profile; the convergence rate is maximum at $\phi_{max}$.

	\newpage

\section{Active Tension Chains and their Networks}

\begin{figure}[H]
	\centering
	\includegraphics[scale=.7]{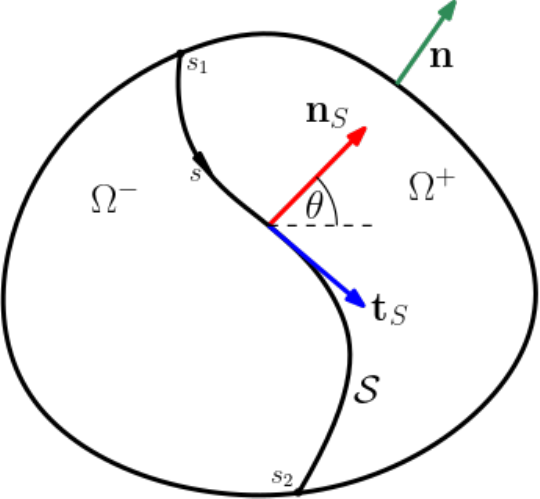}
	\caption{Geometry of the tension chain $\mathcal{S}$ that divides the fixed 2D domain into
	two disjoint parts $\Omega^\pm$. The rest of the symbols are described in the text.}
	\label{sing}
\end{figure}

We consider a moving tension chain in a fixed 2D domain $\Omega$, represented by the smooth evolving curve  $\mathcal{S}$ as shown in Fig.\,\ref{sing}. $\mathcal{S}$ is a material curve consisting of the stronger stresslet species, across which bulk fields of the weaker stresslet species may suffer jump discontinuities.

\subsection{Geometry}

$\mathcal{S}$ has a local parametrization $\mathbf{x}=\boldsymbol{r}(s,t)$, where $s$ is an arc length parameter. The unit tangent field on $\mathcal{S}$ is $\mathbf{t}_S:=\partial_s\boldsymbol{r}$; the unit normal field  $\mathbf{n}_S$ is defined such that $\{\mathbf{t}_S,\mathbf{n}_S\}$ is positively oriented. Then, the normal angle field $\theta(s,t)$ is defined such that $\mathbf{n}_S=(\cos\theta,\sin\theta)$ and $\mathbf{t}_S=(\sin\theta,-\cos\theta)$. The curvature is defined as $H:=\partial_s\theta$. The variation of the normal and tangent along the curve is given by the Serret-Frenet formula\cite{SimhaBhattacharya1998}: $\partial_s\mathbf{n}_S=-H\,\mathbf{t}_S$ and $\partial_s\mathbf{t}_S=H\,\mathbf{n}_S$.

The normal and tangential  speeds are defined as  $V:=\partial_t\boldsymbol{r}\cdot\mathbf{n}_S$ and $W:=-\partial_t\boldsymbol{r}\cdot\mathbf{t}_S$, respectively. $\mathcal{S}$ divides the domain $\Omega$ into two disjoint parts $\Omega^\pm$; we assume that $\mathbf{n}_S$ points into $\Omega^+$. A normal arc-length trajectory at a point $\mathbf{x}$ on $\mathcal{S}$ is a curve $\bar{\alpha}(t)$ such that $\mathbf{t}_S(\bar{\alpha}(t),t)\cdot\frac{d\mathbf{x}(\bar{\alpha}(t),t)}{dt}=0$. Then, the normal time derivative of a smooth field $\psi_S$ on $\mathcal{S}$ is defined as $\mathring{\psi_S}:=\frac{d\phi_S(\bar{\alpha}(t),t)}{dt}$ \cite{SimhaBhattacharya1998}. From this definition, it follows that  $\mathring{\boldsymbol{r}}=V\,\mathbf{n}_S$. 

From these definitions, the following ``transport identities'' can be readily obtained \cite{Gurtinbook93}:
\begin{equation}
	\partial_s W=-H\,V,~~~~\mathring{\theta}=\partial_s V,~~~\mathring{H}=\partial^2_{ss}V+H^2V.
	\label{transportidentity}
\end{equation}

Hence, the evolution of the normal field and the curvature follow
\begin{equation}
	\dot{\mathbf{n}}_S=(WH-\partial_s V)\mathbf{t}_S,~~~~\dot{H}=\partial^2_{ss}V+H^2V-W\partial_sH.
\end{equation}

We emphasize the purely geometric (kinematic) nature of the above equation.


\subsection{Compatibility Conditions at the Singular Structure}

The jump in a bulk discontinuous field, say $\psi(\mathbf{X},t)$, on $\mathcal{S}$ is defined by $\llbracket\psi\rrbracket:=\psi^+-\psi^-$, where $\psi^\pm$ are the limiting values of $\psi$ as one approaches $\mathcal{S}$ from $\Omega^\pm$. We also define the average value of $\psi$ at $\mathcal{S}$ as $\langle\psi\rangle:=(\psi^++\psi^-)/2$.

The bulk displacement field $\boldsymbol{u}(\mathbf{X},t)$ is a  smooth function in $\Omega$. However, the bulk velocity field $\dot{\boldsymbol{u}}$, and the bulk displacement gradient $\nabla\boldsymbol{u}$, are piece-wise smooth functions in $\Omega$, suffering jump discontinuities  $\llbracket\dot{\boldsymbol{u}}\rrbracket$ and $\llbracket\nabla\boldsymbol{u}\rrbracket$, respectively, on $\mathcal{S}$.

Continuity of the bulk displacement field $\boldsymbol{u}$ implies the following well-known Hadamard compatibility condition for the deformation gradient $\mathbf{F}=\mathbf{I}+\nabla\boldsymbol{u}$, and the velocity compatibility condition \cite{SimhaBhattacharya1998}: 
\begin{subequations}
	\begin{align}
		&\llbracket\mathbf{F}\rrbracket\mathsf{I} =\mathbf{0}\Leftrightarrow\llbracket\mathbf{F}\rrbracket=\boldsymbol{a}\otimes\mathbf{n}_S~~~\text{for arbitrary}~\boldsymbol{a}\in\mathbb{R}^2,\label{comp1}\\
		&\llbracket\dot{\boldsymbol{u}}\rrbracket +V\,\llbracket\mathbf{F}\rrbracket \mathbf{n}_S=\mathbf{0}.\label{comp2}
	\end{align}
\end{subequations}

The intrinsic velocity field of the singular structure $\mathcal{S}$ is then defined by  $\mathbf{v}_S:=\langle\dot{\boldsymbol{u}}\rangle+V\,\langle\mathbf{F}\rangle\mathbf{n}_S$, and the intrinsic deformation gradient is $\mathsf{F}:=\mathsf{f}\otimes\mathbf{t}_S$ where $\mathsf{f}:=\langle\mathbf{F}\rangle\mathbf{t}_S$ is the stretch vector at the tension chain $\mathcal{S}$ \cite{SimhaBhattacharya1998}. We can write $\mathsf{f}=\lambda_S\tilde{\mathbf{t}}_S$, where $\lambda_S$ is the local stretching of $\mathcal{S}$ and $\tilde{\mathbf{t}}_S$ is the unit tagent vector field to the deformed $\mathcal{S}$. The Green-Lagrange strain tensor at $\mathcal{S}$ is then given by $\mathsf{E}=(\mathsf{F}^T\mathsf{F}-\mathsf{I})/2=\frac{{\lambda_S}^2-1}{2}\mathbf{t}_S\otimes\mathbf{t}_S$.

\subsection{Divergence  and Transport Theorems}

For a piece-wise smooth bulk vector field $\mathbf{v}$ and a bulk tensor field $\boldsymbol{\sigma}$  on $\Omega$, which suffer jump discontinuities $\llbracket\mathbf{v}\rrbracket$ and $\llbracket\boldsymbol{\sigma}\rrbracket$, respectively, on $\mathcal{S}$, we can write \cite{SimhaBhattacharya1998}
\begin{subequations}
	\begin{align}
		&\int_{\Omega}\nabla\cdot\mathbf{v}\,da=\int_{\partial\Omega}\mathbf{v}\cdot\mathbf{n}\,dl-\int_{\mathcal{S}}\llbracket\mathbf{v}\rrbracket\cdot\mathbf{n}_S\,ds,~\text{and} \\
		&\int_{\Omega}\nabla\cdot\boldsymbol{\sigma}\,da=\int_{\partial\Omega}\boldsymbol{\sigma}\mathbf{n}\,dl-\int_{\mathcal{S}}\llbracket\boldsymbol{\sigma}\rrbracket\mathbf{n}_S\,ds,
	\end{align}
\end{subequations}
where $da$ is the area measure on $\Omega$, $dl$ is the line element on $\partial\Omega$, and $s$ is an arc length parametrization of $\mathcal{S}$.

For a  vector field $\mathbf{v}_S$  defined on $\mathcal{S}$, 
\begin{equation}
	\int_{\mathcal{S}}\Big( \nabla^S\cdot\mathbf{v}_S+H\,\mathbf{v}_S\cdot\mathbf{n}_S\Big)\,ds=\int_{\partial\mathcal{S}}\mathbf{v}_S\cdot\mathbf{t}_S.
\end{equation}

where $\mathbf{t}_S$ is the unit outward normal to $\partial\mathcal{S}$, lying in the tangent plane of $\mathcal{S}$.

For a piecewise smooth bulk scalar field $\psi$ on $\Omega$ which suffers jump discontinuity $\llbracket\psi\rrbracket$ on $\mathcal{S}$, 
\begin{equation}
	\frac{d}{dt}\int_{\Omega}\psi\,da=\int_{\Omega}\dot{\psi}\,da-\int_{\mathcal{S}}V\llbracket\psi\rrbracket\,ds.
\end{equation}

For a singular scalar field $\psi_S$ defined on $\mathcal{S}$ \cite{SimhaBhattacharya1998}, 
\begin{equation}
	\frac{d}{dt}\int_{\mathcal{S}}\psi_S\,ds=\int_{\mathcal{S}}\Big(\mathring{\psi}_S-\psi_S HV\Big)\,ds+\int_{\partial\mathcal{S}}\psi_S\,W.
\end{equation}

\subsection{Mass Balance}

Let the bulk  mass density per unit area be $\rho$ and the singular mass density per unit length be $\rho_S$, the bulk and singular mass  flux be $\mathbf{j}$ and $\mathbf{j}_S$ (that includes both mass diffusion and mass advection by the respective velocity fields), the bulk and interface mass turnover (due to binding and unbinding) be $\Pi$ and $\Pi_S$, respectively. Then, the equation for the global balance of mass reads
\begin{equation}
	\frac{d}{dt}\int_{\Omega}\rho\,da+\frac{d}{dt}\int_{\mathcal{S}}\rho_S\,ds=\int_{\Omega}\Pi\,da+\int_{\mathcal{S}}\Pi_S\,ds-\int_{\partial\Omega}\mathbf{j}\cdot\mathbf{n}\,dl-\int_{\partial\mathcal{S}}\mathbf{j}_S\cdot\mathbf{t}_S+\int_{\partial\mathcal{S}}\rho_S W
\end{equation}
where the final term in the right hand side is due to the fact that the domain $\Omega$ is fixed while the curve $\mathcal{S}$ is evolving \cite{SwainGupta18}.

Using the divergence and transport theorems for bulk and singular fields, the above equation yields
\begin{subequations}
	\begin{align}
		&\int_{\Omega}\dot{\rho}\,da-\int_{\mathcal{S}}V\llbracket\rho\rrbracket\,ds+\int_{\mathcal{S}}\Big(\mathring{\rho}_S-\rho_S HV\Big)\,ds=\nonumber\\
		&\int_{\Omega}\Pi\,da+\int_{\mathcal{S}}\Pi_S\,ds
		-\int_{\Omega}\nabla\cdot\mathbf{j}\,da-\int_{\mathcal{S}}\llbracket\mathbf{j}\rrbracket\cdot\mathbf{n}_S\,ds-\int_{\mathcal{S}}\Big(\nabla^S\cdot\mathbf{j}_S+H\,\mathbf{j}_S\cdot\mathbf{n}_S\Big)\,ds.
	\end{align}
\end{subequations}

Using the arbitrariness of $\Omega$ and $\mathcal{S}$, and localizing, we obtain
\begin{subequations}
	\begin{align}
		&\dot{\rho}=-\nabla\cdot\mathbf{j}+\Pi~~~\text{in}~\Omega\setminus\mathcal{S},\\
		&\mathring{\rho}_S-\rho_S HV-V\llbracket\rho\rrbracket=-\llbracket\mathbf{j}\rrbracket\cdot\mathbf{n}_S-\Big(\nabla^S\cdot\mathbf{j}_S+H\,\mathbf{j}_S\cdot\mathbf{n}_S\Big)+\Pi_S~~~\text{on}~\mathcal{S}.
	\end{align}
\end{subequations}

\subsection{Force Balance}

Let the piecewise smooth bulk stress tensor be $\boldsymbol{\sigma}$, the singular stress vector (internal reaction force acting per unit cross sectional `area' of the curve $\mathcal{S}$) be $\boldsymbol{\sigma}_S$. Then, the global linear momentum balance equation, in absence of inertia of the bulk and the singular structure, is
\begin{equation}
	\int_{\partial\Omega}\boldsymbol{\sigma}\mathbf{n}\,dl+\int_{\partial\mathcal{S}}\boldsymbol{\sigma}_S=\int_{\Omega}\Gamma\dot{\boldsymbol{u}}\,da+\int_{\mathcal{S}}\Gamma_S\mathbf{v}_S\,ds,
\end{equation}
where the right hand side represents the total frictional force. Using the bulk and the singular divergence theorems, we obtain
\begin{equation}
	\int_{\Omega}\nabla\cdot\boldsymbol{\sigma}\,da+\int_{\mathcal{S}}\llbracket\boldsymbol{\sigma}\rrbracket\mathbf{n}_S\,ds+\int_{\mathcal{S}}\partial_s\boldsymbol{\sigma}_S\,ds=\int_{\Omega}\Gamma\mathbf{v}\,da+\int_{\mathcal{S}}\Gamma_S\mathbf{v}_S\,ds.
\end{equation}

Using the arbitrariness of $\Omega$ and $\mathcal{S}$, and localizing, this yields
\begin{subequations}
	\begin{align}
		&\nabla\cdot\boldsymbol{\sigma}=\Gamma\dot{\boldsymbol{u}}~~~\text{in}~\Omega,\\
		&\partial_s\boldsymbol{\sigma}_S+\llbracket\boldsymbol{\sigma}\rrbracket\mathbf{n}_S=\Gamma_S\mathbf{v}_S~~~\text{on}~\mathcal{S}.
	\end{align}
\end{subequations}

On the other hand, the global balance of angular momentum, in absence of distributed moments, reads \cite{Gurtinbook93}

\begin{equation}
	\int_{\partial\Omega}\boldsymbol{r}\times\boldsymbol{\sigma}\mathbf{n}\,dl+\int_{\partial\mathcal{S}}\boldsymbol{r}\times\boldsymbol{\sigma}_S=\int_{\Omega}\boldsymbol{r}\times\Gamma\dot{\boldsymbol{u}}\,da+\int_{\mathcal{S}}\boldsymbol{r}\times\Gamma_S\mathbf{v}_S\,ds,
\end{equation}

Using the divergence theorems, localizing, and then exploiting  the balance of linear momentum yields

\begin{subequations}
	\begin{align}
		&\boldsymbol{\sigma}^T=\boldsymbol{\sigma},\\
		&\mathbf{t}_S\times\boldsymbol{\sigma}_S=\mathbf{0}.\label{singforcetan}
	\end{align}
\end{subequations}
\eqref{singforcetan} implies that the stress vector in the singular structure is purely tangential, i.e., $\boldsymbol{\sigma}_S=\gamma\,\mathbf{t_S}$. Here, $\gamma$ is the capillary tension of the singular structure. Hence, the singular momentum balance yields
\begin{equation}
	\partial_s\gamma\,\mathbf{t}_S+\gamma H\,\mathbf{n}_S+\llbracket\boldsymbol{\sigma}\rrbracket\mathbf{n}_S=\Gamma_S\mathbf{v}_S~~~\text{on}~\mathcal{S}.\label{singmom}
\end{equation}

Normal and tangential projections of this equation onto $\mathcal{S}$ are, respectively,
\begin{subequations}
	\begin{align}
		& \gamma H+\llbracket\boldsymbol{\sigma}\rrbracket\mathbf{n}_S\cdot\mathbf{n}_S=\Gamma_S\mathbf{v}_S\cdot\mathbf{n}_S,~~\text{and}\\
		&\partial_s\gamma+\llbracket\boldsymbol{\sigma}\rrbracket\mathbf{n}_S\cdot\mathbf{t}_S=\Gamma_S\mathbf{v}_S\cdot\mathbf{t}_S.
	\end{align}
\end{subequations}

If we assume, for simplicity, that both the passive and the active parts of the bulk stress are isotropic, i.e., $\boldsymbol{\sigma}^{\pm}=p^\pm\mathbf{I}$, then these equations boil down to
\begin{subequations}
	\begin{align}
		& \gamma H+\llbracket p^e\rrbracket+\llbracket p^a\rrbracket=\Gamma_S\mathbf{v}_S\cdot\mathbf{n}_S,~~\text{and}\\
		&\partial_s\gamma=\Gamma_S\mathbf{v}_S\cdot\mathbf{t}_S.
	\end{align}
	\label{singmonbal}
\end{subequations}


\subsection{Static Tension Chains}

For the stationary case,  \eqref{singmonbal}  implies the active version of Young-Laplace law:
\begin{equation}
	\gamma H+\llbracket p^e\rrbracket+\llbracket p^a\rrbracket=0,~~\text{and}~~\gamma=const.
	\label{younglaplace}
\end{equation}
Hence, (1) the interface is flat ($H=0$) if active pressure jump $		\llbracket p^a \rrbracket$ counter-balances the (passive) elastic pressure jump $	\llbracket p^e \rrbracket$ (i.e., $\llbracket p^e\rrbracket+\llbracket p^a\rrbracket=0$); (2) in absence of passive  pressure jump (i.e., $\llbracket p^e\rrbracket=0$), active pressure jump  gives rise to a curved singularity ($H=-\llbracket p^a\rrbracket/\gamma$).


\subsection{Interaction between Tension Chains}

If two singular structures move towards each other, will they scatter, i.e., cross each other, when they meet (phantom crossing) or will they merge into a single singular structure? Without solving the complete initial-boundary-value-problem, we can approach this question qualitatively, restricting to two cases: when one of the singular structures is curved and the other is straight; and when both are straight and parallel.

\subsubsection{Curved lines: phantom crossings}

\begin{figure}[H]
	\centering
	\includegraphics[scale=.7]{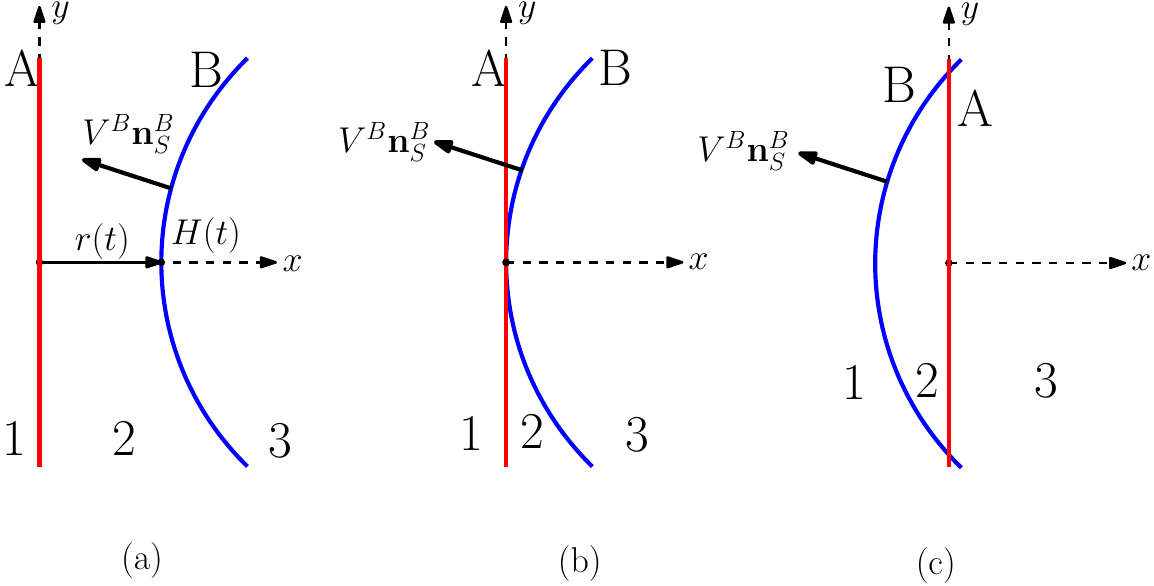}
	\caption{Interaction between a stationary straight A and moving curved B singular lines can be analysed in three possible stages: (a) before encounter B moves towards the left, (b) first time encounter of B with A, and (c) after encounter B continues to move towards the left. The bulk regions in between the lines are labelled 1, 2 and 3. }
	\label{twosing}
\end{figure}

We will consider two singular structures A and B, where A is a static straight line and B is a line of constant curvature $H(t)$ moving towards A, as shown in Fig.\,\ref{twosing}. Let the minimum distance between A and B be $r(t)$.

We now make the assumptions that (1) the tangential speed $W\equiv 0$ for both A and B; (2) the bulk dynamics is fast and bulk deformation is negligible in the time scale of the dynamics of the singular structures, i.e., $\dot{\boldsymbol{u}}=0$ and $\mathbf{F}\approx\mathbf{I}$ $\Rightarrow$ $\mathbf{v}_S=V\mathbf{n}_S$; (3) $\nabla\rho= 0$ in the bulk and $\partial_s\rho_S=0$ on both A and B; and (4) passive bulk pressure is the same across A and B:  $p^e_1=p^e_2=p^e_3$, $\gamma^A=\zeta^S\rho^A_S$,  $\gamma^B=\zeta^S\rho^B_S$,  $p^a_i=\zeta\rho_i$.

Then for configuration Fig.\,\ref{twosing}(a), A is static (i.e., $V^A=0$) and B is moving leftwards, imply
\begin{equation}
	p_1=p_2~~~~\text{and}~~~~\gamma^B H^B-(p_2-p_3)>0\,,
\end{equation}

and for configuration Fig.\,\ref{twosing}(c), A is static and B is moving leftwards imply
\begin{equation}
	p_2=p_3~~~~\text{and}~~~~\gamma^B H^B-(p_1-p_2)>0.
\end{equation}

Hence, if
\begin{equation}
	p_1=p_2=p_3~~~~\text{and}~~~~\gamma^B H^B>0,
\end{equation}
then B will cross A.

The curvature of B evolves according to the kinematic equation,
\begin{equation}
	\dot{H}^B=\partial^2_{ss}V^B+(H^B)^2V^B.
\end{equation}

\subsubsection{Straight lines: mergers}

\begin{figure}[H]
	\centering
	\includegraphics[scale=.7]{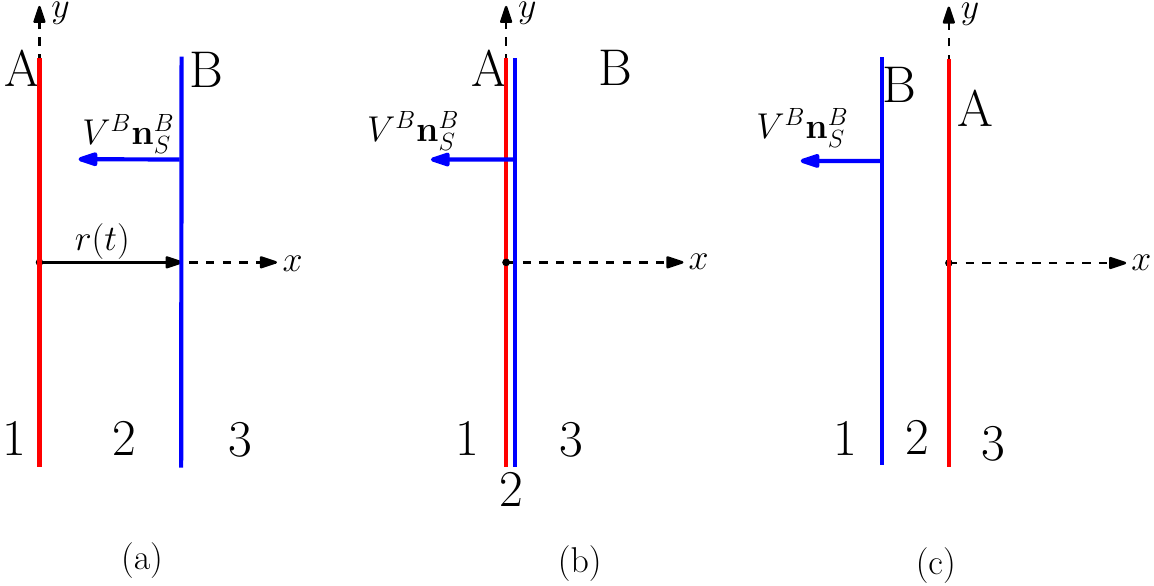}
	\caption{Interaction between two straight singular lines, one stationary A and the other moving B,  can be analysed in three possible stages: (a) before encounter B moves towards the left, (b) first time encounter of B with A, and (c) after encounter B continues to move towards the left. The bulk regions in between the lines are labelled 1, 2 and 3.}
	\label{twosingflats}
\end{figure}

We will now consider two singular structures A and B, where A is a static straight line and B is a straight line moving towards A, as shown in Fig.\,\ref{twosingflats}. Let the minimum distance between A and B be $r(t)$.

For the stage (a), A is static (i.e., $V^A=0$) and B is moving leftwards (i.e., $V^B>0$ as shown in the figure) imply
\begin{equation}
	p_1=p_2~~~~\text{and}~~~~p_3-p_2>0\,,
\end{equation}
and 
for the stage (c), A is static and B is moving leftwards imply
\begin{equation}
	p_2=p_3~~~~\text{and}~~~~p_2-p_1>0.
\end{equation}
These conditions cannot be satisfied simultaneously, which means that the two straight singular structures moving towards each other will merge, giving rise to a single bundled singular structure.

Thus, from the above qualitative analysis, obtained without solving the dynamical equations,  we conclude that 

(1) two straight parallel singular structure moving towards each other will merge to form a bundle.

(2) a convex singular structure (even with infinitesimal curvature) advancing towards a straight singular structure will cross the latter. 


\subsection{Junctions of Tension Chains}

\begin{figure}[H]
	\centering
	\includegraphics[scale=.7]{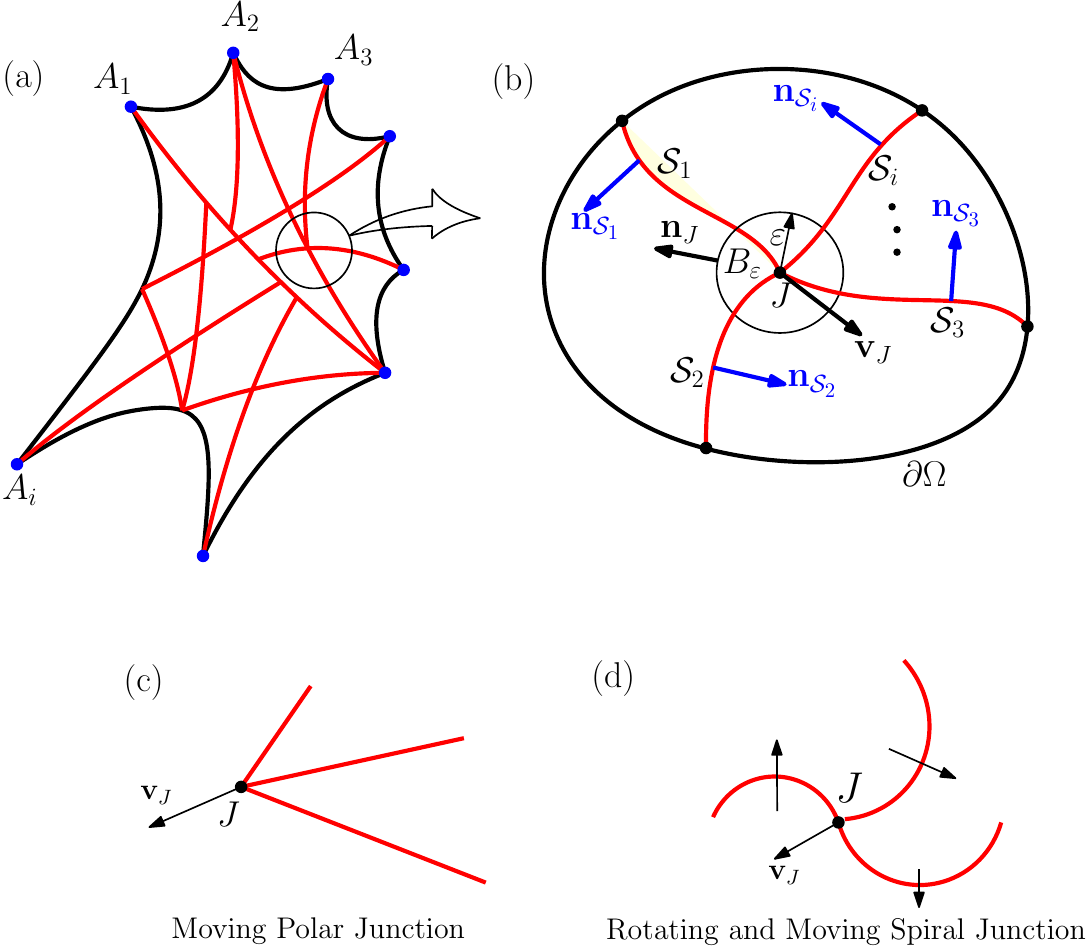}
	\caption{(a) An active web of tension in the cellular interior stabilised by discrete anchoring sites $\{A_i\}$ at the cell boundary. (b) A zoom-up of the geometry of a  junction $J$ supporting  $N$ tension chains $\mathcal{S}_1$, \ldots, $\mathcal{S}_N$. (c) A moving polar junction $J$ with velocity ${\bf v}_J$ and (d) A moving and rotating spiral junction.}
	\label{tcjunction}
\end{figure}

Here we analyse the mass, force and torque balance conditions on a junction $J$ supporting $N$ tension chains, as shown in Fig.\,\ref{tcjunction}(b). For this, we first review the divergence and transport theorems as applied to curves $\mathcal{S}_i$ meeting at a junction $J$  embedded in a 2D domain $\Omega$.

\subsubsection{Divergence and Transport theorems}
The divergence theorem for a piece-wise smooth tensor field $\boldsymbol{\sigma}$ on $\Omega$ which is discontinuous across the curves $\mathcal{S}_i$ and singular at the junction $J$ \cite{SimhaBhattacharya1998} is
\begin{equation}
	\int_{\Omega}\nabla\cdot\boldsymbol{\sigma}\,da=\int_{\partial\Omega}\boldsymbol{\sigma}\mathbf{n}\,dl-\sum_{i=1}^{N}\int_{\mathcal{S}_i}\llbracket\boldsymbol{\sigma}\rrbracket\mathbf{n}_{\mathcal{S}_i}\,ds-	\lim_{\varepsilon\to 0}\int_{\partial B_{\varepsilon}}\boldsymbol{\sigma}\mathbf{n}_J\,dl,
\end{equation}
where $B_{\varepsilon}$ is a small ball of radius $\varepsilon$ containing the junction. The divergence theorem for a piece-wise smooth vector field $\mathbf{v}$ on $\Omega$ which is discontinuous across the curves $\mathcal{S}_i$ and singular at the junction $J$ \cite{SimhaBhattacharya1998} is
\begin{equation}
	\int_{\Omega}\nabla\cdot\mathbf{v}\,da=\int_{\partial\Omega}\mathbf{v}\cdot\mathbf{n}\,dl-\sum_{i=1}^{N}\int_{\mathcal{S}_i}\llbracket\mathbf{v}\rrbracket\mathbf{n}_{\mathcal{S}_i}\,ds-	\lim_{\varepsilon\to 0}\int_{\partial B_{\varepsilon}}\mathbf{v}\cdot\mathbf{n}_J\,dl.
\end{equation}

For a  vector field $\mathbf{v}_{S_i}$  defined on $\mathcal{S}_i$ whose one end $A_i$ lies on $\partial\Omega$ and the other end is at $J$, 
\begin{equation}
	\int_{\mathcal{S}_i}\Big( \nabla^S\cdot\mathbf{v}_{S_i}+H\,\mathbf{v}_{S_i}\cdot\mathbf{n}_{S_i}\Big)\,ds=(\mathbf{v}_{S_i}\cdot\mathbf{t}_{S_i})\bigg|_{A_i}-(\mathbf{v}_{S_i}\cdot\mathbf{t}_{S_i})\bigg|_{J}.
\end{equation}

For a piece-wise smooth bulk scalar field $\psi$ on $\Omega$ which suffers jump discontinuity $\llbracket\psi\rrbracket_i$ on $\mathcal{S}_i$, and is singular at the junction $J$, the transport theorem is \cite{SimhaBhattacharya1998}:  
\begin{equation}
	\frac{d}{dt}\int_{\Omega}\psi\,da=\int_{\Omega}\dot{\psi}\,da-\sum_{i=1}^{N}\int_{\mathcal{S}_i}V_i\llbracket\psi\rrbracket_i\,ds-\lim_{\varepsilon\to 0}\int_{\partial B_\varepsilon}\psi\,\mathbf{v}_J\cdot\mathbf{n}_J\,dl.
\end{equation}

For a  scalar field $\psi_{S_i}$ defined on the curve $\mathcal{S}_i$ whose one boundary $A_i$ lies on $\partial\Omega$ and the other at the junction $J$, 
\begin{equation}
	\frac{d}{dt}\int_{\mathcal{S}_i}\psi_{S_i}\,ds=\int_{\mathcal{S}_i}\Big(\mathring{\psi}_{S_i}-\psi_{S_i} H_iV_i\Big)\,ds+(\psi_{S_i}\,W_i)\bigg|_{A_i}-(\psi_{S_i}\mathbf{v}_J\cdot\mathbf{t}_{S_i})\bigg|_{J}.
\end{equation}
\subsubsection{Mass Balance}

The mass balance for an arbitrary domain $\Omega$ containing a junction $J$ of $N$ tension chains $\mathcal{S}_i$, where $A_i:=\mathcal{S}_i\cap\partial\Omega$, is 
\begin{equation}
	\frac{d}{dt}\int_{\Omega}\rho\,da+\sum_{i=1}^{N}\frac{d}{dt}\int_{\mathcal{S}_i}\rho_{S_i}\,ds=\int_{\Omega}\Pi\,da+\sum_{i=1}^{N}\int_{\mathcal{S}_i}\Pi_{\rho_{S_i}}\,ds-\int_{\partial\Omega}\mathbf{j}_\rho\cdot\mathbf{n}\,dl-\sum_{i=1}^{N}\mathbf{j}_{\rho_{S_i}}\cdot\mathbf{t}_{S_i}\bigg|_{A_i}+\sum_{i=1}^{N}(\rho_{S_i} W_i)\bigg|_{A_i}
\end{equation}

Using the divergence and transport theorems for bulk and singular fields, the above equation yields
\begin{subequations}
	\begin{align}
		&\int_{\Omega}\dot{\rho}\,da-\sum_{i=1}^{N}\int_{\mathcal{S}_i}V_i\,\llbracket\rho\rrbracket_i\,ds-\lim_{\varepsilon\to 0}\int_{\partial B_\varepsilon}\rho\,\mathbf{v}_J\cdot\mathbf{n}_J\,dl\nonumber\\
		&+\sum_{i=1}^{N}\Bigg[\int_{\mathcal{S}_i}\Big(\mathring{\rho}_{S_i}-\rho_{S_i} H_iV_i\Big)\,ds-(\rho_{S_i}\mathbf{v}_J\cdot\mathbf{t}_{S_i})\bigg|_{J}\Bigg]=\nonumber\\
		&\int_{\Omega}\Pi\,da+\sum_{i=1}^{N}\int_{\mathcal{S}_i}\Pi_{\rho_{S_i}}\,ds\nonumber\\
		&-\int_{\Omega}\nabla\cdot\mathbf{j}_\rho\,da-\sum_{i=1}^{N}\int_{\mathcal{S}_i}\llbracket\mathbf{j}_\rho\rrbracket\cdot\mathbf{n}_{S_i}\,ds-\lim_{\varepsilon\to 0}\int_{\partial B_{\varepsilon}}\mathbf{j}_\rho\cdot\mathbf{n}_J\,dl\nonumber\\
		&-\int_{\mathcal{S}_i}\Big(\nabla^S\cdot\mathbf{j}_{S_i}+H_i\,\mathbf{j}_{\rho_{S_i}}\cdot\mathbf{n}_{S_i}\Big)\,ds-\sum_{i=1}^{N}\mathbf{j}_{\rho_{S_i}}\cdot\mathbf{t}_{S_i}\bigg|_{J}.
	\end{align}
\end{subequations}

Localizing, we obtain
\begin{subequations}
\begin{align}
	&\dot{\rho}=-\nabla\cdot\mathbf{j}_\rho+\Pi~~~\text{in}~~\Omega,\\
	&\mathring{\rho}_{S_i}-\rho_{S_i} H_iV_i-V_i\,\llbracket\rho\rrbracket_i=-\llbracket\mathbf{j}_\rho\rrbracket\cdot\mathbf{n}_{S_i}-\Big(\nabla^S\cdot\mathbf{j}_{\rho_{S_i}}+H_i\,\mathbf{j}_{\rho_{S_i}}\cdot\mathbf{n}_{S_i}\Big)+\Pi_{\rho_{S_i}}~~~\text{on}~~\mathcal{S}_i,\\
	&\lim_{\varepsilon\to 0}\int_{\partial B_\varepsilon}\rho\,\mathbf{v}_J\cdot\mathbf{n}_J\,dl+\sum_{i=i}^{N}(\rho_{S_i}\mathbf{v}_J\cdot\mathbf{t}_{S_i})\bigg|_{J}=\lim_{\varepsilon\to 0}\int_{\partial B_{\varepsilon}}\mathbf{j}_\rho\cdot\mathbf{n}_J\,dl+\sum_{i=1}^{N}\mathbf{j}_{\rho_{S_i}}\cdot\mathbf{t}_{S_i}\bigg|_{J}~~~~\text{at}~~J.
\end{align}
\end{subequations}

\subsubsection{Force Balance}

 Neglecting inertia, the global force balance for an arbitrary fixed region $\Omega$ containing a junction is
\begin{equation}
\int_{\partial\Omega}\boldsymbol{\sigma}\mathbf{n}\,dl+\sum_{i=1}^{N}\boldsymbol{\sigma}_{\mathcal{S}_i}\big|_{A_i}=\int_{\Omega}\Gamma\dot{\boldsymbol{u}}\,da+\sum_{i=1}^{N}\int_{\mathcal{S}_i\setminus J}\Gamma_{S_i}\mathbf{v}_{S_i}\,ds+\Gamma_J\mathbf{v}_J,
\end{equation}
where $\{A_i\}$ are points where the curves $\{\mathcal{S}_i\}$ meet $\partial\Omega$. Using the divergence theorem, the local force balance equations  can be obtained as
\begin{subequations}
	\begin{align}
	& \nabla\cdot\boldsymbol{\sigma}=\Gamma\dot{\boldsymbol{u}}~~~~\text{in}~~\Omega;\label{fb-bulk}\\
	&\partial_{s_i}\boldsymbol{\sigma}_{\mathcal{S}_i}+\llbracket\boldsymbol{\sigma}\rrbracket\,\mathbf{n}_{\mathcal{S}_i}=\Gamma_{S_i}\mathbf{v}_{S_i}~~~~\text{at}~~\mathcal{S}_i,~~i=1,2,3,\ldots,N;\label{fb-line}\\
	&\lim_{\varepsilon\to 0}\int_{\partial B_{\varepsilon}}\boldsymbol{\sigma}\mathbf{n}_J\,dl+\sum_{i=1}^{N}\gamma_i\mathbf{t}_{\mathcal{S}_i}\big|_{J}=\Gamma_J\mathbf{v}_J~~~~\text{at}~~J.\label{fb-junc}
	\end{align}
\end{subequations}
Thus the bulk stress field $\boldsymbol{\sigma}$, besides being discontinuous at the curves $\mathcal{S}_i$, is necessarily singular at the junction $J$. However, this point singularity $\mathbf{f}_J:=\lim_{\varepsilon\to 0}\int_{\partial B_{\varepsilon}}\boldsymbol{\sigma}\mathbf{n}_J\,dl$ is finite; it balances the net  force exerted at the junction by the  tension chains $\mathcal{S}_i$ and the junctional frictional force. Note that if $\boldsymbol{\sigma}$ is isotropic, $\mathbf{f}_J=0$.

The compatibility conditions for velocity at the junction $J$ are \cite{Fischeretal2012}
\begin{equation}
V_i\big|_{J}=\mathbf{v}_J\cdot\mathbf{n}_{\mathcal{S}_i}\big|_{J},~~~i=1,2,\ldots,N.
\end{equation}

 Using force balance equations \eqref{fb-line} and \eqref{fb-junc} in the velocity compatibility condition yields

\begin{equation}
	\frac{\Gamma_J}{\Gamma_{S_i}}\gamma_iH_i(J)=\Bigg(\mathbf{f}_J+\sum_{j=1}^{N}\gamma_j\mathbf{t}_{\mathcal{S}_j}\Bigg)\cdot\mathbf{n}_{\mathcal{S}_i}\Bigg|_{J},~~~~i=1,2,\ldots,N,
\end{equation}
since $\llbracket\boldsymbol{\sigma}\rrbracket\mathbf{n}_{\mathcal{S}_i}\cdot\mathbf{n}_{\mathcal{S}_i}\Bigg|_{J}=0$.
Using the identities $(\mathbf{t}_{\mathcal{S}_j}\cdot\mathbf{n}_{\mathcal{S}_i})\big|_{J}=\sin(\theta_j-\theta_i)$ and $\mathbf{f}_J\cdot\mathbf{n}_{\mathcal{S}_i}\big|_{J}=f_J\sin(\theta_J-\theta_i)$ where $\theta_i$ and $\theta_J$ are the angles that $\mathbf{n}_{\mathcal{S}_j}\big|_{J}$ and $\mathbf{f}_J$, respectively, make with a fixed global axis (such that $(\theta_2-\theta_1)+(\theta_3-\theta_2)+\cdots+(\theta_N-\theta_1)=2\pi$), and $f_J:=|\mathbf{f}_J|$, the last equation reduces to

\begin{equation}
	\frac{\Gamma_J}{\Gamma_{S_i}}\gamma_iH_i(J)=f_J\sin(\theta_J-\theta_i)+\sum_{j=1}^{N}\gamma_j\sin(\theta_j-\theta_i),~~~i=1,2,\ldots,N.
	\label{junccond}
\end{equation}
With the limit $\Gamma_J\to 0$, the above equation is the generalized Young-Dupr\'{e} equation for junctions of force chains embedded in an elastic medium. For a triple junction with the limit $\Gamma_J\to 0$, if the bulk stress is isotropic, i.e., $f_J=0$, then equation \eqref{junccond} becomes the standard Young-Dupr\'{e} equation
\begin{equation}
\frac{\gamma_1}{\sin(\theta_2-\theta_3)}=\frac{\gamma_2}{\sin(\theta_3-\theta_1)}=\frac{\gamma_3}{\sin(\theta_1-\theta_2)}.
\end{equation}

On the other hand, for a triple junction, multiplying  equation \eqref{junccond} with $\gamma_i$ and summing over $i$ gives
\begin{equation}
	\sum_{i=1}^{3}\frac{\Gamma_J}{\Gamma_{S_i}}\gamma^2_iH_i(J)=f_J\sum_{i=1}^{3}\gamma_i\sin(\theta_J-\theta_i).
\end{equation}
This equation puts a constraint on the sign of $H_i(J)$: for isotropic bulk stress (i.e., for fluids), $f_J=0$, we have  $\sum_{i=1}^{3}\frac{\Gamma_J}{\Gamma_{S_i}}\gamma^2_iH_i(J)=0$. Hence, all $H_i(J)$s cannot have the same sign, unless $H_i(J)=0$ for all $i$. For a triple junction with curved chains, the typical configuration will be like Fig.\,\ref{tcjunction}(d); one can have a rotating and moving spiral junction, or a moving polar junction. The estimation of the  linear and angular velocity of the junction require a solution of the full dynamical equations, which we will take up in a later study.



\subsection{Complete Initial-Boundary-Value-Problem in 2D}

Here we specify the complete boundary value problem needed to solve the static and dynamic equations
involving $N$ singular lines, each anchored at the cell boundary, and embedded in 2D.

\subsubsection{Equations in the bulk}
The bulk flux density for $\rho$ and $\phi$  have respective advective and diffusive parts.
\begin{subequations}
	\begin{align}
		&\Gamma\dot{\boldsymbol{u}}=\nabla\cdot\boldsymbol{\sigma},\\
		&\dot{\rho}=-\nabla\cdot(\rho\,\dot{\boldsymbol{u}}-D\nabla\rho)+\Pi_\rho(\rho,\phi,\boldsymbol{\epsilon}),\\
		&\dot{\phi}=-\nabla\cdot(\phi\,\dot{\boldsymbol{u}}-D\nabla\phi)+\Pi_\phi(\rho,\phi,\boldsymbol{\epsilon}),
	\end{align}
\end{subequations}
where $\boldsymbol{\epsilon}=(\nabla\boldsymbol{u}+\nabla\boldsymbol{u}^T)/2$ is the linearized strain tensor, along
with constitutive relation for the bulk stress field
\begin{equation}
	\boldsymbol{\sigma}=\hat{\boldsymbol{\sigma}}(\boldsymbol{\epsilon},\dot{\boldsymbol{\epsilon}},\rho,\phi).
\end{equation}

\subsubsection{Equations at the singular structure}

The singular fields $\rho_S$ and $\phi_S$ have the respective advective and diffusive parts.
\begin{subequations}
	\begin{align}
		& \gamma H+\llbracket\boldsymbol{\sigma}\rrbracket\,\mathbf{n}_S\cdot\mathbf{n}_S=\Gamma_S\mathbf{v}_S\cdot\mathbf{n}_S,\\
		&\nabla^S\gamma+\llbracket\boldsymbol{\sigma}\rrbracket\,\mathbf{n}_S\cdot\mathbf{t}_S=\Gamma_S\mathbf{v}_S\cdot\mathbf{t}_S,\\
		&\mathring{\rho}_S-\rho_S HV-V\,\llbracket\rho\rrbracket=-\llbracket\rho\,\dot{\boldsymbol{u}}-D\nabla\rho\rrbracket\cdot\mathbf{n}_S-\Big(\nabla^S\cdot(\rho_S\,\mathbf{v}_S-D\nabla^S\rho_S)+H\,(\rho_S\,\mathbf{v}_S-D\nabla^S\rho_S)\cdot\mathbf{n}_S\Big)\nonumber\\
		&\hspace{80mm}+\Pi_{\rho_S}(\rho_S,{\epsilon}_S)\\
		&\mathring{\phi}_S-\phi_S HV-V\,\llbracket\phi\rrbracket=-\llbracket\phi\,\dot{\boldsymbol{u}}-D\nabla\phi\rrbracket\cdot\mathbf{n}_S-\Big(\nabla^S\cdot(\phi_S\,\mathbf{v}_S-D\nabla^S\phi_S)+H\,(\phi_S\,\mathbf{v}_S-D\nabla^S\phi_S)\cdot\mathbf{n}_S\Big)\nonumber\\
		&\hspace{80mm}+\Pi_{\phi_S}(\phi_S,{\epsilon}_S),
	\end{align}
\end{subequations}

where
\begin{equation}
	\mathbf{v}_S:=\langle\dot{\boldsymbol{u}}\rangle+V(\mathbf{I}+\langle\nabla\boldsymbol{u}\rangle)\mathbf{n}_S.
\end{equation}

With this, we impose the compatibility conditions
\begin{subequations}
	\begin{align}
		&\llbracket\nabla\boldsymbol{u}\rrbracket =\boldsymbol{a}\otimes\mathbf{n}_S,\label{comp11}\\
		&\llbracket\dot{\boldsymbol{u}}\rrbracket +V\,\boldsymbol{a}=\mathbf{0}.\label{comp21}
	\end{align}
\end{subequations}
and announce the constitutive relation for the singular structure
\begin{equation}
	\gamma=\hat{\gamma}(\lambda_S,\theta,\rho_S,\phi_S,\boldsymbol{a},V).
\end{equation}
If $\gamma$ depends on the slope $\theta$, then we call $\gamma$ anisotropic. In that case, the tension chain $\mathcal{S}$ may have corners.

\subsubsection{Boundary Anchoring Conditions}

At the anchoring point $A_j$ on the boundary, we have mass balance equations

\begin{equation}
    \lim_{\varepsilon\to 0}\int_{\partial B_{\varepsilon}}\mathbf{j}_\rho\cdot\mathbf{n}_{A_j}\,dl+\sum_{i=1}^{N}\mathbf{j}_{\rho_{S_i}}\cdot\mathbf{t}_{S_i}\bigg|_{A_j}=\mathbf{j}_{\rho_{A_j}}\cdot\mathbf{n}_{A_j},~~~~\lim_{\varepsilon\to 0}\int_{\partial B_{\varepsilon}}\mathbf{j}_\phi\cdot\mathbf{n}_{A_j}\,dl+\sum_{i=1}^{N}\mathbf{j}_{\phi{S_i}}\cdot\mathbf{t}_{S_i}\bigg|_{A_j}=\mathbf{j}_{\phi{A_j}}\cdot\mathbf{n}_{A_j};
\end{equation}
and the force balance equation
\begin{equation}
    \lim_{\varepsilon\to 0}\int_{\partial B_{\varepsilon}}\boldsymbol{\sigma}\mathbf{n}_{A_j}\,dl+\sum_{i=1}^{N}\gamma_i\mathbf{t}_{\mathcal{S}_i}\big|_{A_j}+\gamma_{\partial \mathcal{B}}\mathbf{t}_{\partial\mathcal{B}}=\mathbf{P}_{A_j}.
\end{equation}

At $A_i$, we impose the anchoring conditions $\boldsymbol{u}=0$, $\nabla\rho\cdot\mathbf{n}=0$ and $\nabla\phi\cdot\mathbf{n}=0$. On $\partial\mathcal{B}\setminus\cup_iA_i$, we impose $\boldsymbol{\sigma}\mathbf{m}=\mathbf{0}$, $\nabla\rho\cdot\mathbf{n}=\rho'_{A_i}$ and $\nabla\phi\cdot\mathbf{n}=\phi'_{A_i}$.

\subsubsection{Initial Conditions}
The initial conditions $\boldsymbol{u}(\mathbf{x},0)$, $\rho(\mathbf{x},0)$ and $\phi(\mathbf{x},0)$ are white noise about the homogeneous unstrained uniform steady state which respect the boundary conditions.





\section{Segregation with Wetting at a Substrate}

In this section, we discuss the segregation of a mixture of stresslets in the vicinity of a substrate (line or surface/membrane) located at $x=0$, when one of the species preferentially wets the substrate. We model this by introducing an attractive Lennard-Jones potential for the species that preferentially wets the substrate. The dynamics of the mixture of stresslets becomes
\begin{subequations}
	\begin{align}
		&\dot{\rho}_1+\partial_{x}(\rho_1\,v)=D\,\partial^2_{xx}\rho_1-k^b_1(x)\,\rho_a-k^u_1(\epsilon)\,\rho_1\\
		&\dot{\rho}_2+\partial_{x}(\rho_2\,v)=D\,\partial^2_{xx}\rho_2-k^b_2\,\rho_a-k^u_2(\epsilon)\,\rho_2
	\end{align}
\end{subequations}
where $k^b_1(x)=k^b_{10}\,f(x)$,  $f$ is a Lennard-Jones wall potential at $x=0$; and $\rho_a=\rho_a^0-(C/A)\epsilon$ as before. We see that
\begin{subequations}
	\begin{align}
		&k^b_{\text{avg}}(x)=\frac{k^b_1(x)+k^b_2}{2}=\frac{k^b_{10}+k^b_2}{2}+\frac{k^b_{10}}{2}\bigg(f(x)-1\bigg),\\
		&k^b_{\text{rel}}(x)=\frac{k^b_1(x)-k^b_2}{2}=\frac{k^b_{10}-k^b_2}{2}+\frac{k^b_{10}}{2}\bigg(f(x)-1\bigg).
	\end{align}
\end{subequations}
We redefine the characteristic time scale by $t^\star:=\frac{2}{k^b_{10}+k^b_2}$, and let $k^{b}_{\text{rel}}:=t^\star\frac{k^b_{10}-k^b_2}{2}$. Other non-dimensionalizations remain the same. Define $p(x):=\frac{k^b_{10}}{k^b_{10}+k^b_2}\bigg(f(x)-1\bigg)$.  

With this, the non-dimensionalized governing equations, for elastic interaction between the stresslets embedded in an elastomer, become
\begin{subequations}
	\begin{align}
		& v=\dot{u} = \partial_x\,\sigma,~~\epsilon=\partial_{x}u \label{u-nondim-wet}\\
		& \dot{\rho}+\partial_x(\rho\,\dot{u})=D\,\partial^2_{xx} \rho +\bigg(1 -\frac{C}{A}\,\epsilon\bigg)\bigg(1+p(x)\bigg)\nonumber\\ &\hspace{45mm}-(k_1+k_3\,\epsilon+\cdots)\,\rho -(k_2+k_4\,\epsilon+\cdots)\,\phi, \label{rho-nondim-wet}\\
		& \dot{\phi}+\partial_x(\phi\,\dot{u})=D\,\partial^2_{xx} \phi + \bigg(1-\frac{C}{A}\,\epsilon\bigg)\bigg(k^{b}_{\text{rel}}+p(x)\bigg) \nonumber\\
		&\hspace{45mm}- (k_1+k_3\,\epsilon+\cdots)\,\phi - (k_2+k_4\,\epsilon+\cdots)\,\rho. \label{phi-nondim-wet}
	\end{align}
	\label{dyn-nondim-wet}
\end{subequations}

 We integrate this system of equations numerically to show the preferred localization of the stresslets near the substrate at $x=0$. 

When the {\it stronger} contractile stresslet $\rho_1$ interacts with the substrate via the attractive potential, and when both stresslets show {\it catch-bond} response, then both the species adhere onto the subtrate, with their 
density peaks co-localizing at the substrate. This is displayed in \href{https://drive.google.com/file/d/1YTP1iSPCpkZuJG1GfY-_M6ORvpvc1n45/view?usp=sharing}{Movie S8}.

On the other hand, when the {\it weaker} contractile stresslet $\rho_2$ interacts with the substrate via the attractive potential, and when both stresslets show {\it slip-bond} response, then only the weaker stresslet adheres onto the subtrate, leading to a distinct stratification at the substrate. This is displayed in \href{https://drive.google.com/file/d/1NAcaYcbJCWnnJa0AjEAuaB6D47Ndi-QX/view?usp=sharing}{Movie S9}.

Together this implies that the presence of a substrate(s) with preferential binding automatically leads to differential localisation and stratification.
\\

The above analysis goes through when we take the mixture of stresslets to be embedded in a fluid. The analysis follows along the lines outlined in Sect.\,\ref{sect:singlefluid}. 
For fluidic interaction between the stresslets, $v$ is the fluid velocity which, at the stationary state, follows
\begin{equation}
	\eta\,\partial_x v-\Pi(\rho_a,\rho_1,\rho_2)=0
\end{equation}
where $\Pi(\rho_a,\rho_1,\rho_2)$ is the osmotic pressure consisting of the passive contribution of the actin filaments and the individual active contributions of the non-interacting stresslets: $\Pi(\rho_a,\rho_1,\rho_2)=\Pi(\rho_a)+\zeta_{\text{avg}}(\rho_1)+\zeta_{\text{rel}}(\rho_2)$. For large contractile activity, $\Pi(\rho_a,\rho_1,\rho_2)\approx \zeta_{\text{avg}}\,(\rho_1)^3+\zeta'_1\,(\rho_1)^4+\zeta_{\text{rel}}\,(\rho_2)^3+\zeta'_2\,(\rho_2)^4$, with $\zeta_{\text{avg}},\zeta_{\text{rel}}<0$. The non-monotonic $\Pi(\rho_a,\rho_1,\rho_2)$ is the route to clustering of $\rho_1$ near $x=0$,  leading to a stratification at the substrate.




%
\newpage
\section{Movie Captions}
\vspace{10mm}

\subsection{Segregation of a binary mixture of stresslets with density peak co-localization} 

\noindent
\href{https://drive.google.com/file/d/137avxjvNrZ2WnVWZMiJcew6dv_PjA2r2/view?usp=sharing}{Movie S1.}
In this movie we observe segregation, followed by the formation of singularities in the density profiles $\rho_1$ and $\rho_2$ of the stresslets in finite time, where the density peaks co-localize, starting from a small perturbation about the homogeneous unstrained uniform state. Both stresslets are of catch bond type. The parameters are set at
$B=10$, $C=1$, $D=1$, $k_1=1$, $k_2=0$, $k^b_{\text{rel}}=0$, $k_3=1$, $\zeta_{\text{avg}}=3$, $\zeta_{\text{rel}}=2$, $k_4=2$, $\chi(\rho_a^0)=1$, $\chi'(\rho_a^0)=0.01$, $\chi''(\rho_a^0)=0.001$, $\chi'''(\rho_a^0)=-0.001$, $\eta= 1$.

\vspace{10mm}

\subsection{Segregation of a binary mixture of stresslets with density peak separation} 

\noindent
\href{https://drive.google.com/file/d/1SGvCRVNLZQ_KcbroFJ4WCzA07wu7iXje/view?usp=sharing}{Movie S2.}
In this movie we observe segregation, followed by the formation of singularities in the density profiles $\rho_1$ and $\rho_2$ of the stresslets in finite time, where the density peaks separate, starting from a small perturbation about the homogeneous unstrained uniform state. Both stresslets are of slip bond type. The parameters are set at
$B=10$, $C=1$, $D=1$, $k_1=1$, $k_2=0$, $k^b_{\text{rel}}=0$, $k_3=-2$, $\zeta_{\text{avg}}=2.5$, $\zeta_{\text{rel}}=-4.15$, $k_4=-4.8$, $\chi(\rho_a^0)=1$, $\chi'(\rho_a^0)=0.01$, $\chi''(\rho_a^0)=0.001$, $\chi'''(\rho_a^0)=-0.001$, $\eta=1.1$.

\vspace{10mm}

\subsection{Travelling wave in a binary mixture of stresslets}

\noindent
\href{https://drive.google.com/file/d/1taNDBEVuwdG5PSU6NgP_QmAbmG9t508w/view?usp=sharing}{Movie S3.}
In this movie we see a travelling wave train propagating towards the left in a binary mixture of stresslets. The profiles of the compartments are asymmetric; propagation direction is towards the higher slope of the density profiles. The parameters are set at
$B=4$, $C=1$, $D=1$, $k_1=1$, $k_2=0$, $k^b_{\text{rel}}=0$, $k_3=1$, $\zeta_{\text{avg}}=2.4$, $\zeta_{\text{rel}}=-2$, $k_4=3$, $\chi(\rho_a^0)=1$, $\chi'(\rho_a^0)=0.01$, $\chi''(\rho_a^0)=0.001$, $\chi'''(\rho_a^0)=-0.001$.

\vspace{10mm}

\subsection{Swapping in a binary mixture of stresslets}

\noindent
\href{https://drive.google.com/file/d/1PYB37ATjee2QveQQ05l4hUCd1XBvqUmP/view?usp=sharing}{Movie S4.}
In this movie we see swapping of the two stresslets in a mixture. The profiles of the pulses are symmetric; there is a small to and fro oscillation of the pulsatile  profiles. The parameters are set at
$B=4$, $C=1$, $D=1$, $k_1=1$, $k_2=0$, $k^b_{\text{rel}}=0$, $k_3=1$, $\zeta_{\text{avg}}=2.2$, $\zeta_{\text{rel}}=-2$, $k_4=3$, $\chi(\rho_a^0)=1$, $\chi'(\rho_a^0)=0.01$,  $\chi''(\rho_a^0)=0.001$, $\chi'''(\rho_a^0)=-0.001$.

\vspace{10mm}

\subsection{Temporal coexistence of swap and travelling wave in a binary mixture of stresslets}

\noindent
\href{https://drive.google.com/file/d/1VXxqBFuo5Deq1-xESXc-qcML_YBWR6Dc/view?usp=sharing}{Movie S5.}
In this movie we see a temporal coexistence of the swap and the travelling wave phase. The slow to and fro oscillations of the symmetric pulses observed at earlier times, transitions to a travelling state where the whole wave train moves to the left (in the direction set by the higher slope of the density profile). This gives rise to a pulsatile travelling wave phase. The parameters are set at
$B=4$, $C=1$, $D=1$, $k_1=1$, $k_2=0$, $k^b_{\text{rel}}=0$, $k_3=1$, $\zeta_{\text{avg}}=2.35$, $\zeta_{\text{rel}}=-2$, $k_4=3$, $\chi(\rho_a^0)=1$, $\chi'(\rho_a^0)=0.01$,  $\chi''(\rho_a^0)=0.001$, $\chi'''(\rho_a^0)=-0.001$.

\vspace{10mm}

\subsection{Segregation of single stresslet on elastomer}

\noindent
\href{https://drive.google.com/file/d/1KDbiGwtyOtWpiFQGyXurWQ44JwraGiT0/view?usp=sharing}{Movie S6.}
In this movie we see segregation between high density and low density phases of a single species of stresslets on an elastomer. The high density regime eventually forms singular structures. The parameters are set at
$B=6$, $C=1$, $D=1$, $k_u=1$, $\zeta=2$, $\chi(\rho_a^0)=1$, $\chi'(\rho_a^0)=1$, $\chi''(\rho_a^0)=0.001$, $\chi'''(\rho_a^0)=-0.001$, $\eta=1$, $\alpha=3$.

\vspace{10mm}

\subsection{Segregation of single stresslet in fluid}

\noindent
\href{https://drive.google.com/file/d/1pp3cNh6lbD8cQk3Kwiox_jqrvffA9n8i/view?usp=sharing}{Movie S7.}
In this movie we see segregation between high density and low density phases of a single species of stresslets in a fluid. The high density regime eventually forms singular structures. The parameters are set at
$D=1$, $\zeta_{\text{avg}}=19.76$, $\zeta_{\text{rel}}= -0.54$, $\zeta_3= 0.64$, $k_{1}^b=2$, $k_{1}^u= 0.5$, $\eta=1$.

\vspace{10mm}
\subsection{Segregation of contractile stresslets with the stronger wetting a substrate}

\noindent
\href{https://drive.google.com/file/d/1YTP1iSPCpkZuJG1GfY-_M6ORvpvc1n45/view?usp=sharing}{Movie S8.}
In this movie we see segregation of two species of contractile stresslets with stronger stresslet interacting with the substrate with a Lennard-Jones wetting potential ($f(x)$). Because of the catch-bond response of the two stresslets, their density peaks co-localize. The parameters are set at
$B=10$, $C=1$, $D=1$, $k_1=1$, $k_2=0$, $k^b_{\text{rel}}=0$, $k_3=1$, $\zeta_{\text{avg}}=3$, $\zeta_{\text{rel}}=2$, $k_4=2$, $\chi(\rho_a^0)=1$, $\chi'(\rho_a^0)=0.01$, $\chi''(\rho_a^0)=0.001$, $\chi'''(\rho_a^0)=-0.001$, $\eta= 1$, and $f(x)= \frac{1}{2}(\frac{1}{x^{12}}-\frac{1}{x^6})$.

\vspace{10mm}
\subsection{Segregation of contractile stresslets with the weaker wetting a substrate: stratification}

\noindent
\href{https://drive.google.com/file/d/1NAcaYcbJCWnnJa0AjEAuaB6D47Ndi-QX/view?usp=sharing}{Movie S9.}
In this movie we see segregation of two species of contractile stresslets with weaker stresslet interacting with the substrate with a Lennard-Jones wetting potential ($f(x)$). Because of the slip-bond response of the two stresslets, their density peaks separate. The parameters are set at
$B=10$, $C=1$, $D=1$, $k_1=1$, $k_2=0$, $k^b_{\text{rel}}=0$, $k_3=-2$, $\zeta_{\text{avg}}=2.5$, $\zeta_{\text{rel}}=-4.7$, $k_4=-4.8$, $\chi(\rho_a^0)=1$, $\chi'(\rho_a^0)=0.01$, $\chi''(\rho_a^0)=0.001$, $\chi'''(\rho_a^0)=-0.001$, $\eta= 1.1$, and $f(x)= \frac{1}{2}(\frac{1}{x^{12}}-\frac{1}{x^6})$.




